*Manuscript accepted for publication in Meteoritics & Planetary Science (2019)*

# The Sariçiçek howardite fall in Turkey: Source crater of HED meteorites on Vesta and impact risk of Vestoids

Ozan UNSALAN[1,2], Peter JENNISKENS[3,4,*], Qing-Zhu YIN[5], Ersin KAYGISIZ[1], Jim ALBERS[3], David L. CLARK[6], Mikael GRANVIK[7,8], Iskender DEMIRKOL[9], Ibrahim Y. ERDOGAN[9], Aydin S. BENGU[9], Mehmet E. ÖZEL[10], Zahide TERZIOGLU[11], Nayeob GI[6], Peter BROWN[6], Esref YALCINKAYA[12], Tuğba TEMEL[1], Dinesh K. PRABHU[4,13], Darrel K. ROBERTSON[4,14], Mark BOSLOUGH[15], Daniel R. OSTROWSKI[4,16], Jamie KIMBERLEY[17], Selman ER[12], Douglas J. ROWLAND[5], Kathryn L. BRYSON[4,16], Cisem ALTUNAYAR-UNSALAN[2], Bogdan RANGUELOV[18], Alexander KARAMANOV[18], Dragomir TATCHEV[18], Özlem KOCAHAN[19], Michael I. OSHTRAKH[20], Alevtina A. MAKSIMOVA[20], Maxim S. KARABANALOV[20], Kenneth L. VEROSUB[5], Emily LEVIN[5], Ibrahim UYSAL[21], Viktor HOFFMANN[22,23], Takahiro HIROI[24], Vishnu REDDY[25], Gulce O. ILDIZ[26], Olcay BOLUKBASI[1], Michael E. ZOLENSKY[27], Rupert HOCHLEITNER[28], Melanie KALIWODA[28], Sinan ÖNGEN[12], Rui FAUSTO[29], Bernardo A. NOGUEIRA[29], Andrey V. CHUKIN[20], Daniela KARASHANOVA[30], Vladimir A. SEMIONKIN[20], Mehmet YEŞILTAŞ[31,32], Timothy GLOTCH[32], Ayberk YILMAZ[1], Jon M. FRIEDRICH[33,34], Matthew E. SANBORN[5], Magdalena HUYSKENS[5], Karen ZIEGLER[35], Curtis D. WILLIAMS[5], Maria SCHÖNBÄCHLER[36], Kerstin BAUER[36], Matthias M. M. MEIER[36], Colin MADEN[36], Henner BUSEMANN[36], Kees C. WELTEN[37], Marc W. CAFFEE[38], Matthias LAUBENSTEIN[39], Qin ZHOU[40], Qiu-Li LI[41], Xian-Hua LI[41], Yu LIU[41], Guo-Qiang TANG[41], Derek W. G. SEARS[16,4], Hannah L. MCLAIN[42], Jason P. DWORKIN[43], Jamie E. ELSILA[43], Daniel P. GLAVIN[43], Philippe SCHMITT-KOPPLIN[44,45], Alexander RUF[44,45], Lucille LE CORRE[25], & Nico SCHMEDEMANN[46]
(The Sariçiçek Meteorite Consortium)

[1] University of Istanbul, 34134 Vezneciler, Fatih, Istanbul, Turkey.

[2] Ege University, 35100 Bornova, Izmir, Turkey.

[3] SETI Institute, Mountain View, CA 94043, USA.

[4] NASA Ames Research Center, Moffett Field, CA 94035, USA.

[5] University of California at Davis, Davis, CA 95616, USA.

[6] Western University, London, Ontario, N6A 5B7, Canada.

[7] University of Helsinki, FI-00014 Helsinki, Finland.

[8] Luleå University of Technology, S-98128 Kiruna, Sweden.

[9] Bingöl University, 12000 Bingöl, Turkey.

[10] Fatih Sultan Mehmet Vakif University, Halicioglu, 34445 Istanbul, Turkey.

[11] Ankara University, 06100 Tandogan, Ankara, Turkey.

[12] Istanbul University Cerrahpasa, 34320 Avcilar, Istanbul, Turkey.

[13]Analytical Mechanics Associates Inc.

[14] Science & Technology Corp.

[15] Sandia National Laboratories, Albuquerque, NM 87185-130, USA.

[16] Bay Area Environmental Research Institute, Petaluma, CA 94952, USA.

[17] New Mexico Institute of Mining and Technology, Socorro, NM 87801-4796, USA.

[18] Institute of Physical Chemistry, B. A. S., 1113 Sofia, Bulgaria.

[19] Namik Kemal University, 59030 Merkez, Tekirdağ, Turkey.

[20] Ural Federal University, Ekaterinburg, 620002, Russian Federation.

[21] Karadeniz Technical University, 61080 Trabzon, Turkey.

[22] University of Munich, D-80333 Munich, Germany.

[23] University of Tübingen, D-72076 Tübingen, Germany.

[24] Brown University, Providence, RI 02912, USA.

[25] Planetary Science Institute, Tucson, AZ 85719, USA.

[26] Istanbul Kultur University, 34156 Bakirkoy, Istanbul, Turkey.

[27] NASA Johnson Space Center, Houston, TX 77058, USA.

[28] Mineralogical State Collection Munich (SNSB), D-80333 Munich, Germany.

[29] CQC, Dep. of Chemistry, University of Coimbra, P-3004-535 Coimbra, Portugal.

[30] Institute of Optical Materials and Technologies, B. A. S., Sofia 1113, Bulgaria.

[31] Kirklareli University, 39100 Kirklareli, Turkey.

[32] Stony Book University, Stony Brook, NY 11794, USA.

[33] Fordham University, Bronx, NY 10458, USA.

[34] American Museum of Natural History, New York, NY 10024, USA.

[35] University of New Mexico, Albuquerque, NM 87131, USA.

[36] ETH Zürich, CH-8092 Zürich, Switzerland.

[37] University of California Berkeley, Berkeley, CA 94720, USA.

[38] Purdue University, West Lafayette, IN 47907, USA.

[39] Istituto Nazionale di Fisica Nucleare, Laboratori Nazionali del Gran Sasso, I-67100 Assergi (AQ), Italy.

[40] National Astronomical Observatories, C. A. S., Beijing 100012, China.

[41] State Key Laboratory of Lithospheric Evolution, C. A. S., Beijing 100029, China.

[42] Catholic University of America, Washington, DC 20064, USA.

[43] NASA Goddard Space Flight Center, Greenbelt, MD 20771, USA.

[44] Helmholtz Zentrum München, D-85764 Neuherberg, Germany.

[45] Technische Universität München, D-85354 Freising-Weihenstephan, Germany.

[46] Freie Universität Berlin, D-12249 Berlin, Germany.

*Corresponding author. Email: Petrus.M.Jenniskens@nasa.gov

**Abstract** – The Sariçiçek howardite meteorite shower consisting of 343 documented stones occurred on 2 September 2015 in Turkey and is the first documented howardite fall. Cosmogenic isotopes show that Sariçiçek experienced a complex cosmic ray exposure history, exposed during ~12–14 Ma in a regolith near the surface of a parent asteroid, and that an ~1 m sized meteoroid was launched by an impact 22 ± 2 Ma ago to Earth (as did one third of all HED meteorites). SIMS dating of zircon and baddeleyite yielded 4550.4 ± 2.5 Ma and 4553 ± 8.8 Ma crystallization ages for the basaltic magma clasts. The apatite U-Pb age of 4525 ± 17 Ma, K-Ar age of ~3.9 Ga, and the U,Th-He ages of 1.8 ± 0.7 and 2.6 ± 0.3 Ga are interpreted to represent thermal metamorphic and impact-related resetting ages, respectively. Petrographic, geochemical and O-, Cr- and Ti-isotopic studies confirm that Sariçiçek belongs to the normal clan of HED meteorites. Petrographic observations and analysis of organic material indicate a small portion of carbonaceous chondrite material in the Sariçiçek regolith and organic contamination of the meteorite after a few days on soil. Video observations of the fall show an atmospheric entry at 17.3 ± 0.8 $kms^{-1}$ from NW, fragmentations at 37, 33, 31 and 27 km altitude, and provide a pre-atmospheric orbit that is the first dynamical link between the normal HED meteorite clan and the inner Main Belt. Spectral data indicate the similarity of Sariçiçek with the Vesta asteroid family (V-class) spectra, a group of asteroids stretching to delivery resonances, which includes (4) Vesta. Dynamical modeling of meteoroid delivery to Earth shows that the complete disruption of a ~1 km sized Vesta family asteroid or a ~10 km sized impact crater on Vesta is required to provide sufficient meteoroids ≤4 m in size to account for the influx of meteorites from this HED clan. The 16.7 km diameter Antonia impact crater on Vesta was formed on terrain of the same age as given by the $^{4}He$ retention age of Sariçiçek. Lunar scaling for crater production to crater counts of its ejecta blanket show it was formed ~22 Ma ago.

# INTRODUCTION

The link from asteroid (4) Vesta to howardite-eucrite-diogenite (HED) meteorites has been thoroughly covered in the literature since the work by McCord et al. (1970) demonstrated a shared (V-class) visible-to-near-infrared spectrum. Since that time, studies of HED's petrographical and geochemical properties support an origin from a Vesta-like protoplanet (e.g., Consolmagno and Drake 1977; Mittlefehldt 2015). More recently, the Dawn mission (especially the Gamma Ray and Neutron Detector instrument spectroscopic results of Prettyman et al. 2015), further cemented the link between the main group (normal) HEDs and Vesta by showing good agreement in the concentration of K and Th within Vesta's regolith to that of eucrite-rich howardites.

Most normal HED meteorites fall from 0.1 to 4 m sized meteoroids that were excavated in an impact in the last ~100 Ma. The cosmic-ray exposure age distribution of HED meteorites is broad (Eugster and Michel 1995), meaning more than one collision is responsible for the meteorites collected at Earth. However, about one-third of all measured non-anomalous HED meteorites have a distinct cosmic-ray exposure age of 22 Ma (Llorca et al. 2009; Welten et al. 2012; Cartwright et al. 2014).

The source crater of the 22 Ma clan of HED meteorites remains to be identified. That impact may have occurred on Vesta itself. Now Dawn has visited Vesta and mapped in detail a great many craters; efforts have begun to date the most recently formed craters using crater size-frequency distributions on their ejecta blankets. Because the population of small impactors is unknown in the asteroid belt, two different chronology systems have been developed that result in different age estimates (O'Brien et al. 2014; Schedemann et al. 2014).

Alternatively, the impact may have involved the disruption of one of the larger members of Vesta's associated asteroid family (the Vestoids), which likely originated from the impacts that formed the Rheasilvia impact basin and the smaller and partially overlapping (older) Veneneia impact basin (Marchi et al. 2012, Ivanov and Melosh 2013). Astronomers studying the Vesta asteroid family in the 1990's demonstrated that the distribution of V-class asteroids stretches to the 3:1 delivery resonance that can bring V-class asteroids to near-Earth orbit (e.g., Cruikshank et al. 1991; Binzel et al. 2002). The much smaller HED meteoroids are more affected by radiation-driven forces and may follow a different pathway, a pathway that can only be probed from the arrival orbits of documented HED falls.

Not all HED meteorites originate from Vesta or its asteroid family. There is a group of isotopically anomalous eucrites (Sanborn and Yin 2014; Mittlefehldt 2015; Sanborn et al. 2016). One example is the meteorite Bunburra Rockhole (Bland et al. 2009; Benedix et al. 2017), the only documented HED fall with precise orbital information (Spurny et al. 2012). Bunburra Rockhole is an anomalous eucrite with isotopic signatures suggesting it originated from a different source than most HED meteorites (Bland et al. 2009; Benedix et al. 2017).

Well documented HED falls are important also because they shed light on the damage caused by larger Vestoid impacts on Earth. These basaltic achondrites represent a distinctly different type of material than ordinary chondrites. Only ~4% of the 20 m to 2 km sized near-Earth asteroids in danger of impacting Earth are of V-class, but half would have relatively high entry speed and impact energies of 1–1000 Mt (Reddy et al. 2011; Brown et al. 2016).

On 2 September, 2015, an eucrite-rich howardite fell in Turkey. Here, we present results from a consortium study of what proved to be the first documented normal-clan HED meteorite fall. We determined the approach trajectory and orbit of the meteoroid, its size and impact speed, and studied a few of the recovered meteorites in great detail to determine its material properties and collisional history. In this paper, we will focus on results that further the study of the impact risk and the search for the normal 22 Ma clan HED source crater.

## METHODS

A bolide of ~0.13 kT initial kinetic energy was detected by U.S. Government satellite sensors at +39.1ºN and 40.2ºE, near the town of Bingöl, Turkey, at 20:10:30 UTC on September 2, 2015 (https://cneos.jpl.nasa.gov/fireballs). Small meteorites fell on corrugated roofs in the nearby village of Sariçiçek. "Sariçiçek" (Turkish for "Yellow-Flower") is now the official name of the meteorite (Bouvier et al. 2016).

### Field study of the fall

A field expedition to the area was conducted by the University of Istanbul and Bingöl University on Sep. 29 – October 4, 2015. Video camera recordings and data from seismic sensors in the area were used to reconstruct the meteor trajectory and its airburst. The bolide's entry speed and direction were derived from direct imaging of the meteor and shadows cast using techniques described in Popova et al. (2013) and Borovicka et al. (2013). From all available video security footage, seven sites were selected that offered the best calibration opportunities (Table 1), with three redundant sites at Bingöl University and Muş Alparslan University to recognize systematic errors, and two sites at Kiği and Karliova that offered a perpendicular perspective to the line connecting Bingöl and Muş (Fig. 1). This provided six independent pairs of perspectives, from which the uncertainty in the direction of the trajectory and entry speed was determined.

For calibration, suitable sun-shadow images were obtained from those same video cameras at different times in the day. The height of shadow obstacles was measured in the field. In addition, calibration images were taken with a digital camera, from the perspective of the video security cameras, with a number of 50 cm markers scattered in the field of view to assist the correction for perspective. The shadow of the Bingöl rectorate building was traced (Fig. 1A), with azimuth angles of the front tip of the shadow measured relative to the position under the tip of the overhanging building. At the soccer court, the shadow of a fence was traced (Fig. 1B). Uncertainty in determining the exact position on the ground below the lamp head at the third site in Bingöl University (Fig. 1C) proved responsible for a small systematic error of –3º azimuth and +1.2º elevation compared to the other two sites.

Muş Alparslan was far enough from the trajectory to capture the final part of the meteor itself in two videos. The foreground in one street-view scene (Fig. 1D) has significant perspective with nearby buildings and distant lights, and required only a small warp to remove the lens distortions. Star-background images were taken with the digital camera just in front of the video security camera, providing a good match to foreground features. Stars aligned to a precision of 1.6’ observed-calculated (O-C). Before the meteor itself entered the frame, video frames were recorded only every 0.17s, making the earlier illumination of the landscape less suitable for light curve reconstruction. Camera #64 (Fig. 1E) reproduced the star field to 5.4’ O-C, but the observed images of the meteor were in fact internal reflections in the camera or camera housing, not a direct image of the meteor itself. This was confirmed by a third site at Muş based on shadows from a lantern pole (Fig. 1F).

Photographs of the building in Kiği (Fig. 1H) from different viewpoints, as well as the height of the shadow obstacle in Karliova, were provided by the Bingöl video security center. The photographs helped determine the ground-projected point below the building's roof tip. In calibrating the depth scale at Kiği, we took into account that the street slopes down in a direction away from the camera.

At the meteorite fall site near Sariçiçek, a grid search (within the survey bands marked in Fig. 5 below) was conducted perpendicular to the trajectory in the densest part of the meteorite strewn field. Following this demonstration, local inhabitants of Sariçiçek (led by Nezir Ergun) with assistance from Bingöl University staff tracked new meteorite recoveries and collected positional information. In total, 343 meteorites were documented (Table 2).

**Analysis of the meteorites**

Several meteorites were made available for this study. Figure 2 shows Sariçiçek samples SC12 and SC14, which are the main focus of the work presented here. Samples were broken and fragments of each stone were distributed to the international community of researchers participating in this consortium study. Additional samples of Sariçiçek became available later and were used for comparison studies (last column of Table 2). In the remainder of this section, the methods used for each analysis are described in the order in which results are later presented.

At NASA Ames Research Center in Moffett Field, California, first a small tip of SC12 was removed for classification (sample SC12a). Subsequently, the bulk volume density of SC12b and whole stone SC14 were determined using a NextEngine 3D laser scanner. The samples were rotated eight times for a full 360° image, taking 3300 polygons per rotation. The measurement was repeated after rotating the sample 90° to scan the poles, and all were aligned for a full 3D image the grain volume densities were determined with a Quantachrome gas pycnometer, using nitrogen.

Next, SC12 (SC12b) and SC14 were broken to distribute samples in the consortium (Fig. 2). At NASA Ames, the quasistatic compression strength was measured in unconfined compression at ~6 $MPas^{-1}$ (load rate of 33 $Ns^{-1}$), using a SouthWark-Emery Tensile Machine to measure the load at which uncut meteorites developed the first crack. SC12 failed at a load of 218 kg (480 lbs). SC14 already failed at a load of 100 kg (220 lbs), but was further compressed to a load of 530 kg (1170 lbs), creating more fragments (Fig. 2). Aluminum foil between meteorite and press was used to determine the surface area. In the same manner, at the Geology Department of the University of Istanbul, Turkey, other meteorites (SC50, 54, 57 and 239) were broken using a Yüksel Kaya Makina press (model YKM071 and press390 software by Teknodinamik Co.) and a load rate of 100 $Ns^{-1}$

At the New Mexico Institute of Mining and Technology (New Mexico Tech.) in Socorro, New Mexico, small samples of SC12 and SC14 were cut into nominal 5 x 5 x 5 mm cubes. These are smaller than the 10 x 10 x 10 mm samples typically employed, but in this case the material was fine grained and cracked on a small scale. Sample SC14 broke in the final preparation step, but sample SC12 was suitable for measurement. The sample was compressed to failure at a constant displacement rate of 0.01 $mms^{-1}$ (corresponding strain rate $2x10^{-3}$ $s^{-1}$) using a MTS Landmark Load Frame. Images of the sample were recorded during compression to track the evolution of failure in the sample.

The larger Sariçiçek sub-samples of SC12 (SC12b) and SC14 were imaged with high-resolution X-ray computed tomography at the Center for Molecular and Genomic Imaging of the University of California Davis. Each sample arrived with a small fragment broken off (SC12b-a1 and SC12b-a2). SC14-a1 was further broken into two equal pieces that were aligned and imaged together. X-ray tomographic images were obtained on a MicroXCT-200 specimen CT scanner (Carl Zeiss X-ray Microscopy). The CT scanner has a variable X-ray source capable of a voltage range of 20–90 kV with 1–8 W of power. Once the source and detector settings were established, the optimal X-ray filtration was determined by selecting among one of 12 proprietary filters for optimal contrast (90 kV and 88 microAmp). 1600 projections were obtained over a 360° rotation. The camera pixels were binned by 2 to increase signal to noise in the image and the source-detector configuration resulted in a voxel size of 28.3 μm for SC12b-a1, 20.3 μm for SC14-a1, and 5.5 μm for SC12b-a2 and SC14-a2.

The distribution of fracture lengths was measured from the microCT images using the ImageJ software and the Ride Detection plugin (Steger 1998). Fractures are planes. Each microCT scan provided multiple two-dimensional views of the fractures. We assume that fracturing follows the Weibull distribution (Weibull 1951), that they are randomly distributed through the target, and that the likelihood of encountering a fracture increases with distance. This results in a relationship:

$$\sigma_l = \sigma_s(n_s/n_l)^{\alpha}$$

where $\sigma_s$ and $\sigma_l$ refers to stress in the small and large object, $n_s$ and $n_l$ refer to the number of cracks per unit volume of the small and large object, and $\alpha$ is the shape parameter called the Weibull coefficient. A relationship exists between the distributions of measured trace length and actual fracture plane size (Piggott 1997), where the slope of a log–log plot of trace length versus fracture density is proportional to $\alpha$. The value for $\alpha$ remains mostly unknown in meteorites (Asphaug et al. 2002), while terrestrial rocks like concrete, granite and basalt have an $\alpha$ of ~0.20, ~0.16 and ~0.11, respectively.

Petrographic analyses of the small tip broken from sample SC12 and of a subsample of the broken SC14 were carried out using the Cameca SX100 electron microprobe at the E-beam laboratory of the Astromaterials and Exploration Science (ARES) Division, NASA Johnson Space Center in Houston, Texas. A 15 kV focused beam was used, and the following natural mineral standards: kaersutite, chromite, rutile, apatite, rhodonite, troilite, orthoclase and oligoclase. Pure metals were used as standards for Ni and Co.

At the Department of Earth & Environmental Sciences of the University of Munich, Germany, analyses of sample SC182 were carried out using a Cameca SX100 electron microprobe. It was operated at 15 keV acceleration voltage and 20 nA beam current. Synthetic wollastonite (Ca), natural olivine (Fe in silicates, Mg, Si), hematite (Fe in oxides, metals and sulfides), corundum (Al), natural ilmenite (Mn), fluorapatite (P), orthoclase (K), sphalerite (S), synthetic NiO (Ni), synthetic $Cr_2O_3$ (Cr) and albite (Na) were used as standards. A matrix correction was performed by using the PAP procedure (Pouchou and Pichoir 1984).

At the Ural Federal University in Ekaterinburg, Russian Federation, a search of xenolithic clasts in a thin section of SC181 was conducted under normal and polarized light using an Axiovert 40 MAT microscope, and by Scanning Electron Microscopy (SEM) analysis using an Auriga CrossBeam SEM with an X-max 80 energy dispersive X-ray spectroscopy (EDS) device (Oxford Instruments). At Istanbul University, a thin section of SC18 was studied with a Leitz OrthoplanPol optical microscope, while SC18 was studied by SEM at Namik Kemal University in Merkez, Turkey. And at the University of Coimbra, Portugal, a small fragment of SC239 was studied using a Horiba LabRam HR Evolution micro-Raman system, with He-Ne laser excitation at 632.8 nm, spectral resolution 1.5 $cm^{-1}$, and spot size 0.85 micrometer. Spectra were collected with an acquisition time of 20 s, 10 accumulations, and laser power ~4 mW.

The next analysis techniques pertain to composition measurements and are described in more detail. At the University of California at Davis, fusion crust free material was selected from several small ~0.1g fragments of SC12b. A subsample of the crushed, homogenized powder (40.21 mg) was placed into a PTFE Parr bomb along with a mixture of ultraclean concentrated HF and $HNO_3$ acids in a 3:1 ratio. The PTFE bomb was sealed in a stainless steel jacket and heated in a 190°C oven for 96 h to ensure complete dissolution of refractory phases. After 96 h, the resulting solution was dried down and treated with alternating treatments of concentrated $HNO_3$ and 6 N HCl to dissolve any fluorides formed during the dissolution procedure. The resulting sample solution was divided into two aliquots: one for major/minor/trace bulk composition measurements (10% of the sample), and the other for Cr isotopic analysis (90% of the sample). The aliquot for the major/minor/trace bulk composition measurements was brought up in a 2% $HNO_3$ solution and prepared in two dilutions (2000× for trace elements and 40000× for major elements). The sample solutions and calibration standard solutions were spiked with an internal standard composed of Re, In, and Bi to account for drift in the mass spectrometer during the analytical session. As a result, we do not report Re, In and Bi in the meteorite samples. A calibration curve, a fit line of counts per second versus concentration $R^2$ = 0.999 or better was generated for each element using the well-characterized Allende Smithsonian standard reference material to determine abundances of individual elements of Sariçiçek. A separate aliquot of the CM chondrite Murchison was measured as an unknown to check for accuracy during the analytical session. Both Allende and Murchison were processed using the same dissolution and dilution procedures as the Sariçiçek sample. Measurements were made using a Thermo Element XR high-resolution inductively coupled plasma–mass spectrometer (HR-ICP-MS) at UC Davis, at the low, medium or high resolution needed for a particular element.

At Fordham University in Bronx, New York, bulk chemical analysis was conducted on chips and powder of SC14. The material was separated into five individual aliquots for replicate measurements. The mass (mg) of each aliquot are as follows: (n = 5, 137.9, 115.5, 106.4, 106.1, and 92.1). Dissolution and inductively coupled plasma–mass spectrometry (ICPMS) analyses are based on a matrix-matching scheme (Friedrich et al. 2003; Wolf et al. 2012). In short, each sample aliquot was ground to <100 mesh in a clean agate mortar and pestle. Those powders were placed in Teflon bombs with 1 mL HF and 5 mL $HNO_3$ and placed in a microwave digestion system. The resulting solution is taken to incipient dryness in Teflon beakers on a specially constructed drybath incubator at 75 °C. $HClO_4$ is then added and again the solution is taken to incipient dryness at 75 °C. The samples are then taken up to a total of 50 mL of ~1% $HNO_3$ solution after adding internal standards (Be, Rh, In, Tl) used to correct for potential mass-dependent drift during ICP-MS analysis. These solutions were used for trace element analysis; fivefold dilutions of portions of those solutions were used for major element analyses akin to the method of Wolf et al. (2012). A Thermo Scientific X Series II ICP-MS was used for all analyses. During ICP-MS analysis, the Allende Standard Reference Meteorite (Jarosewich et al. 1987), USGS basaltic standards BIR-1 and BCR-1, and the NIST 688 basalt standard were used for an external calibration scheme for quantification of the individual elemental analytes. Standards and procedural blanks were digested using the same method outlined above.

Two samples of Sariçiçek, SC12b and SC14, were analyzed for triple oxygen isotopes at the University of New Mexico in Albuquerque. The two samples were gently crushed with a mortar and pestle. A few fragments of interior material were selected under a stereoscopic microscope to avoid any possible contamination from fusion crust. The bulk fragments were pretreated by an acid-wash with weak HCl and subsequent rinsing in distilled water (removal of possible terrestrial weathering products). One portion of SC14 was not acid-treated, and several subsamples of this portion were also analyzed. Two large feldspar (plagioclase) grains were picked from SC14. Oxygen isotope analyses of several subsamples of the two stones were performed by laser fluorination at UNM (Sharp 1990). Samples were pre-fluorinated ($BrF_5$) in a vacuum chamber in order to clean the stainless steel system and to react residual traces of water or air in the fluorination chamber. Molecular oxygen was released from the samples by laser-assisted fluorination (20W far-infrared $CO_2$ laser) in a $BrF_5$-atmosphere, producing molecular $O_2$ and solid fluorides. Excess $BrF_5$ was then removed from the produced $O_2$ by reaction with hot NaCl. The oxygen was purified by freezing onto a 13Å molecular sieve at –196°C, followed by elution of the $O_2$ from the first sieve at ~300 °C (heat gun) into a He-stream that carries the oxygen through a CG column (separation of $O_2$ and $NF_3$, a possible interference with the $^{17}O$ measurement) to a second 13Å molecular sieve at −196°C. After removal of the He, the $O_2$ is then released directly into a dual inlet isotope ratio mass spectrometer (Thermo Finnigan MAT 253). The oxygen isotope ratios were calibrated against the isotopic composition of San Carlos olivine. Each sample analysis consisted of 20 cycles of sample-standard comparison. Olivine standards (~1–2 mg) were analyzed daily. Oxygen isotopic ratios were calculated using the following procedure: The $\delta^{18}O$ values refer to the per-mil deviation in a sample ($^{18}O/^{16}O$) from SMOW, expressed as $\delta^{18}O$ = $[(^{18}O/^{16}O)sample/(^{18}O/^{16}O)_{SMOW}-1] * 10^3$. The delta values were converted to linearized values by calculating: $\delta^{18}/^{17}O' = \ln([\delta^{18/17}O + 10^3]/10^3) * 10^3$ in order to create straightline mass-fractionation curves. The $\delta^{17}O'$ values were obtained from the linear δ–values by the following relationship: $\delta^{17}O' = \delta^{17}O' - 0.528 * \delta^{18}O'$, $\Delta^{17}O'$ values of zero define the terrestrial mass-fractionation line, and $\Delta^{17}O'$ values deviating from zero indicate mass-independent isotope fractionation. Typical analytical precision of the laser fluorination technique is better than ± 0.02‰ for $\Delta^{17}O'$.

At the University of California Davis, bulk rock powders were generated from a fusion crust free portion of a subsample of Sariçiçek SC12 and the howardite Bholghati by crushing in an agate mortar and pestle. The bulk rock powders were homogenized and an aliquot of 40.21 mg and 15.24 mg were taken of the Sariçiçek and Bholghati powders, respectively. The powders were combined with a 3:1 solution mixture of concentrated HF and $HNO_3$ and sealed in PTFE Parr bomb capsules within stainless steel jackets. The Parr bombs were heated in a 190 °C oven for 4 days. After dissolution was complete, the solutions were dried down, acid-treated with 6 M HCl and concentrated $HNO_3$ to remove fluorides, then brought up in 1 mL of 6 M HCl. Chromium was separated using a three-column chromatography following a procedure described by Yamakawa et al. (2009). The isotopic composition of the purified Cr separate was determined using a Thermo Triton Plus thermal ionization mass spectrometer at UC Davis. A total of 3 µg of Cr was combined with 3 µl of an Al-boric acid-silica gel activator solution and loaded onto an outgassed W filament. A total of four filaments were prepared for each sample (total Cr load of 12 µg). Each set of four sample filaments were bracketed with two filaments before and after loaded with the NIST SRM 979 terrestial chromium isotopic standard, prepared in the same manner and with the same Cr load as the samples. Each filament analysis was made up of 1200 ratio measurements with an 8 s integration time. A gain calibration was performed at the start of each filament and a baseline was measured every 25 ratios. The amplifiers were rotated between each 25-ratio block to eliminate any bias due to differing cup efficiencies. Instrumental mass fractionation was made using the $^{50}Cr/^{52}Cr$ ratio ($^{50}Cr/^{52}Cr = 0.051859$; Shields et al. 1966) and corrected using the exponential law. The signal intensity for $^{52}Cr$ was set to 10 V (±15%) for the duration of the run.

At ETH Zürich, Switzerland, high precision Ti isotope data were obtained using an ion exchange procedure for Ti separation from the sample matrix followed by measurements on a Neptune MC-ICP-MS. The analytical method follows that of Williams (2015) with a modification based on Zhang et al. (2011). In brief, two subsamples (SC14-Z1 and SC14-Z3) were crushed and dissolved using the Parr Bomb digestion procedure described in Schönbächler et al. (2004). For the first step of the chemical separation, the procedure of Zhang et al. (2011) using TODGA resin was adapted. This was followed by an ion exchange column using anion exchange resin (Bio-Rad AG1-X8), in which the samples are loaded in 4 M HF, followed by matrix elution in 4 M HF, 0.5 M HCl + 0.5 M HF and the collection of Ti in 6 M HCl + 1M HF (Schönbächler et al. 2004; Williams 2015). This column was carried out twice to achieve an improved Ti separation from interfering elements such as Ca, Cr and V. Blanks for the Parr Bomb digestion were 0.69 ng Ti and for the column chemistry 6.65 ng Ti. Considering the total Ti amount in the sample (> 20 µg), the blanks are negligible. The isotopic analyses were performed on a Neptune MC-ICP-MS and corrections for isobaric interferences from Ca, Cr and V on Ti isotopes were applied. The samples were bracketed by an ETH in-house Ti wire standard solution. Each analysis consisted of one block with 40 integrations of 8.39 s for the main and 4.19 s for the second cycle. On-peak background correction was applied and samples were analysed in medium- and high-resolution mode. The measured ratios were internally normalized to $^{49}Ti/^{47}Ti$ =0.749766 (Niederer et al. 1981) and are reported in the epsilon notation (the deviation from the Ti wire standard expressed in parts per $10^4$).

In addition to the Sariçiçek samples processed at ETH Zürich, a separate aliquot was processed at UC Davis. The aliquot was the same sample from which Cr was previously separated. The column separation and mass spectrometry followed the procedures described in Zhang et al. (2011). Titanium isotope ratios were measured on the Thermo Neptune Plus ICP-MS at UC Davis.

At NASA Goddard Space Flight Center in Greenbelt, Maryland, two separate amino acid measurements were made of a ~155 mg aliquot of a crushed fragment of SC12 and a 1.72 g aliquot of SC14. As controls, a 150 mg sample of a pebble collected from the fall location of SC14 and an 830 mg sample of soil collected from the fall location of the Sariçiçek SC16 meteorite were also extracted and analyzed for amino acids. The SC12 meteorite sample and SC14 recovery site pebble were powdered separately in a ceramic mortar and pestle, transferred to a borosilicate glass test tube, flame-sealed with 1 ml of Millipore Milli-Q Integral 10 (18.2 MΩ, < 1 ppb total organic carbon) ultrapure water and heated at 100 °C for 24 h. The soil sample was fine-grained and did not need to be powdered prior to hot water extraction. A procedural blank (glass tube with 1 ml Millipore water) was carried through the identical extraction protocol. After heating, one half of the water extract was transferred to a separate glass tube, dried under vacuum, and the residue subjected to a 6 M HCl acid vapor hydrolysis procedure at 150 °C for 3 h to determine total hydrolyzable amino acid content. The acid-hydrolyzed water extracts were desalted using cation-exchange resin (AG50W-X8, 100–200 mesh, hydrogen form, BIO-RAD), and the amino acids recovered by elution with 2 M $NH_4OH$ (prepared from Millipore water and $NH_3$(g) (AirProducts, in vacuo). The remaining half of each water extract (non-hydrolyzed fraction) was taken through the identical desalting procedure in parallel with the acid-hydrolyzed extracts to determine the free amino acid abundances in the meteorites and soil sample. The amino acids in the $NH_4OH$ eluates were dried under vacuum to remove excess ammonia; the residues were then redissolved in 100 µl of Millipore water, transferred to sterile microcentrifuge tubes, and stored at –20°C prior to analysis. Based on our analysis of amino acid standards taken through the entire extraction and acid hydrolysis procedure, we found no evidence of significant decomposition, racemization, or thermal degradation of the amino acids during the extraction procedure. The amino acids in the $NH_4OH$ eluates were derivatized with *o*-phthaldialdehyde/*N*-acetyl-L-cysteine (OPA/NAC) for 15 min at room temperature. The abundance, distribution, and enantiomeric compositions of the two- to six-carbon aliphatic amino acids present in the non-hydrolyzed and acid-hydrolyzed water extracts of SC12 and controls were then determined by ultra performance liquid chromatography fluorescence detection and time of flight mass spectrometry (hereafter LC-FD/ToF-MS) using a Waters ACQUITY H Class UPLC with fluorescence detector and Waters Xevo G2 XS. The instrument parameters and analy<tical conditions used were similar to those described elsewhere (Glavin et al. 2006, 2010). For the Xevo mass calibrations, an automatically applied lockmass of a fragment of Leucine Enkephalin (278.1141 Da) with a scan time of 1 s every 60 s is used. The capillary voltage was set to 1.2 kV. The amino acids and their enantiomeric ratios were quantified from the peak areas generated from both fluorescence detection and from the mass chromatogram of their OPA/NAC derivatives as described previously (Glavin et al. 2006). The reported amino acid abundances in the Sariçiçek SC12 meteorite sample and controls below are the average value of three separate LC-FD/ToF-MS measurements. The errors given are based on the standard deviation of the average value of three separate measurements.

The concentrations of short-lived cosmogenic radionuclides, as well as long-lived cosmogenic $^{26}Al$ and natural radioactivity, were measured using non-destructive gamma ray spectroscopy. The complete stone SC26 (131.88g) was measured in the STELLA (SubTErranean LowLevel Assay) facility of underground laboratories at the Laboratori Nazionali del Gran Sasso (LNGS) in Italy, using a high-purity germanium (HPGe) detector of 370 $cm^3$ (Arpesella 1996). The counting time was 7.8 days. The counting efficiencies were calculated using a Monte Carlo code. This code was validated through measurements and analyses of samples of well-known radionuclide activities and geometries. The uncertainties in the radionuclide activities are dominated by the uncertainty in the counting efficiency, which is conservatively estimated at 10%. The density and composition were taken from the measurements performed on other specimens of this meteorite and presented in this paper.

For the analysis of the long-lived cosmogenic radionuclides $^{10}$Be (half-life = 1.36 × $10^6$ yr), $^{26}$Al (half-life = 7.05 × $10^5$ yr) and $^{36}$Cl (half-life = 3.01 × $10^5$ yr), samples of 52.0 and 58.5 mg of SC12 and SC14 were dissolved in a mixture of concentrated HF/$HNO_3$ along with ~2.8 mg of Be and ~3.6 mg of Cl carrier. After dissolution, Cl was isolated as AgCl, and the remaining solution was evaporated to dryness. The residue was dissolved in dilute HCl and a small aliquot was taken for chemical analysis by inductively coupled plasma optical emission spectroscopy (ICP-OES) using an iCAP 6300 instrument. The elements Mg, Al, K, Ca, Ti, Mn, Fe, Co, Ni were analysed.

After measuring the Al content of the dissolved sample, we added 5.0 and 5.4 mg of Al carrier to the main solution of SC12 and SC14, respectively. We separated Be and Al using procedures described previously (e.g., Welten et al. 2001, 2012) and measured the concentrations of $^{10}$Be, $^{26}$Al and $^{36}$Cl by accelerator mass spectrometry (AMS) at Purdue University in West Lafayette, Indiana (Sharma et al. 2000). The measured $^{10}$Be/Be, $^{26}$Al/Al and $^{36}$Cl/Cl ratios are corrected for blank levels (which are <1% of the measured values) and normalized to AMS standards (Sharma et al. 1990; Nishiizumi 2004; Nishiizumi et al. 2007).

At the Helmholtz Zentrum München, Germany, an extract of SC12 for negative mode electrospray Fourier transform ion cyclotron resonance mass spectrometry (ESI(-)-FT-ICR-MS, 12 Tesla) analysis was prepared as described previously in Schmitt-Kopplin et al. (2012). Briefly, an intact fragment of about 80 mg weight was first washed with methanol (rapid contact with 1 ml methanol that was subsequently discarded) and immediately crushed in an agate mortar with 0.5 mL of LC/MS grade methanol and further transferred into an Eppendorf tube within an ultrasonic bath for 1 min. The tube was then centrifuged for 3 min. The supernatant (methanolic extract) was directly used for infusion FT-ICR-MS. Three thousand scans were accumulated with 4 million data points. The conversion of the exact masses into elementary composition is based on exact mass differences and shown in more detail in Tziotis et al. (2011). The average mass resolution ranged near 1,000,000 at nominal mass 200, 400,000 at mass 400 and 300,000 at mass 600. Prior the sample extraction, great care was used to clean the agate pillar with solvent in ultrasonic bath. A "blank" sample was produced by following the same extraction procedure without any meteorite fragment, and analysed before and after the meteorite analysis. No significant mass peaks in the mass range of the meteorite extract were observed. In order to fully exploit the advantages of FT-ICR-MS, we routinely control the instrument performance by means of external calibration on arginine clusters prior to any analysis. Relative *m/z* errors were usually < 100 ppb across a range of 150 < *m/z* < 1,500.

At ETH Zurich, noble gases were measured in two samples of SC12 (SC12-Z1 and SC12-Z2) with masses of 44.6 and 92.0 mg, respectively, and two samples of SC14 (SC14-Z2.1 and SC14-Z2.2) with masses of 32.5 and 19.7 mg, respectively. Samples were weighed (uncertainty <0.05 mg), wrapped into Al-foil and loaded into a custom-built single-collector sector-field noble gas mass spectrometer equipped with a Baur–Signer source. The samples were then exposed to ultra-high (~$10^{-10}$ mbar) vacuum for about 2 weeks, before being analyzed according to a protocol described by Meier et al. (2017). Blank contributions to the total signal were negligible (<0.02%) for all He, Ne, and <2% for Ar isotopes.

Before in-situ U-Pb analysis, datable minerals were searched in a polished section of Sariçiçek meteorite SC12a polished mounts. Backscattered electron (BSE) images were obtained by the field emission scanning electron microscope (FESEM) of Carl Zeiss SUPRA-55 at the National Astronomical Observatories (NAO), Chinese Academy of Sciences (CAS) in Beijing, China. U-bearing mineral grains, including zircon, baddeleyite, and apatite, were identified and located with an energy dispersive spectrometer (EDS). Cathodoluminescence and corresponding BSE images for zircon grains were taken by a Nova NanoSEM FESEM at the Institute of Geology and Geophysics (IGG), CAS, in Beijing.

Subsequently, micro-Raman spectra were taken of the identified grains to confirm the mineral assignments. Raman spectra were collected using a laser Raman spectrometer of Horiba LabRaM HR800 connected to a Olympus BX41 microscope at IGG, CAS. The 532 nm wavelength of a solid-state laser was used, with the beam focused on a ~1 micrometer spot. The laser Raman spectrometer was calibrated to the peak at 520 $cm^{-1}$ with a single-crystal silicon standard. Raman spectral mapping scanned from 120 to 800 $cm^{-1}$, covering the characteristic peaks of baddeleyite.

In-situ isotopic analysis of U-Pb was performed on a large-geometry, double-focusing secondary ion mass spectrometer, CAMECA IMS-1280HR ion microprobe at the Institute of Geology and Geophysics of the Chinese Academy of Sciences. A polished section of the SC12 was carbon-coated prior to SIMS analysis. U-Pb dating for zircon and baddeleyite in Sariçiçek was conducted with a small primary beam of $O^-$ with a diameter of ~ 4 × 5 μm under dynamic multi-collector mode, slightly modified from procedure of Liu et al. (2015). The analytical method here is described only briefly. The "oxygen flooding technique" with a working $O_2$ gas pressure of 4~5 $\times 10^{-6}$ Torr was used to greatly enhance Pb ion yield and suppress the baddeleyite crystal orientation effect (Wingate and Compston 2000; Li et al. 2010). The primary ion beam of $O^-$ was accelerated at –13 kV potential, with an intensity of ~ 0.8 nA. Mass resolving power is set at 8000 (50% peak height definition). Before analysis, each spot was pre-sputtered using a ~ 3 nA primary beam on a square area of 25×25 $\mu m^2$ for 120 s to remove the surface contamination and to enhance the secondary ions yield. For zircon and baddeleyite analyses, we used the $^{207}Pb/^{206}Pb$ ratio of M257 zircon standard to calibrate the EM yields. The data acquisition includes five sequences. The $^{90}Zr_2{}^{16}O^+$ was measured as matrix peak. $^{180}Hf\,^{16}O^+$ peak was used for peak centering. $^{204}Pb^+$, $^{206}Pb^+$ and $^{207}Pb^+$ were obtained simultaneously during the third sequence on L2, L1 and C detectors, and $^{238}U^+$, $^{232}Th^{16}O^+$ and $^{238}U^{16}O^+$ in the fourth sequence. $^{238}U^{16}O_2{}^+$ was detected in the final sequence. Each measurement for U-Pb dating consists of seven cycles, taking nearly 14 min. Pb/U fractionation was calibrated with the empirically established power law relationship between $^{206}Pb/^{238}U$ and $^{238}U^{16}O_2/^{238}U$ against standard RM M257 (Nasdala et al. 2008). Uranium concentrations were calibrated against zircon M257 with U ~840 ppm (Nasdala et al. 2008). Correction of common Pb was made by measuring the amount of $^{204}Pb$ and the CDT Pb isotopic compositions ($^{206}Pb/^{204}Pb$ = 9.307, $^{207}Pb/^{206}Pb$ = 1.09861, Tatsumoto et al. 1973).

U-Pb dating was performed for apatites in Sariçiçek with a 20 x 30 μm beam spot size under dynamic multi-collector mode as well. The $O_2^-$ primary ion beam was used with an intensity between 9 and 12 nA. The detector configuration is similar to that of zircon and baddeleyite. The only difference is that $^{40}Ca_3{}^{31}P_2{}^{16}O_2{}^{+}$ peak was used as matrix peak and for peak centering, which was measured in the first sequence. Accurate Pb isotopic composition in NIST610 glass was used to calibrate the relative yields among different electron multipliers. Each measurement for apatite U-Pb dating consists of 10 cycles, taking nearly 18 min.

Pb/U ratios were calibrated with a power law relationship between Pb/U and $UO_2$/U relative to an apatite standard of NW-1 (1160 Ma) that comes from the same complex of Prairie Lake as that of the Sano et al. (1999) apatite standard (PRAP). U concentration is calibrated relative to the Durango apatite which has U ~9 ppm (Trotter and Eggins 2006). Correction of common Pb was made by measuring the amount of $^{204}$Pb and the CDT Pb isotopic compositions (Tatsumoto et al. 1973).

At NASA Ames Research Center, the natural and induced thermoluminescence (TL) were measured using a modified Daybreak Nuclear and Medical Inc. Thermoluminescence Analyzer. One chip of ~40 mg was taken from Sariçiçek SC12, being greater than ~6 mm from clearly visible fusion crust. This was gently crushed, the magnetic fraction removed, and then gently crushed again to produce ~200 μm grains. A 140 Ci $^{90}$Sr beta source was used for the irradiations in the determination of induced TL. Two aliquots, removed from the homogenized powder, each of 4 mg were measured. Natural TL is determined by the "equivalent dose" method since the anomalous fading prevents the use of the better (internally normalized) peak height ratio method. The dose administered (calculated from 25 krad in 1987; Hasan et al. 1987) was 12.74 krad.

At Brown University in Providence, Rhode Island, reflectance spectra measurements were made directly on a fragment of SC12 and on ground material taken from the surface of this sample. Before grounding the sample, any fragments with fusion crusts were separated, and only the interior portions were ground. The ground particulate sample was dry-sieved into three size fractions: <25, <125, and 125–500 μm. Their bidirectional UV-Vis-NIR reflectance spectra were measured at NASA Reflectance Experiment Laboratory (RELAB) from 0.3 to 2.6 μm at every 5 nm under the viewing geometry of 30° incidence and 0° emergence angles while each sample was spun at a rate of 1.5 s/rotation. Biconical Fourier transform reflectance spectra of the same samples were measured from 1.5 to 100 μm and were scaled to and spliced with the UV-Vis-NIR spectra at 2.5 μm. Near-IR absorbance measurements were performed on meteorites numbered SC51, 55, 239 and 327 at the University of Istanbul, Turkey. Each sample was crushed in an agate mortar with pestle to make fine powders. A Nicolet 6700 FT-IR Spectrometer with Nicolet NIR Smart Updrift unit was used, with a spectral resolution of 4 $cm^{-1}$ in the 0.9–2.5 μm wavelength region. For each measurement 256 scans were added.

Finally, the fusion crust and melting properties of SC239 were analyzed at the Laboratories of the Institute of Physical Chemistry and Institute of Optical Materials and Technologies at Bulgarian Academy of Sciences, Sofia, using both SEM (JEOL 6390) and TEM (JEOL 2100). A mesh was placed on a set of SEM images of the crust, which was found rich in bubbles. The center of each bubble was manually identified, after which the diameter and volume of each bubble was calculated. To improve the volume measurements, fragments of SC239 were scanned by X-ray computed tomography (Bruker SkyScan 1272 microtomograph). A larger fragment with fusion crust size 65 × 33 × 27 mm was scanned at voxel size of 4 micron, and a smaller piece with size 1.2 × 1.0 × 0.9 mm was scanned at a voxel size of 0.4 micron. To determine the temperature at which the material started to melt, forming the bottom of fusion crust, one small sample of SC239 was studied by means of in-situ hot stage optical microscopy (a horizontal optical dilatometer model Misura ODLT), by heating the sample at a rate of 5º/min, up to 1593 K and observe the changes in the sample's morphology.

## RESULTS

### Meteoroid and Atmospheric Entry

#### *Trajectory and Orbit*

The results of linear trajectory reconstructions are presented in Table 3 and Fig. 3. The precision of the final trajectory solution was evaluated based on the range of solutions for individual pairs of perspectives and how the solution changed when one of the stations was removed from the combined least-squares fit. Different station combinations showed that the position of the trajectory is uncertain by ±0.6 km for fits assuming a constant speed (an approach least sensitive to random measurement errors near the end of the trajectory). In that case, the direction of the radiant is uncertain by ±0.8º. When we, instead, assume a Jacchia-type deceleration profile along the trajectory (Jacchia et al. 1967), then the entry speed is uncertain by ±0.8 $kms^{-1}$. As a final check, the average speed was calculated for short altitude-sections of the trail (assuming no deceleration in each section) and the result was compared to the velocity fit from all data combined and found in agreement (large open circles compared to other symbols in Fig. 3), except for the first point based on faint shadows seen at Kiği.

The combination of all data provides the apparent radiant position at R.A. = 276.5 ± 1.4º, Decl. = +59.7 ± 0.8º, near the star ξ–Draconis, and apparent entry speed at $V_\infty$ = 17.1 ± 0.8 $kms^{-1}$, assuming a Jacchia et al. (1967) deceleration profile (Fig. 3). If the strongly decelerated final part of the meteor trajectory, captured only in the Muş Alparslan street camera, is ignored, and the speed is assumed constant, instead, then the best-fit solution is a constant $V_\infty$ = 16.9 ± 0.4 $kms^{-1}$ over the entire trajectory and an apparent radiant R.A. = 276.4 ± 0.9º, Decl. = +59.6 ± 0.7º, in good agreement.

The meteor was first detected as a faint shadow in Kiği (Fig. 1H), when it was at ~60.2 km altitude. Only at ~58.4 km was the roof top shadow well enough defined to give an accurate direction. The final fragments of the meteor were seen to fade in the Muş Alparslan street camera (Fig. 1D) when it penetrated to 21.3 km, with strong deceleration in the final 4–6 km, especially in the final 2 km of the visible flight.

*The Light Curve*

The lightcurve of the meteor is shown in Fig. 4, both as a function of time and as a function of altitude. The lightcurve was determined from the brightness of surfaces illuminated by the meteor. Pixel intensity curves (in arbitrary units, a.u.) were corrected for range to the meteor (to a standard distance of 100 km) and aligned vertically on a logarithmic scale as a function of time, assuming all remaining factors that translate flux to pixel brightness are multiplicative. When aligned in altitude, instead, the light curves from individual stations do not perfectly overlap (Fig. 4). Note how Kiği and Karliova are slightly shifted relative to the Bingöl rectorate site and the #66 camera at Muş, for example. This implies that small systematic errors are still present in the trajectory solution. Taking this uncertainty into account, we determined that the initial fragmentation occurred at 36.5 ± 1.0 km altitude, followed by flares at 33.0 ± 1.0, 31.0 ± 1.2, and 27.4 ± 1.4 km altitude.

The absolute brightness was calibrated against that of the Moon, casting a shadow of the roof on the street in Kiği. At the time of the fireball, the Moon had an apparent brightness of –11.3 magnitude, defined in the visible V Johnson pass band, with zero magnitudes corresponding to $F_v = 3.67 \times 10^{-11}$ W $m^{-2}$ $nm^{-1}$ (Jenniskens 2006). The black-and-white camera pass band was broader, presumably covering the range of about 400–700 nm. By comparing meteor shadows to those cast by the Moon, it was determined that an apparent visual magnitude of –12.7 ± 0.7 caused the first roof tip shadows measured in this video (at 66 km from the meteor path). From this calibration, the meteor reached an absolute (at 100 km-distance) peak visual magnitude of $M_v = -16.8 \pm 0.7$ (a peak flux of $F_v = 1.9 \times 10^{-4}$ W $m^{-2}$ $nm^{-1}$).

*Meteorite Strewn Field*

Table 2 gives the assigned meteorite numbers and mass of 343 meteorites, 168 of those with find coordinates. For masses >10g (below which the distribution is not fully sampled), the distribution has a differential mass index of s = 1.77 ± 0.05 (corresponding to a magnitude size distribution index of $\chi = 2.04 \pm 0.09$ if they would be observed as independent meteors). Most mass is in the larger fragments. That distribution is more shallow than that expected for catastrophic fragmentation and steeper than expected for a collisionally relaxed distribution. The value is that expected for a collisional cascade, where bigger particles break up into smaller pieces, and then those smaller pieces become the parent of even smaller pieces, etc. (Jenniskens 2006).

A total of 24.78 kg of documented falls has been collected, the largest fragment weighing 1.47 kg (Fig. 5). The finder of an additional ~4.5 kg of reported finds (bringing the total to 446 meteorites) could not be verified, making it uncertain that some of these are not already in the list.

Figure 5 shows the distribution of find locations relative to the ground-projected meteoroid trajectory. Open symbols in Fig. 5 show the position where we calculated that masses of different size would have fallen if they were released during the first flare at 36.5 km (diamonds), or during the final one at 27.4 km altitude (squares). The atmospheric wind sounding data from stations 17351 Adana and 17130 Ankara for 12h UTC September 2 and 0 h UTC September 3 (http://weather.uwyo.edu/upperair/sounding.html) were interpolated to estimate the prevailing winds at 20h10m UTC over Bingöl. The strewn field is compact, with small stones being blown towards the larger meteorites (Fig. 5). The calculated positions are in reasonable agreement with the actual find locations, the difference suggesting that the actual trajectory over the fall location was ~0.7 km further west than that extrapolated from the meteor trajectory. This is within bounds of the ±0.6 km uncertainty of the trajectory at the position of the meteor and ±0.8º uncertainty in direction of the radiant.

The dispersion of the strewn field is most consistent with meteorites having fallen from the final disruption at 27.4 km. If material survived from the early breakup, then small masses should have fallen farther north of the known strewn field.

*Infrasound and Seismic Signals*

Infraround signals from the fireball were detected on the arrays I31, I48 and I46 of the International Monitoring System (IMS) (Christie and Campus 2010). Signals were identified based on an increased signal correlation across the array, with the corresponding best beam azimuths consistent with arrival from the bolide and showing celerities near 0.28 $kms^{-1}$ as expected for stratospheric arrivals (Ens et al. 2012). The signals measured at I46 and particularly I48 are quite weak, the latter being virtually at the noise level. Other stations located within 4,000 km range of the estimated terminal burst location (39.1N, 40.2E) included I19, I26 and I43, which did not record the fireball.

The Sariçiçek multi-station period average is 2.6 sec, which using the corresponding Ens et al. (2012) relation provides a yield of 0.12 kT (kiloton equivalent TNT = 4.184 x $10^{12}$ J). However the confidence bounds are comparable to the value itself (i.e., 0–0.20 kT). For this event, the periods are internally consistent for I31 and I46, but are much higher for I48 where the SNR is small. The I48 period (~5 s) is near the middle of the microbarom band (Garces et al. 2010) and the pre- and postsignal microbaroms at this station are well defined and emanate from within a few tens of degrees of the bolide arrival azimuth. This makes the resulting signal suspect, as we cannot clearly distinguish the bolide signal from microbaroms at the station given the low SNR. Taking the I31 and I46 periods near ~1.8 s gives a 0.03 kT yield, using either the AFTAC period-yield relation (ReVelle 1997) or the Ens et al. (2012) single station period with a formal uncertainty upper limit of <0.06 kT.

The amplitude and signal at I31 are sufficiently high that amplitude-based yields might also be expected to produce reasonable values (Edwards et al. 2006). Using the wind-corrected amplitude-yield in Ens et al. (2012) produces an independent estimate of ~0.05 kT. The small dominant periods at the stations with strong, clear signals is consistent with a modest (~0.1 kT) yield and certainly not the type of infrasound signal normally found from a larger, kT-class bolide. A larger (>0.2 kT) event would have shown significantly more high-frequency content than detected at the stations.

Turkey itself has a dense seismic network that monitors a seismically active area. Among the usual seismic signatures detected by nearby stations, we searched for a consistent wave pattern that arrived at the stations at about the expected time for an airburst to couple to the ground after the event time at 20:10:30.15 (36.5 km altitude) UTC (Cansi 1995). Fourteen stations recorded a signal that traced the airwave generated by the meteor during the travel in the atmosphere, although they are covered by noise at some stations. Most energy in the airwave was at frequencies higher than 3 Hz.

Based on expected travel times, first arriving at most stations were the airwaves emanating from the lower final airburst at 27.4 km altitude. The Bingöl (BNGB) and Solhan (SLHN) seismic stations were nearest to this final flare at 38.9623N, 40.5289E. The time difference between the arrival times of airwaves at these stations is 63 s, which corresponds to a wave velocity of ~330 m $s^{-1}$, which is consistent with the mean sound wave velocity (Fig. 6). The airburst propagated slower to the stations in backward direction relative to the meteor path, where they displayed a sharp waveform (Fig. 6), and arrived at these stations slightly later than the estimated arrival time from a constant 330 m $s^{-1}$.

## Meteorite Physical Properties

*Density and Strength of Recovered Meteorites.*

The mean bulk volume density is 2.929 ± 0.003 g cm$^{-3}$ for SC12 (SC12b) and 2.890 ± 0.004 g/cm$^3$ for SC14, based on 3D volumes of 9.232 ± 0.011 and 6.822 ± 0.008 cm$^{-3}$, respectively. We adopted an average value of 2.910 ± 0.020 g cm$^{-3}$. The average value of grain density from multiple measurements of the same fragments is 3.221 ± 0.015 g cm$^{-3}$. From the bulk and grain densities, an average porosity of 9.4 ± 0.9% follows, in agreement with values in Macke et al. (2008).

A cube-shaped sample of SC12 (5.09 ± 0.01 mm along the load direction, and 4.93 ± 0.01 mm and 5.12 ± 0.01 mm in directions mutually perpendicular to this) was compressed to an initial peak of ~75 MPa, followed by a slight drop corresponding to some failure. The sample then reloaded to an absolute maximum stress level of 79.1 ± 0.3 MPa, which is taken to be the compressive strength of this sample. All strength values were measured at load/displacement rates low enough to correspond to quasistatic measurements of the sample strength. The range of strengths observed in this work are within the range of compressive strengths previously observed in ordinary chondrite samples (6.2–420 MPa), as summarized by Kimberley and Ramesh (2011). Some laboratory strength measurements have been performed on metal meteorite samples (e.g. Johnson & Remo 1974; Furnish et al. 1995) and carbonaceous chondrites (Cotto-Figueroa et al. 2015), but there are no published strength measurements of achondritic stony meteorites, making the above reported measurement unique.

The whole-stone compression strength measurements are more uncertain, because of possible shear stresses and an uncertain surface area. The combined data are shown in Table 4 and serve mainly to point out that SC12 may be representative for other recovered meteorites, but SC14 was significantly weaker.

*Fractures and Grain Orientation*

Figure 7 shows representative microCT images of Sariçiçek SC12 and SC14. The larger fragment SC12b-a1 shows the rich clast texture of this meteorite. Interesting features to note are the FeNi grains surrounded by a dark silicate in Fig. 7B, the large FeNi grain (~1mm) in Fig. 7C, and a grain consisting of FeNi and FeS in Fig. 7H. Because SC14-a1 separated into two halves before imaging, the largest fracture line in Figs. 7E–F shows how the two halves were aligned for imaging.

The fractures in these samples were mostly caused by sample preparation, and may include intrinsic fractures and perhaps also fractures caused by the fall. The fracture length distribution in these samples may be representative for fracturing of the meteoroid during atmospheric entry. The density of fracture lengths from the highest resolution microCT images of SC12b-a2 and SC14-a2 are displayed in Fig. 8. The fracture distribution is not a power law over the entire range of sizes because of limitations in counts at the smaller end, and limitations of the sample size at the larger end. In between, the slope corresponds to $\alpha = 0.130 \pm 0.008$ and $0.144 \pm 0.011$ for the SC12b-a1 and SC14-a1, respectively. This value for α is lower than the value of 0.16 commonly used in models of atmospheric entry of ordinary chondrites. Instead, the value more similar to $\alpha \sim 0.11$ of terrestrial basaltic rock.

Analysis of the alignment of metal and metal sulfide grains in fragment SC12b-a1 show weak evidence of grain shape orientation (Fig. 9). Each graph displays the orientation direction of the major axis of the grains. There is a common band in both graphs that may be interpreted as flattening due to impact or perhaps as an effect of settling of soil particles. The axis of flattening is close to the center (0,0) direction in the graph. This analysis in 3D scans is the first of its kind in a howardite sample to our knowledge. Previously, Gattacceca et al. (2008) measured the anisotropy of the magnetic susceptibility of HED achondrites. They found, too, that the average petrographic fabric of magnetic grains in eucrites and howardites is oblate.

**Meteorite Petrography, Mineralogy, and Cosmochemistry**

*Petrography and Mineralogy*

SC12a (Fig. 10) and SC14 (Fig. 11) contain petrographically heterogeneous rock consisting of lithic and mineral clasts in a fine-grained matrix of crushed material. The matrix consists of high-Ca pyroxene (probably clinopyroxene) from the eucritic component (having a composition range of $Fs_{12–68}Wo_{6–45}$, FeO/MnO= 21.0–45.9, with average $Fs_{47}Wo_{9}$ and Percent of Mean Deviation (PMD) = 19.1%, *N=29*), low-Ca pyroxene (probably orthopyroxene) from the diogenitic component ($Fs_{22–70}Wo_{1–5}$, FeO/MnO= 25.2–52.2, average $Fs_{34}Wo_1$, PMD=35.5%, *N=29*), plagioclase ($An_{83.5–89.6}Or_{0.2–4.0}$), silica, kamacite, troilite, chromite, ilmenite and rare olivine ($Fa_{18-21}$, FeO/MnO= 30.1–43.1, average $Fa_{80}$, PMD=1.6%, *N=4*). Clino- and ortho-pyroxene are present in approximately equal amounts. Single mineral grains and clasts of orthopyroxene ($Fs_{22–70}Wo_{1–2}$), clinopyroxene ($Fs_{28–60}$ $Wo_{6–36}$)–some compositionally zoned, plagioclase ($An_{85.5–89.6}Or_{0.2–0.6}$), and rare olivine ($Fa_{21}$), all to a maximum size of 1 mm. Sariçiçek also contains rare grains of zircon, baddeleyite, and merrillite.

Crystals of clinopyroxene frequently contain exsolution lamellae of orthopyroxene, both containing oriented chromite inclusions. Three types of rock clasts are distinguished (see details in Fig. 11). (1) Clasts consisting of plagioclase and silica, the former containing inclusions of chromite and ilmenite, and the latter containing blebs of troilite; (2) clasts consisting of an intergrowth of plagioclase and silica, with both phases hosting large blebs of troilite; (3) ophitic to subophitic basalt clasts consisting of an intergrowth of plagioclase ($An_{85}Or_1$) laths and zoned clinopyroxene ($Fs_{33–55}$ $Wo_{6–12}$), in some cases with troilite blebs situated along the boundaries of the plagioclase crystal laths. Single mineral grains and clasts show different degrees of shock deformation, including irregular fractures and kinked pyroxene lamellae, and a significant fraction of the matrix is so fine-grained that it appears opaque in thin section. The total abundance of diogenitic material exceeds 10 vol % (Fig. 12A), which classifies the meteorite as a howardite.

In sample SC181, xenolithic clasts detected comprised of metallic particles, a 2 mm sized clast rich in finely dispersed metallic iron grains similar to material found in ordinary chondrites, a 350 micron sized chondrule, and a carbonaceous chondrite like clast. The metal grains sometimes had troilite (FeS) inclusions. Individual grains of troilite and chromite ($FeCr_2O_4$) were also found as inclusions in the silicate matrix. Metal grains consisted of kamacite with Ni content of 4–7 wt.% and Co of ~1 wt.%. Some grains with Ni content of 8–9 wt.% may have been martensite. One had associated Cu inclusions. Only one grain was found with both kamacite (4 wt% Ni) and taenite (43 wt% Ni). Two kamacite grains had exceptionally low Ni abundance of 0 and 1 wt%, respectively. Similar low Ni abundances were seen in metal veins in the silicate matrix. Many ilmenite and chromite inclusions were observed. 3–5 wt.% of Al was present, suggesting some amount of hercynite ($FeAl_2O_4$). Two silicate particles, of size 0.5 and 0.2 mm respectively, had mosaic structure. The iron content was 2–6 wt% in black particles, 6–7 wt% in dark gray particles, and 14–17 wt% in light gray particles. One particle had an associated ilmenite inclusion ($FeTiO_3$) at one side of the particle and many troilite inclusions bordering the other side.

*Bulk ElementalCcomposition: Major, Minor and Trace Element Abundances.*

The abundances for 58 major, minor, and trace elements are presented in Table 5. Thirty-eight elements were quantified by both laboratories in the consortium investigating the bulk composition of Sariçiçek. University of California at Davis (UCD) results are for the measurement of a single aliquot with a typical reproducibility of ≤5% based on repeat measurements of samples. Fordham University results are presented as a mean of the results for the five analyzed aliquots (see Methods section above). For the Fordham results, inter-aliquot errors in percent relative standard deviation are ≤14% for all elements except for the following (K: 18%, Ni: 17%, Cs: 20%, Ir: 24%), which is typical considering sample homogeneity.

The measured aliquots are large enough for the interlaboratory analyses of most elements to agree within measurement error. Larger differences are found for: Na (28% difference), Al (28%), Co (70%), Ni (factor 2.5 difference), Ga (50%), Ru (80%), and Pt (40%), which indicate variation in the amounts of plagioclase and FeNi alloys between the samples measured. In general, the refractory platinum group and siderophile elements show larger discrepancies than lithophile elements, which may be attributed to a nugget effect during sampling or because of intrinsic differences between stones SC12 (measured at UCD) and SC14 (Fordham). Because of the general agreement of the values generated by the two laboratories, when discussing our results we use the mean value of the two. We note that the conclusions would be the same if either laboratory's results would have been used separately.

As may be expected from their origins as mixtures of eucritic and diogenitic end members, the howardites generally possess compositions between those established for the eucrites and diogenites. Compared to the very complete database of Warren et al. (2009), Sariçiçek has Al, Co, Ni, Sc, Sm, and V abundances as a function of Ca and Mg that indicate a howardite composition (e.g., $Al_2O_3$ versus MgO in Fig. 12B). The compositions indicate a slightly greater affinity to the eucrites than the diogenites. This is borne out in the mineralogy of Sariçiçek, which shows greater eucrite material than diogenite material.

Warren et al. (2009) postulated that a subset of the howardites possess higher than typical siderophile element abundances and higher than typical amounts of noble gases (Fig. 13). They called these the "regolithic howardites" and this subgroup differs from typical howardites in that they are probably true surface material rather than simply eucrites–diogenite mixtures. Sarıçiçek has a relatively high iridium abundance of 8–10 ng $g^{-1}$ (Table 5). This relatively high siderophile element content of Sarıçiçek and noble gas abundance ($^{20}Ne$ = 1.8 nL $g^{-1}$ STP), see below, would place Sariçiçek in the "regolithic howardite" region defined by Warren et al. (2009). The Meteoritical Bulletin has 352 entries for howardites (17 falls) (https://www.lpi.usra.edu/meteor/metbull.php, last accessed April 5, 2018). Including Sariçiçek, 15 are now known to be regolithic howardites (Cartwright et al. 2013, 2014).

*Oxygen Isotope Analysis*

Oxygen isotope values are listed in Table 6. The $\delta^{18}O$ values of bulk Sariçiçek rock chips are typical of other HED meteorite (Clayton and Mayeda 1996, 1999; Scott et al. 2009). The $\Delta^{17}O'$ values are on the more negative side of typical HEDs, but overlapping the normal HED ranges of $\Delta^{17}O'$ = –0.247 ± 0.050 and $\delta^{18}O$ = 3.74 ± 0.56 (2s). The prime symbol here points to differences based on isotope ratios being plotted on a natural log scale. In contrast, Bunburra Rockhole had $\Delta^{17}O'$ = –0.127 ± 0.044 (Bland et al. 2009). The portion of SC14 that was not acid-treated yielded the same oxygen isotope results, suggesting that terrestrial alteration in this stone was absent to minimal. Stone SC14 was larger than SC12, and was possibly only affected by rain at its outermost areas. Sample SC12 was found 16 days after the fall, sample SC14 28 days (Table 2).

The two monomineralic feldspar analyses yielded the most positive $\delta^{18}O$ values of this entire data set. The variation in three-oxygen isotope space of the bulk rock subsamples are a function of their modal abundances of pyroxenes and feldspars, as expected from intermineral isotope fractionations in magmatic systems which show a temperature-dependent difference between pyroxene $\delta^{18}O$ values (more negative) and feldspar $\delta^{18}O$ values.

*Chromium and Titanium Isotope Analysis*

The Cr isotopic results are shown in Table 7. The Cr isotopic composition is reported in ε-notation, deviations of the $^{53}Cr/^{52}Cr$ and $^{54}Cr/^{52}Cr$ ratios from the measured NIST SRM 979 terrestial Cr isotopic standard in parts per 10,000. Figure 14 shows the combined $\Delta^{17}O$'-$\varepsilon^{54}Cr$ systematics for Sariçiçek. While previous studies have documented the composition of eucrites and diogenites (Trinquier et al. 2007; Sanborn and Yin 2014, Sanborn et al. 2016), prior to this study there have been no howardites investigated. To this end, in addition to Sariçiçek, another classified howardite, Bholghati, was also analyzed. In terms of both $\Delta^{17}O$' and $\varepsilon^{54}Cr$, Sariçiçek SC14 and Bholghati overlap in their composition. While the $\Delta^{17}O$' for Sariçiçek plots marginally lower than the bulk of the normal eucrites and diogenite field, the $\varepsilon^{54}Cr$ is indistinguishable with that of the normal eucrites and diogenites. The small offset in $\Delta^{17}O$' (Fig. 14) is probably due to the regolithic nature of Sariçiçek (see above), a small admixture of xenolith clasts, or a cryptic component that were thoroughly mixed with the howardite. Both isotopes could be displaced from the normal HEDs towards the lower right by carbonaceous chondrite components, most likely CM/CV types, which plot outside and to the lower right of Fig. 14 at around $\varepsilon^{54}Cr \sim +0.9$ and $\Delta^{17}O' \sim -3$ (Sanborn et al. 2016).

The Ti isotope results are listed in Table 8. In addition to Sariçiçek, dissolutions from three different eucrites (Béréba, Juvinas, and Pasamonte) were also analyzed in this study. The average of the Ti isotope data for Sariçiçek agrees with the average obtained from the three eucrites. Moreover, the Ti isotope composition of Sariçiçek and the three eucrites overlap with previous data obtained on eucrites and the howardite Kapoeta (Zhang et al. 2012; Williams 2015) within the analytical uncertainties. Since each meteorite group displays its own characteristic Ti isotope compositions (Trinquier et al. 2009; Zhang et al. 2012; Williams 2015), this is further evidence that Sariçiçek belongs to the HED meteorite clan.

*Xenolithic Organic Matter*

The inner face white/light grey regions of meteorite SC239 are plagioclase feldspars (major component, anorthite) and quartz, while the Raman spectra of the probed dark grey and black regions are essentially identical, showing signatures of pyroxene type minerals (clino- and orthopyroxene) and traces of olivine, mixed with carbonaceous materials (D- and G-bands) (Fig. 15). The carbonaceous material is thought to be xenolithic.

No visible carbonaceous inclusions were found in SC12a, but a carbonaceous chondrite clast was observed in SC181 (see above). SC12 did contain a cryptic component of organics. SC12 was found to be highly depleted in amino acids with a total amino acid abundance of 45 ppb (Table 9). SC12 contained only trace amounts (~0.2–10 ppb) of the common protein amino acids glycine, aspartic and glutamic acids, serine, alanine and valine, as well as the nonprotein amino acids, α-, β- and γ-amino-*n*-butyric acid, α-aminoisobutyric acid and ε-amino-*n*-caproic acid. In contrast to most carbonaceous chondrites which contain several hundred to several thousand ppb of extraterrestrial amino acids in the free form (Glavin et al. 2010), no free amino acids above a level of 0.1 ppb were detected in SC12. This suggests the parent body of Sariçiçek was highly depleted in amino acids and amino acid precursors.

Trace amounts of predominately the L-enantiomers of the protein amino acids aspartic and glutamic acids, alanine, serine, and valine were detected above procedural blank levels in SC12, which indicates that this sample was exposed to some terrestrial amino acid contamination before it was recovered on September 18, 2015, during collection, or during subsequent handling. Sariçiçek SC14 had a much higher amino acid abundance (~ 666 ppb total), but it was recovered 12 days later, after rain, and the enantiomeric composition of amino acids was similar to the terrestrial pebble and soil collected at the meteorite recovery locations, pointing to direct exposure to terrestrial amino acids at the landing site (Table 9).

Sariçiçek SC12 was also poor in polar soluble organic matter, but did contain compounds intrinsic to meteorites. Signals covered the whole mass range up to 750 amu (atomic mass unit). All mass peaks resulted in 4000 elementary compositions in the C-H-N-O-S-Mg elemental space. Thermostable compounds containing C, H, and O, as well as those containing C, H, O and Mg, including the dihydroxymagnesium carboxylates described by Ruf et al. (2016), were equally most important in number and covered almost 75% of the total number of elementary compositions, followed by sulphur- and nitrogen-containing compounds respectively. Terrestial impurity signals were minimized to ubiquitous alkyl benzene sulfonates and biogenic fatty acids.

*Cosmogenic Randionuclides and Pre-Atmospheric Size*

The AMS results for $^{10}$Be, $^{26}$Al, and $^{36}$Cl of samples SC12 and SC14 are listed in Table 10. Calculated production rates from the model of Leya and Masarik (2009) use the measured elemental concentrations from Table 5. Figure 16 compares the measured concentrations for $^{26}$Al, to model calculations for objects with radii of 12–100 cm. Similar calculations were done for $^{10}$Be and $^{36}$Cl. These concentrations are not only affected by shielding conditions, but also by variations in target element composition. The main target elements for $^{36}$Cl production are Ca and Fe. Since the relative production rate of $^{36}$Cl from Ca is a factor of 10 higher than from Fe, we normalized the measured $^{36}$Cl concentrations to total Fe +10 × Ca to account for variations in the two main target elements for $^{36}$Cl production in Sariçiçek. The higher $^{36}$Cl concentration in SC14 is partly due to a higher Ca content and partly due to a higher shielding depth of this fragment. The high measured concentrations clearly indicate that $^{10}$Be, $^{26}$Al and $^{36}$Cl levels represent saturation values, as expected for a meteorite with the CRE age of ~22 Ma derived from the cosmogenic noble gases.

There are two degrees of freedom in fitting the models to the measurements: the production rates have an absolute uncertainty of about 10% (Leya and Masarik 2009), which permits a 10% vertical shift of the model diagram, and the pre-atmospheric depth of the two Sariçiçek samples is not known a priori, which permits a horizontal shift of the measured points in Fig. 16. However, the pre-atmospheric depth of a given fragment needs to be the same for all elements considered. Consistent results are achieved for a shielding depths of 5 ± 5 cm (SC12) and 15 ± 5 cm (SC14), respectively, and an object with a pre-atmospheric radius of ~50 cm. With those shielding depths, the nominal production rate model of $^{10}$Be by Leya and Masarik (2009) needs to be increased by 8% (shifting the model up 8%), consistent with our previous findings that the $^{10}$Be production rates are systematically on the low side (Welten et al. 2012), while $^{36}$Cl production rates were increased by 10%.

The example of Fig. 16 shows that the $^{26}Al$ concentration for both SC12 and SC14 are in the expected range of calculated production rates for howardites with pre-atmospheric radii of 40–60 cm (mass of 800–2600 kg). All radionuclide concentrations would permit irradiation in an object with a radius up to ~80 cm, but this would not yield a consistent shielding depth for SC14. For example, the $^{26}Al$ concentration would also permit irradiation at a depth of 15–20 in an object with 80 cm radius, whereas the $^{10}Be$ and $^{36}Cl$ concentrations would favour a depth of >40 cm in the same object. We thus favour a radius of 40–60 cm.

The stone SC26 (Fig. 24, below) was measured at the Laboratori Nazionali del Gran Sasso by gamma ray spectroscopy 35 days after the fall. The concentrations of the natural radionuclides $^{232}Th$ and $^{238}U$ as well as for $^{40}K$ in the meteorite specimens are listed in Table 11. All of them are in accordance with the measurements presented in Table 5.

In Table 11 are given also the measured activity concentrations for the positively identified short- and medium-lived cosmogenic radionuclides ($^{7}Be$, $^{46}Sc$, $^{54}Mn$, $^{22}Na$, $^{26}Al$). Only upper detection limits are reported for $^{48}V$, $^{51}Cr$, $^{56}Co$, $^{57}Co$, $^{58}Co$, $^{60}Co$ and $^{44}Ti$. The activity of the very short-lived radionuclides $^{52}Mn$ (half-life = 6 days), $^{48}V$ (half-life = 16 days) and $^{51}Cr$ (half-life = 28 days) was below the detection limit. The radionuclides of cobalt are expected to be low, as the composition of the howardite shows this element only in trace quantities. The given activities are the ones calculated back to the date of fall following the simple decay law, taking into account the time that passed between the fall of the meteorite and its measurement. In order to estimate the depth of origin of the specimen within the meteoroid, the data of $^{54}Mn$ and $^{22}Na$ were used. The $^{22}Na$ data was compared to the calculations of Bhandari et al. (1993) for H chondrites, renormalizing the measured concentrations and taking into account the different concentrations of the target elements (Al, Mg, Si) in the howardite with respect to an average H chondrite (Table 5). If we take the measured activity as saturation value the resulting possible range in the radius is 5–15 cm. If we assume that it came from anywhere in the meteoroid the range cannot be determined. The rather low specific activity shows only that the specimen most probably comes from the surface of the parent body.

The data of $^{54}$Mn was normalised to the concentration of its main target Fe (as reported in the other sections). Comparing the measured activity of $^{54}$Mn to the calculations of Kohman and Bender (1967) would give a range for the radius of <20 cm in case SC26 came from the central part, or a larger radius if it came from near the surface of a larger body. Indeed, the measured value for $^{26}$Al of (79.0 ± 6.1) dpm kg$^{-1}$ plots in the top few centimeters of the surface of a larger meteoroid (Fig. 16).

*Solar Wind Derived Noble Gases (He, Ne, Ar) and Cosmic-Ray Exposure Ages*

Regolithic howardites are defined primarily by their high content of solar wind noble gases (see previous Fig. 13; Warren et al. 2009; Cartwright et al. 2014). Results from noble gas measurements are listed in Table 12. From the three noble gases He, Ne, and Ar measured, Ne is the most straightforward to interpret. The Ne isotopic composition of all Sariçiçek samples (Fig. 17) can be understood as a two-component mixture between a solar wind component fractionated during implantation into the regolith (Grimberg et al. 2006) and a cosmogenic component with $^{22}$Ne/$^{21}$Ne = ~1.11.

This $^{22}$Ne/$^{21}$Ne ratio–according to a production rate model by Leya and Masarik (2009) for a radius of R = 50 cm meteoroid with target chemistry measured in Sariçiçek–is compatible with shielding depths of 5 ± 5 cm and 15 ± 5 cm for SC12 and SC14, respectively, deduced from cosmogenic radionuclides. The concentration of cosmogenic $^{21}$Ne ($^{21}Ne_{cos}$) resulting from the two-component deconvolution between solar and cosmogenic endmembers is given in Table 12. The elemental composition of Sariçiçek SC12 and SC14 (Tab. 5) and the cosmogenic $^{22}$Ne/$^{21}$Ne ratio (~1.11) was used to determine production rates of cosmogenic nuclides ($^{3}$He, $^{21}$Ne, $^{38}$Ar), applying the empirical formulas given by Eugster and Michel (1995) for howardites. The calculated production rates, nominal and regolith-contribution-corrected cosmic-ray exposure ages are given in Table 13.

The He isotopic composition of Sariçiçek is a mixture of at least three components: solar wind He, cosmogenic He (addopting $^{4}$He/$^{3}$He = 5.2 as suggested by Eugster and Michel 1995), and radiogenic $^{4}$He from the decay of U and Th. So there are three components, but only two isotopes to resolve them. This degeneracy can partially be broken by using the variable concentration of solar noble gases in the samples. The non-cosmogenic $^{20}Ne_{non\text{-}cos}$, i.e., the measured $^{20}Ne$ minus the cosmogenic $^{20}Ne$ calculated from $^{21}Ne_{cos}$, can plausibly be assumed to be pure solar wind, based on the Ne three isotope diagram (Fig. 17). By plotting the non-cosmogenic $^{4}He$ (where $^{4}He_{cos} \sim 5.2 \times {}^{3}He_{meas}$, since the $^{3}He/^{4}He$ ratio suggests ~90% of the $^{3}He$ is cosmogenic) against $^{20}Ne_{non\text{-}cos}$, extrapolation to $^{20}Ne_{non\text{-}cos} = 0$ should reveal any radiogenic $^{4}He$ excess as *y*-axis intercept (Fig. 18). Using a trend line through all four samples, the radiogenic $^{4}He$ excess corresponds (with SC26 whole stone U = 57 ± 3 ppb [Table 11] and typical Th/U = 3.5 [Nittler et al. 2004]) to a radiogenic retention age of $2.3^{+0.8}_{-0.9}$ Ga (1σ). If the samples are fitted individually using the measured U and Th abundances in Table 5 (assuming 3% uncertainty in absolute concentrations) the resulting radiogenic $^{4}He$ concentration corresponds to radiogenic retention ages (1σ) of $1.76^{+0.62}_{-0.75}$ Ga for SC12 and 2.6 ± 0.3 Ga for SC14, respectively (Table 13).

All four samples yield a consistent $^{40}Ar$ concentration of 10.30 ± 0.15 (1σ) × nL g$^{-1}$ STP (Table 12), suggesting that the $^{40}Ar$ is mainly radiogenic. The concentration of adsorbed atmospheric $^{40}Ar$ would be expected to vary with the sample surface area and mass. The measured concentration corresponds (at measured K = 248 ± 14 and 220 ± 10 ppm, see Table 5) to a retention age of 3.50 ± 0.07 and 3.7 ± 0.1 Ga for SC12 and SC14, respectively.

The nominal cosmic-ray exposure ages of Sariçiçek SC12 (given in brackets in Table 13) are 20, 29, and 29 Ma for cosmogenic He, Ne, and Ar, respectively, and 24, 30, 30 Ma for fragment SC14. The He ages are lower than the Ne and Ar ages, which is often observed for meteorites and usually explained by partial diffusive loss of He. These exposure ages have to be corrected for potential pre-exposure of the samples in the regolith, as shown by Wieler et al. (1989) for Fayetteville (H4) and more recently by Meier et al. (2014) for Ghubara (L5). Grains with higher residence times have a higher average probability to pick up solar wind, and will also experience, on average, a longer exposure to cosmic rays.

This means that for regolith breccias, the nominal cosmic-ray exposure age is not representative of the time the meteoroid actually spent in space on its way to Earth (4π exposure), but also contains a pre-exposure signature from irradiation in the parent body regolith (2π exposure). When extrapolating the Ne (or Ar) data down to a concentration of zero solar gas, the 4π exposure (which affects all samples invariably) can be resolved from the 2π exposure, which affects samples as a function of their solar wind content (Fig. 19). The resulting 4π exposure ages (corrected for shielding at the respective positions of the two samples) are in a range of 20–23 Ma, in excellent agreement with a peak in the exposure age distribution of howardites (Welten et al. 1997). A cosmic-ray exposure age at the peak of the 22 Ma clan distribution suggests that Sariçiçek did not further break after ejection prior to impacting Earth. Since the production rates fall roughly by a factor of ~2 when going from 4π to 2π exposure (e.g., Leya et al. 2001), the 2π cosmogenic noble gases correspond to regolith exposure ages of ~12–14 Ma. This regolith exposure must have happened within a few meters of the surface, at an arbitrary time before ejection ~22 Ma ago.

*U-Pb Chronology*

The U-Pb dating results for zircon (some examples shown in both cathodoluminescence and backscattered electron images in Fig. 20A) and baddeleyite (shown in Fig. 20B) are listed in Tables 14 and 15, respectively, where uncertainties for individual isotopic data analyses are reported as 1σ.

The U-Pb dating results for apatite (individual grains shown in Fig. 21) are listed in Table 16, where uncertainties for individual isotopic data analyses are reported as 1σ. The intercept age and Pb-Pb ages in Fig. 22, quoted at the 95% confidence level, were calculated using ISOPLOT 3.0 (Ludwig 2003). The new solar system average $^{238}U/^{235}U$ ratio of 137.794 ± 0.027 (Goldmann et al. 2015) was used in the age calculations and data reduction.

Raman results show that the baddeleyite in Sariçiçek is monoclinic, which implies that later impact events did not disturb its U-Pb isotopic system (Niihara 2012, Zhou et al. 2013). The U-Pb ages of zircons, baddeleyites and apatites revealed a concordant ages of 4550.4 ± 2.5 Ma, 4553.5 ± 8.8 Ma, and 4525 ± 17 Ma, respectively, consistent within error to their corresponding Pb-Pb ages of 4551.1 ± 2.8 Ma, 4558.0 ± 8.2 Ma, and 4524 ± 12 Ma. Using the new solar system average $^{238}U/^{235}U$ ratio of 137.794 ± 0.027 (Goldmann et al. 2015) in place of old value of 137.88 results in a downward age adjustment of –0.88 Ma. It is still within the quoted uncertainties of SIMS results.

*Thermoluminescence*

The natural thermoluminescence (TL) of Sariçiçek (Table 17) is at the peak in the histogram for HED meteorites at 10 krad equivalent dose (Sears et al. 1991; Takeda and Graham 1991). This reflects the recent recovery of the meteorite and an orbit with a "normal" perihelion. The measured perihelion distance of Sariçiçek ($q$ = 1.0087 AU) is therefore characteristic of other HED falls.

If Sariçiçek were an eucrite, the induced TL sensitivity values would place it in the petrologic type 3 class defined by Takeda and Graham (1991), which have TL sensititivy of 250–400. The presence of diogenite will lower the TL sensitivity by dilution. A 10% diogentic component would place Sariçiçek still in the petrologic type 3 class.

*Spectral Reflectance*

Reflectance spectra of all the samples are plotted in Fig. 23. The samples are reasonably fresh for the optical spectroscopic purposes based on the 1 and 2 μm band shapes (Table 18) and the strength of the 3 μm hydration (terrestrial weathering) band. According to the Band I and Band II center and band area ratio plots (Fig. 23), the reflectance spectra are consistent with a howardite that is relatively rich in eucrite material. The spectra are also broadly consistent with Vesta, given that Vestan regolith is richer in eucrite material than diogenite material (Ruesch et al. 2015).

The near-IR reflection and absorption spectra show two major bands at 0.9 μm (Band I) and 2.0 μm (Band II). Two weak bands were observed at 1.19 μm and 1.31 μm, which are attributed to the M1 site in pyroxenes (Klima et al. 2008) or small amounts of plagioclase (Karr et al. 1975; Cloutis et al. 2013), respectively. Plagioclase contains minor amounts of $Fe^{+2}$. The Band I center was observed at 0.93 μm for each sample while Band II center was observed at 1.97 μm, 1.97 μm, 1.98 μm and 2.00 μm for SC55, SC239, SC327 and SC51, respectively (Table 18). Ruesch et al. (2015) reported that a correlation exists between the wavelengths of Band I, Band II centers and ferrosilite (Fs), wollastonite (Wo) contents. Table 18 lists the calculated Fs and Wo contents. The spectroscopic ranges obtained overlap the compositional range measured with the electron microprobe.

*Heating and Surface Ablation*

In this sample SC26, we have a rare surviving fragment of surface material, presumably from the back of the original meteoroid (Fujiwara et al. 1989). SC26 has features that show its predominant orientation during the late stages of flight. The back side (Fig. 24A) shows a melt rim, where melt flowing from the front side of the meteorite was blown away. The front side (Fig. 24B) has flow lines and more rounded features. This side has the smooth relatively flat surfaces that are created in a late breakup relatively deep in the atmosphere, presumably during the 27.4 km altitude breakup.

The back side (Fig. 24A) is much more irregular in shape. The regmaglypts (thumbprint-like indentations) point to a different manner of ablation. With less than 2 cm overburden from its center, it is unlikely that the regmaglypt pattern is due to turbulent flow. Instead, the pattern may have originated from spallation of low-melting-point minerals near the surface, or from features in the surface that were pre-atmospheric.

All samples have a distinct sheen, which was identified as due to abundant vesicles (from vapor bubbles) in the fusion crust, presumably trapped in a melt with higher viscosity than that of ordinary chondrites (Fig. 25A). The number of vesicles decreases exponentially with increasing vesicle size in the 6–50 μm range, with most vesicle volume at small 6–10 μm size scales (Fig. 25B). Just above the unmelted meteorite material is a layer of initially modified material, which appears homogenized with some scattered vesicles (Fig. 25A). At other places, the bottom of the fusion crust showed a pattern of parallel fractures filled with silicate melt (not shown). At places, melted troilite fills pre-existing cracks. At the bottom of the melt layer are metal grains. A small grain of gold was detected here in the Ural Federal University study of SC181. Above that layer, bubbles form in a densely stacked pattern throughout a melt layer as a result of the evaporation of volatile components such as sulfides (layer labeled "viscous melt" in Fig. 25A). At the top of the glass layer, large bubbles are found, some of which have broken the surface (Fig. 25A).

The heating experiments at ambient pressure of SC239 showed that the sample remained unchanged up to 1300 K. At 1490 K, a notable amount of liquid phase is formed on the surface, causing a foaming. This confirms that gasses remained trapped in the melt. At 1499 K, bubbles migrate to the surface and pressure was released. At 1513 K, the meteorite showed increased melting and started to deform into a droplet-like shape. When heated at 5 K $min^{-1}$, the apparent viscosity of the sample reached a value of about $10^{10}$ poises (the sintering point) in the temperature range 1370–1410 K and a value of about $10^{5.5}$ poises (at the half sphere point) at 1510–1530 K. It is not clear that the sample was totally molten at these temperatures. Tomography of the heat-treated sample shows inhomogeneity, suggesting that some parts did not melt completely at the temperatures applied. The porosity of the heat-treated sample was 45%, suggesting significant gas evolution during heat treatment.

The front (Fig. 24B) side of SC26 has fusion crust that is slightly more red in color compared to the yellowish back side (Fig. 24A). This coloration is also seen in other oriented Sariçiçek meteorites. The redder color signifies a stronger oxidation at higher temperatures: The melted sample SC239 had changed colour from its original dark brown to a more reddish-brown. At high magnification in SEM images obtained by the Bulgarian Academy of Sciences institutes, crystals were seen scattered in a glassy matrix. The glassy matrix has a composition similar to the bulk meteorite based on EDS analysis. Triangular crystals have a composition rich in iron and oxygen, suggesting the formation of octrahedral magnetite, while some other crystals suggest the formation of iron carbonates based on EDS analysis. Such crystals are also found inside the vesicles.

## DISCUSSION

### Chronology of Primary Igneous Differentiation and Thermal Metamorphism in Sariçiçek

The peak age of basaltic magmatism on Vesta is 4552 ± 7 Ma (2σ), based on combining our U(Pb)-Pb dating results for zircons and baddeleyites in Sariçiçek with the compilation of published U(Pb)-Pb zircon data from non-cumulate eucrites (e.g., Zhou et al. 2013). Although the U-Pb ages for zircon and baddeleyite are similar to each other within analytical error, it seems that the age of baddeleyite is slightly but systematically older than zircon. This is reasonable because (1) a baddeleyite core was seen surrounded by a zircon mantle (Fig. 20B, upper right panel); (2) baddeleyite ($ZrO_2$) forms during early stages of magma crystallization, and zircon ($ZrSiO_4$) forms later when silica ($SiO_2$) saturation is achieved during magmatic differentiation and evolution. The age difference is likely to be small, depending on the longevity of the magma chamber. However, zircon could also be formed by metamorphic or metasomatic replacement of primary baddeleyite in contact with silica-rich partial melt or fluid. Age difference can be large or small in this case. This has been observed in terrestrial magmatic systems (e.g., Davidson and Van Breemen 1988; Amelin et al. 1999).

Younger apatite ages clearly post date the crystallization ages of zircons and baddeleyites by ~27 Ma. This could be due to impact resetting, or alternatively due to parent body metamorphism. Evidence from eucrites indicates that most mafic material on Vesta underwent metamorphism on a global scale (Yamaguchi et al. 1996, 1997). If so, the lower closure temperature for Pb diffusion in this phase allowed those grains to continue to equilibrate during metamorphism after the zircon phases had had their U-Pb ages frozen in.

**Meteoroid Size, Fragmentation and Impact Hazard**

Sariçiçek is the first 22 Ma clan HED meteorite fall that was observed by instrumental techniques. The manner of energy deposition of the meteoroid is ground truth for models that simulate the damaging effects from impacts of larger 20–100 m sized V-class asteroids.

*Light Curve Simulation From Simple Fragmentation Models*

The meteor lightcurve was modeled using the one-dimensional triggered progressive fragmentation model (TPFM) (ReVelle 2003, 2004; Ceplecha and ReVelle 2005), which calculates the bolide's light curve and deceleration profile based on an assumed fragmentation behaviour rooted in empirical modeling of bolides. The initial breaking strength, strength increase (via a strength multiplier), and porosity are free parameters. The strength parameter controls the peak of the calculated lightcurve, the porosity value the drop off.

Best-fit curves are shown in Fig. 4. The assumed initial breaking strength of 0.05 MPa for the meteor does replicate the fragmentation features observed in the high-altitude portion of flight; the replication of one or more of the features in the light curve only requires the strength of the fragments to increase by a factor of 1.5 over each fragmentation event. This appears to hold good up to an altitude of 55 km. From an altitude of 55 km down to an altitude of 40 km, the strength increase is more than an order of magnitude. The breakup altitudes at approximately 36, 33, 31, 27 km correspond to dynamic strengths of 0.05, 0.075, 0.11, 2.0 MPa. Although the code cannot replicate multiple flares, the break up of the meteoroid from 40 km down to roughly 27 km is replicated in a mean sense by allowing the ablation coefficient to vary with altitude.

We conclude that solutions for meteoroid diameters of 0.8–1.2 m can fit the lightcurve based on the proposed calibration of absolute intensity of the fireball, in agreement with the diameter derived from cosmogenic nuclides of ~1.0 m. With an entry speed of 17.1 ± 0.8 km $s^{-1}$, the impact energy was 0.025–0.101 kT. This energy estimate agrees well with the 0.03–0.05 kT estimates from analysis of the infrasound signals of the Sariçiçek event. These results are a factor of two lower than the 0.13 kT impact energy reported from U.S. Government satellites.

*Fragmentation in 2-D Hydrodynamical Simulation*

To gain more insight into the manner of fragmentation, a set of hydrocode simulations were run on the Sariçiçek meteoroid entry. Simulations were conducted on the NASA Pleiades supercomputer at NASA Ames Research Center, using the hydrocode ALE3D from Lawrence Livermore National Laboratory (Robertson and Mathias 2015). Larger asteroids are thought to be rubble piles, with rubble following a size frequency distribution N ~ $1/r^3$, deduced from imaging of asteroid Itokawa (Sanchez and Scheeres 2014). Small asteroids such as Sariçiçek could be either a monolithic rock of relatively high strength or a weakly consolidated collection of boulders held together by Van der Waals forces (e.g., Hirabayashi et al. 2015, Campo Bagatin et al. 2018).

The high altitude emission from Sariçiçek points to exposed weakly consolidated materials or efficient ablation of rock at the exposed surface due to melting and vaporization in the intense heat of the bow shock. The smaller a meteoroid, the more important ablation is. Small ~0.1 m sized meteoroids in meteor showers ablate high in the atmosphere. For large meteoroids, ablation is less important because it is only a surface effect and thermal penetration is typically only centimeters deep, or less. The approximate 1 m diameter Sariçiçek meteoroid is at the boundary between where ablation is dominant and where it is negligible. Here, we investigated the effects of a weakly consolidated layer. The investigation of heat transfer and ablation is postponed to a future study.

To account for the flares at 36–27 km altitude, there are two options. The first is a monolithic rock with internal fractures, voids, or other weaknesses of a particular strength that would fail when the dynamic pressure exceeds those strengths. Under this scenario, the rock would initially fracture when the dynamic pressure overcomes the strength of the monolith. In the fragment-cloud model of Wheeler et al. (2017), it would fracture into two or more pieces plus a cloud of dust. The dust cloud would be responsible for a flare and the resulting fragments would continue until they in turn reach a dynamic pressure that causes them to fail.

One of our simulations studied a "cherry" structure asteroid with a thick strengthless outer layer of material and a core with randomized strengths of 5, 10, and 20 MPa. The outer layer does blow off high in the atmosphere and even appears to roughly match the peak in energy deposition seen at 56 km, but does not sustain the energy deposition rate observed at lower altitudes. The simulation does show distinct flares, but at higher altitude than observed. When the core starts to fail at 43 km altitude, smaller pieces are swept away into the flow and rapidly decelerated, while some material persists to lower depth as large fragments. Significantly higher strengths would be required to match the flares, but this simulation showed it is possible to create distinct flares from the disruption of a monolithic object with a distribution of internal strengths.

Figure 26 shows the results for an alternative interior structure consisting of a rubble pile of four monolithic boulders embedded in weaker fine-grained material. Integrating the energy under the light curve gives energies of 149 GJ for material deposited above 38 km, and 116, 88, 78, 94 GJ for the flares. Using the density of recovered meteorites (2910 kg $m^{-3}$) and a flight speed of 17.1 km $s^{-1}$, the boulders can be estimated as being of diameters 0.81, 0.73, 0.70, and 0.75 m within 1.27 $m^3$ of weaker material at the surface, assuming all the weaker material blows off in the upper atmosphere and each boulder deposits all of its energy in each flare. The simulation shows the exterior weak layer blowing off at high altitude and as expected shows four flares. With sufficient tuning, this could also provide a match to the observed lightcurve.

Interestingly the simulation also shows, at around 33 km altitude, the weaker leading and trailing boulders have structurally failed but are not immediately dispersed into the flow (Fig. 26). Instead the stronger unbroken boulder and the bow shock hold the leading debris in place, and the trailing debris drafts behind. In this way the intact boulder delays the dispersion of some of the weaker material, and a significant amount of it remains entrained until the strongest boulder finally fails then pancakes and flares. The simulation was axi-symmetric 2D, and the extent to which failed material can draft in front or behind an intact boulder in a 3D simulations remains to be seen.

However, the observed distribution of fragments on the ground implies that most of the meteoroid held together until the final flare at 27 km altitude. This suggests that the lighter, weaker pieces contributed to the higher flares, and were reduced to dust creating few meteorites on the ground, but the stronger, larger pieces remained together until the final flare. The rubble pile simulation shows a mechanism through which that could happen by drafting structurally failed material in front or behind an intact boulder.

If the rubble pile model of the approximate 1 m diameter Sariçiçek is scaled up to the 20 m diameter meteoroid the size of Chelyabinsk (Brown et al. 2013; Popova et al. 2013) the strengths should be reduced according to the Weibull distribution of fragment length scales. Terrestrial analogs show that the Weibull modulus holds constant for many rocks from centimeter size samples up to large rock masses tens or hundreds of meters across (e.g., National Research Council 1996; Bonnet et al. 1999), so it is not unreasonable to scale a 1 m rubble pile up to a 20 m rubble pile using a Weibull modulus to decrease the strength. For the measured modulus of 0.137, this gives a strength ratio of about a third (0.325), implying strengths of 1, 3, 10, and 30 MPa should be used.

Without the Weibull scaling, the energy deposition curve of the 3–100 MPa rubble pile of Sariçiçek material on a Chelyabinsk trajectory shows a peak energy deposition at an altitude of 15 km instead of 30 km, with approximately the same peak energy deposition rate (Fig. 27). The point on the ground directly below the point of maximum energy deposition would be moved west and the overpressure at that point would be higher by a factor of 4–8 due to being twice as close to the ground and assuming either cylindrical or spherical shock wave expansion. However, 30 km away from the ground track, such as was seen at the city of Chelyabinsk, the overpressure would have been very similar due to a similar amount of energy being deposited, a similar peak deposition rate, and a similar distance from the peak deposition location.

*Simulation of the Airburst*

The consequences on the ground are a function of the impact angle. The Sariçiçek airburst was modeled using the multimaterial, multiphysics, multidimensional CTH suite of computer codes that are designed for a wide range of shock wave propagation and material motion applications (McGlaun et al. 1990; Boslough and Crawford 2008). CTH uses finite-volume analogs of the Lagrangian equations of momentum and energy with remapping at every time step for Eulerian differencing. It includes the option of adaptive mesh refinement (AMR) to increase resolution in regions of interest.

The observed light curve (in 94 steps) was used as a proxy for energy deposition along the known flight path with known velocity and timing (e.g., Brown et al. 2013). For each time interval, a quantity of energy was deposited along a known segment of the trajectory into a cylindric volume containing a mass of air 10 times the mass ablated during the interval (assumed for this purpose to be proportional to the fraction of energy lost). The mass of the cylindrical parcel was increased by this amount, it was given an increment of internal energy, and it was inserted with a velocity along the path to conserve momentum.

For these simulations, we exploited bilateral symmetry, with a rectangular half-space domain 160 km along the ground track, 80 km in the lateral direction, and 80 km high. Six levels of refinement were used, with a minimum zone size of 125 m for the shock wave at the surface. We used a gravitationally stabilized US Standard Atmosphere from sea level up to 80 km, and inserted the hot cylinders sequentially for the first 3.14 s of simulation time.

Seismograms were calculated at the position and altitude of the seismic stations, but without including actual topography of the terrain. Reflections from coupling to the ground are not in the model. Thus the records are not intended to be synthetic seismograms, but are only useful for determining first arrival times for comparison to actual data.

Calculated results are in generally good agreement to observations (Fig. 6). Nearby stations have slightly earlier arrival times calculated than observed, but stations further out are a good match. The over pressure pattern on the ground (Fig. 28) shows an asymmetry similar to that observed in the treefall pattern at Tunguska and the damage pattern at Chelyabinsk, with stronger shock propagation in the uprange and lateral directions and a weaker shock downrange. This is the result of the angle of entry and the long section along the entry trajectory along which energy is deposited, with shock arrival times being similar at those locations for different deposition altitudes.

It is also instructive to look at the shape of the shock wave in the vertical symmetry plane, which can be seen in the distribution of kinetic energy (Fig. 28). This figure shows that the reason for the stronger uprange shock is because the shock in that direction has a cylindrical shape that is diverging less (waves closer together) than the spherically shaped shock in the downrange direction of motion.

**Trajectory and Orbit**

The Sariçiçek fall provided the first pre-atmospheric impact orbit of a 22 Ma clan HED meteoroid. The measurements brought to light significant errors in the velocity vectors reported from U.S. Government (USG) satellite observations, enabled a search for pre-impact images of the meteoroid in NEO survey programs, and provided insight into the source region of the 22 Ma clan HED meteoroids in the asteroid main belt.

*Comparison to Satellite Data.*

Large errors in USG-detected fireball directions were recently reported (Jenniskens et al. 2018). The pre-entry velocity vector of Sariçiçek in ECF coordinates was given initially as VX = 5.1 km $s^{-1}$, VY = –6.3 km $s^{-1}$, VZ = –16.0 km $s^{-1}$, corresponding to a 17.9 km s–[1] entry speed (Fig. 3), later modified to VX = 10.3 km $s^{-1}$, VY = –12.2 km $s^{-1}$, VZ = –18.0 km $s^{-1}$, or 24.1 km $s^{-1}$entry speed (http://neo.jpl.nasa.gov/fireballs/). Both solutions have the bolide arriving on a shallow path from the NE. The video-derived direction of the trajectory is ~74º rotated in azimuth to the west. Our derived entry angle of 53.4º is also steeper than the 28º calculated from the ECF vector components. Our calculated entry speed of 17.1 ± 0.8 km $s^{-1}$ agrees with the 17.9 km $s^{-1}$ initially reported, not with the later value, but that may be coincidental given that the actual direction of the trajectory is quite different than reported from USG satellite observations.

*Dynamical Orbit and Search for Pre-Impact Observations*

Because the orbital elements for the Sariçiçek meteoroid were derived from serendipitous video observations, they are necessary less precise than the atmospheric trajectories and pre-atmospheric orbits of meteorites that fall in dedicated photographic all-sky camera networks. However, the results are sufficiently accurate to investigate the possible serendipitous pre-impact imaging of Sariçiçek. Assuming the estimated diameter of 1.0 m and an albedo of 0.15, the Sariçiçek meteoroid would have been an H = 32.7 magnitude object, which can be detected by NEO survey programs just prior to impact. Using methods described in Clark and Wiegert (2011), we generated 1000 test meteoroids using Gaussian distributions across the means and uncertainties of the Sariçiçek first detection state documented in Table 3. The test meteoroid probability cloud members were integrated back in time using a RADAU integrator (Everhart 1985). The trajectory and expanse of the cloud was then matched against our 10,000,000+ sky survey image database which includes major ground-base surveys (e.g. Catalina, PanSTARRS, CFHT), all surveys contributing to the Minor Planet Centre Sky Coverage database, and space-based surveys (HST, WISE, NEOWISE). Visibility of the object was calculated using the asteroidal diameter-magnitude relationship of Bowell et al. (1989).

Back integration of the contact state and uncertainties yields the 60-day prior to contact dynamical orbit shown in Table 3.  Sixty days was chosen as a somewhat arbitrary cutoff ensuring that the meteoroid is outside of Earth's influence, and any ongoing orbital changes are primarily due to precession. The small discrepancy in longitude of the ascending node is due to gravitational impact on the node being missed by analytic methods.

The object’s near 23° inclination combined with its greater orbit speed result in near equal rate of change in solar longitude, and apparent vertical decent of Sariçiçek from the northern ecliptic direction. As a result, the meteoroid phase angle was near 90° for much of its approach. 15 min prior to contact Sariçiçek was an apparent magnitude $m$ = 15.9 object with a phase angle Ø of 103.9°. Its magnitude dropped quickly at greater distance and decreasing phase angle: m = 17.2, Ø = 98.6° at 30 min, $m$ = 18.6, Ø = 94.9° at 1 h, $m$ = 20.0, Ø = 92.7° at 2 h; reducing in to magnitude $m$ = 24 (the approximate limiting magnitude of large surveys) and Ø = 90.5° 13 h prior to contact at just over 600,000 km (less than two lunar distances) from the Earth.

Because of uncertainties in initial position and direction, the probability cloud is quite large. The cloud quickly expands to a 7° apparent width as we move 1 h back in time from contact, remaining approximately that apparent size for the preceding days as ongoing expansion is counteracted upon by distance. Our image database contains 62,729 images captured during the 2 months prior to the Sariçiçek impact. Only 12 images were found to contain any members of the probability cloud, these being images taken by Pan-STARRS on August 29 when the object was at m = 28.5 and 32.1 respectively, well beyond the limiting magnitude of PanSTARRS. Based on counts of probability cloud elements that intersect these images, the probability that these images actually contained the object is low, ranging from 0.2% to 4.0%.

It is difficult to make any statement on the visibility of Sariçiçek on prior passages through the ecliptic. Uncertainties in meteoroid contact velocity manifest themselves as uncertainties in orbit semimajor axis and period. Twenty-one months prior to contact, the first opportunity for a prior near approach, the probability cloud has dispersed a full 270° in true anomaly, making circumstances of a prior fly-by indeterminable. In the most opportune (and theoretical) of cases, the object would have been detectable for just over 1 day.

*Constraints on the Source Region from the Pre-Atmospheric Orbit*

Based on the Sariçiçek orbit, we estimate the likelihood for its entrance route (ER) into the NEO region using the NEO model presented in Granvik et al. (2016). The probabilities vary as a function of the absolute magnitude (a proxy for the physical size), throughout the modelled diameter interval ranging from about 2 km to 30 m. The most likely ERs are in the Hungaria population and the inner Main Belt, with delivery via the $\nu_6$ resonance (Fig. 29). The other ERs are substantially more unlikely with an estimated combined probability of about 5%. Among inner Main Belt ERs, the relatively high inclination suggests that a high-inclination source such as Vesta or the Vesta asteroid family is the most probable if the Sariçiçek or its parent escaped the Main Belt through the $\nu_6$ resonance.

The high probability for a source in the Hungaria population is a consequence of the Hungaria-like orbital inclination of Sariçiçek. However, the Hungaria population has very different spectral properties than Vesta (Kelley and Gaffey 2002), making it unlikely Sariçiçek originated from the Hungaria family.

**Constraints on the Source Crater of Normal 22 Ma Clan HED**

*Can the Source Be the Disruption of a Vesta Family Asteroid?*

HED meteorites are often thought to be debris from the ongoing collisional disruption of Vesta's asteroid family, the Vestoids, members of which come to Earth preferably via the 3:1 resonance (Binzel et al. 2002; Davis et al. 2002, Moskovitz et al. 2008). This family consists of 0.8–8 km sized fragments from the formation of the large Rheasilvia basin and is the source of most V-class near-Earth asteroids (Ivanov and Melosh 2013).

The 22 Ma clan of HED meteorites do not appear to originate directly from the V-class near-Earth asteroids, although these can experience collisions in the asteroid belt or disrupt in processes other than collisions. Such disruptions would put meteoroids on Earth-crossing orbits almost immediately and would not require another ~22 Ma to arrive at Earth.

In a Main Belt population of V-class asteroids in collisional equilibrium with their surroundings, small fragments tend to represent a variety of CRE ages. Based on a collisional equilibrium size frequency distribution (Ivanov and Melosh 2013), extrapolation of the observed population of Vestoids into the small-size regime would result in about $3–7 \times 10^7$ asteroids of $D \geq 1$ m diameter, after taking into account that Yarkovsky and YORP thermal recoil forces quickly remove the smallest asteroids with an escape rate proportional to the semi-major axis drift rate, which is proportional to $1/D$.

Only collisions involving asteroids with $D \geq 0.3$ km would be capable of producing sufficient meteoroids with similar CRE age to account for the 22 Ma clan of HED at Earth (see below). This would make the 525 km diameter Vesta a more likely place of that impact, because it has at least a five times larger cross section for collisions than that of all >0.3 km sized Vesta family members combined.

*Possible Source Crater on Vesta.*

NASA's Dawn mission imaging of Vesta provides crater retention ages that constrain the location of possible source craters. Unfortunately, crater count derived ages have a factor of 2–5 systematic uncertainties, depending on the crater size range considered, due to the unknown population of small impactors. The distribution of projectile diameters below ~3 km is extrapolated either based on computer models involving collisional evolution, resulting in the asteroid flux model chronology system (Marchi et al. 2012; O'Brien et al. 2014), or based on the observed lunar cratering record of small craters, called the lunar surface chronology system (Schedemann et al. 2014).

Several craters with well-defined ejecta blankets overlaying old terrain have been dated by the Dawn mission team by counting impacts on these ejecta blankets. Figure 30 summarizes the estimated formation age of Vesta craters as far as they are known now. Results are given in both chronology systems.

The results show that in the lunar surface chronology system (black dots in Fig. 30), many of the smaller craters have ages in the range of CRE ages of HEDs. In contrast, the asteroid flux model chronology system (grey dots in Fig. 30) assigns ages >100 Ma for small craters, which is significantly older than the 22 Ma clan of HED meteorites. In this scheme, none of the dated craters on Vesta could have produced these HED.

The lunar surface chronology system provides a ~3.5 Ga formation age of the Rheasilvia basin (Yingst et al. 2014) that is contemporary with the ~3.6 Ga K-Ar age of Sariçiçek (Fig. 30). Most other non-anomalous HED meteorites experienced such collision 4.1–3.4 Ga ago based on published Ar-Ar ages (Bogard and Garrison 2003; Bogard 2011) and the 3.8–3.3 Ga old impact melt clasts in howardites (Cohen 2013). The asteroid derived chronology system suggests ~1.2 Ga, instead (Marchi et al. 2012). The dynamical age of the Vestoids is at least 1 Ga, but an age of up to 3.8 Ga is needed to account for outliers at semi-major axis 2.3–2.5 AU. The observed combined Vestoid mass is consistent with a collisional and dynamical mass loss over ~3.5 Ga (Moskovitz et al. 2008).

Of all craters on Vesta, the relevant craters in each size bracket are those young enough to be sharp-rimmed with a well defined ejecta blanket. Those happen to be also the ones studied by the Dawn team. Of all craters shown in Fig. 30 in the lunar surface chronology system, the relevant candidate source crater for the 22 Ma HED clan in each age bracket is the largest crater, which likely produced most fragments. Larger craters tend to be older because the impact frequency increases sharply with decreasing impactor size. The formation age of the crater should be less than the highest HED cosmic ray exposure age of ~100 Ma, but not be so young that debris can not yet have reached Earth.

The larger craters Marcia, Oppia and Octavia (and Publicia–personal communication) were all found 100–200 Ma old in the lunar surface chronology system, much older than the CRE age of most HED (Ruesch et al. 2014; Williams et al. 2014a, 2014b). Licinia is the youngest known large crater with a formation age in the range of measured CRE ages (Fig. 30). If the 49.5 ± 4.6 Ma date of Ruesch et al. (2014) is accurate, Licinia is not responsible for the 22 Ma clan of HED.

The next biggest crater is the 16.75 km diameter Antonia impact crater, which has an asymmetric ejecta blanket directed mostly downhill (Fig. 31). It is located on a slope in the Rheasilvia impact basin, where old terrain howardites overlay diogenite materials as a result of landslides and impact gardening. Only howardites contain the recent solar wind implanted noble gases, suggesting that a deposit of howardites covered eucrite- and diogenite-rich units at the time of the impact. Based on the scaled systematics of lunar surface chronology, crater size-frequency distributions on two different parts of the ejecta blanket of Antonia provided the ages of 18.5 ± 1.2 Ma (Antonia Ejecta East) and 23.7 ± 1.1 Ma (Antonia Ejecta West) (Kneissl et al. 2014), in good agreement with the measured CRE ages of 22 Ma clan HED and that of Sariçiçek (Fig. 31). Antonia Ejecta South gave an age of 12.9 ± 1.4 Ma, which is significantly younger, but this age was based on a smaller area with much less craters (97 versus 477 and 348, respectively). While consistent in the lunar surface chronology system, the asteroid-derived chronology gives a factor of four higher ages (Fig. 30).

Crater size-frequency distributions of the underlying terrain provided an age of 1.45 ± 0.64 Ga (Antonia ejecta West) in the same system (Kneissl et al. 2014). We noticed that this age is similar to the relatively low radiogenic retention ages derived for $^{4}He$ in Sariçiçek SC12 (1.8 ± 0.7 Ga) and SC14 (2.6 ± 0.3 Ga). However, the $^{4}He$ content are necessarily derived from different aliquots of SC12 and SC14 than used for the composition measurement (both are destructive measurements). Measured values of the U,Th concentrations are given in Tables 5 and 11, with U concentrations ranging from 34 ppb (SC12) to 57 ppb (SC26). Figure 32 shows how the derived U,Th-He age changes if the U concentration in the measured aliquot was different from that measured in another part of the same stone. Unless both the SC12 and SC14 aliquots had significantly lower U (and Th) concentrations than measured for each stone, both ages fall below the common ~3.7 Ga peak of Ar-Ar ages in HED (Bogard 2011). The measured U,Th-He resetting ages are upper limits, because not all accumulated radiogenic He may have been lost from minerals in the smaller impacts that determine the terrain age. Indeed, SC12 could have been shocked more than the more fragile SC14. The terrain age around Antonia is also similar to young ~1.4 Ga $^{40}Ar$-$^{39}Ar$ ages measured for feldspar grains close to a glass vein in the howardite Kapoeta (Lindsay et al. 2015).

During the formation of that terrain, solar wind noble gases were implanted into the outer micron of individual soil particles of loose regolith on the surfaces of airless bodies (McKay et al. 1991). Impact gardening mixes those grains in the regolith, gradually accumulating noble gases. When that regolith is lithified by shocks to form a regolithic breccia assembly, only some of that gas is released. Sariçiçek's high content of solar wind implanted noble gases implies it was excavated from material that accumulated from soil particles at the surface of Vesta. Those particles included exogenous matter and shocked material from impacts elsewhere on Vesta (shocked enough to cause flattening of Fe and FeS). Sariçiçek's regolith exposure age of ~12–14 Ma may signify the regolith mixing timescale at the depth from which this meteoroid was excavated.

*Dynamical Arguments for a Source Crater on Vesta.*

Using an ejecta scaling model (Housen and Holsapple 2011), 3–9 × $10^{11}$ kg of material can have escaped the gravity of Vesta from the Antonia impact. Assuming a Weibull size-frequency distribution with the same 2.0 shape parameter exponent as the Vestoids (Ivanov and Melosh 2013), but a scale parameter of 1.0 m, puts most mass in the size range of 0.25–1.25 m, with ~7 × $10^7$ fragments $D \geq$1-m in size and the largest fragment ~4 m.

To estimate the fraction of Earth impactors among Vestoids that escape the main asteroid belt through the $\nu_6$ secular resonance, we reanalyzed integrations of 4153 test asteroids with initial osculating orbital elements similar to the Vesta family $2.26 < a < 2.48$ AU, $0.035 < e < 0.162$, and $5.0 < i < 8.3^o$ until they end up in one of the predefined sinks: a collision with a planet or the Sun, or escape from the inner solar system (Granvik et al. 2016, 2017). Vesta itself is located near the center of the Vesta family, with proper elements $a = 2.362$ AU, $e = 0.099$, and $i = 6.36^o$. Granvik et al. 2016, 2017) used an augmented version of the RMVS3 integrator implemented in the SWIFT package (Levison and Duncan 1994) that allows modeling of Yarkovsky drift in semimajor axis. The Yarkovsky force causes the test asteroids to drift towards the $\nu_6$ secular resonance, the 3:1 mean motion resonance and a variety of minor resonances (Nesvorny et al. 2008), which then increases their eccentricity until the test asteroids reach the near-Earth space, defined as a perihelion distance $q < 1.3$ AU. The short semimajor axis of Sariçiçek during impact suggests it drifted towards the $\nu_6$ secular resonance. At this stage, Granvik et al. (2016) turned off the Yarkovsky drift because it becomes negligible compared to the perturbations caused by close planetary encounters. For modeling the Yarkovsky drift they assumed that all test asteroids have the same diameter $D = 100$ m. The assumption for diameter has a negligible effect on the orbital distribution for test asteroids entering the near-Earth space (Granvik et al. 2017), and the results can therefore be assumed representative also for 1 meter diameter class objects. Finally we find that 84 test asteroids (2.0 ± 0.2%) eventually impact the Earth.

Yarkovsky and YORP forces determine the transfer efficiency from Vesta to the $\nu_6$ resonance. Based on the integrations, it takes 79 Ma on average for 100 m sized Vestoids to evolve from their source in the Vesta asteroid family into a perihelion distance q = 1.3 AU orbit and another 14 Ma to impact Earth (total 93 Ma). The manner in which the ejection conditions put Antonia ejecta into a slightly different regime of semi-major axis and eccentricity has little effect. The final phase is not size dependent, because it is dominated by gravitational perturbations. The early phase is size dependent, with smaller asteroids evolving faster into resonances. The canonical Yarkovsky drift rate of semimajor axis is 0.0002 AU/Ma for $D$ = 1 km, and inverse scaling with D. For a 1 m meteoroid, that time is about 1.3 Ma. We conclude that it takes Antonia debris about 1.3 + 14 = 15.3 Ma to reach Earth by way of the $\nu_6$ resonance that delivered Sariçiçek.

The 1.3 Ma time can be longer if YORP cycling is important. The idea behind YORP cycling is that YORP will spin-up a body until a deformation of the shape or fragmentation of the body will counteract further spin-up or even dramatically slow down the rotation rate (Bottke et al. 2015, Granvik et al. 2017). The obliquity of the spin axis is more easily changed, by, e.g., impacts from small meteoroids, when the rotation rate has slowed down. A change in the obliquity of the spin axis has a substantial effect on the drift rate in semimajor axis caused by the Yarkovsky effect. In the most extreme cases it may even change the direction of the Yarkovsky drift thereby leading to a Brownian-like drift in semimajor axis. YORP will eventually spin up the body again and the cycle will repeat. In recent simulations (Granvik et al. 2017), it was found that accounting for YORP cycles will lead to a net Yarkovsky drift rate that is comparable to an instantaneous Yarkovsky drift rate for asteroids two orders of magnitude larger. That is, the instantaneous Yarkovsky drift rate for 100 m sized objects (used in the simulations discussed above) corresponds to the net drift rate for 1 m sized objects with YORP cycling. If YORP cycling is important even for small 1 m sized meteoroids like Sariçiçek, then Antonia debris would take about 93 Ma to impact Earth. Such a long timescale is unlikely (cf. Granvik et al. 2017), given that this would produce much higher CRE ages than observed, but it does suggest that YORP may increase the travel time to somewhat higher than 15.3 Ma.

The observed rate of impacts (about one asteroid with $D \geq 1$ m every 4 years globally) is understood if Earth now experiences the wave of Antonia debris at its peak, and the duration of the wave is only 5–10 Ma.

*Possible Source Craters of Other HED Meteorites.*

Meter-sized debris from Marcia, Oppia, Octavia, and Publicia have long passed. If Licinia's debris of age 49.5 ± 4.6 Ma (Ruesch et al. 2014) accounts for the CRE peak at ~54 Ma (Fig. 31), then some of Licinia's debris may still be on its way to Earth. Licinia could account for HED meteorites with CRE ages in the range 45–55 Ma (Fig. 30). The relatively large amount of debris that was excavated could explain why these meteoroids are still arriving at Earth in significant numbers. This would imply that the arrival time distribution at Earth has a long ~50 Ma tail, as perhaps expected from variable YORP cycling.

After Antonia, the next biggest craters with ejecta blankets, 14.9 km Cornelia (dated to 9–14 Ma by Krohn et al. 2014), 11.6 km Fabia, and 11.3 km Canuleia can be responsible for other peaks in the CRE age distribution of Fig. 31. If so, these groups sample different terrains and may have different material properties.

If 10.3 km wide Rubria (14–24 Ma according to Krohn et al. 2014) is contemporary with Antonia, it would add ~23% of ejecta to the Antonia peak based on the relative crater size. Rubria is located on the Rheasilvia ejecta blanket, but has reddish ejecta that may be unlike Sariçiçek. The reddish material has been interpreted as Rheasilvia impact melt or HED material rich in exogenic (non HED) material (Le Corre et al. 2013, 2015).

Material from the young (~2.5 Ma: Ruesch et al. 2014) 10.5 km Arruntia crater is likely still on its way to the $\nu_6$ resonance (Fig. 30). The next biggest, 8.1 km Sossia, and all smaller craters with ejecta blankets together ejected only ~60% of the ejecta mass of Antonia. Some of those impacts will be too young for material to have arrived at Earth.

## CONCLUSIONS

The Sariçiçek meteoroid arrived at Earth on a 23º inclined orbit with a semimajor axis of ~1.44 AU. It approached Earth in an apparent vertical descent from northern ecliptic latitudes, reaching +16 magnitude brightness with a 104º phase angle 15 min prior to contact with Earth's atmosphere, but in an area of the sky for which no archived images were found.

The impact created an airburst sufficient to trigger seismic sensors, albeit 300 times less powerful than the one that swept the city of Chelyabinsk. The 53º inclined entry angle in Earth's atmosphere and extended duration of energy deposition caused a more pronounced overpressure and sharper onset in the backward direction, where travel times from different sources along the trajectory coalesced.

The Sariçiçek meteorites consist of mostly lithic and mineral clasts of eucritic composition in a fine-grained matrix of crushed material. The diogenitic component of this eucrite-diogenite mixture exceeds 10 vol%, which defines this as a howardite.

Oxygen, chromium (Cr), and titanium (Ti) isotopic compositions confirm that Sariçiçek originated from the same parent body as the HED clan of meteorites, excluding the anomalous eucrites. Bulk Sariçiçek (SC) rock chips of stones SC12 and SC14 have a mean $\Delta^{17}O' = -0.31 \pm 0.08$ and $-0.31 \pm 0.06$ ($2\sigma$), respectively, with $\delta^{18}O = 3.54 \pm 0.20$ and $3.60 \pm 0.14$, overlapping the normal HED ranges of $\Delta^{17}O' = -0.247 \pm 0.050$ and $\delta^{18}O = 3.74 \pm 0.56$ ($2\sigma$). Cr and Ti isotope compositions of Sariçiçek are also normal, with. $\varepsilon^{54}Cr = -0.66 \pm 0.07$, $\varepsilon^{46}Ti/^{47}Ti = -0.19 \pm 0.16$, $\varepsilon^{48}Ti/^{47}Ti = +0.05 \pm 0.16$, and $\varepsilon^{50}Ti/^{47}Ti = -1.26 \pm 0.21$, compared to Bholghati's $\varepsilon^{54}Cr = -0.63 \pm 0.10$, and Kapoeta's $\varepsilon^{46}Ti/^{47}Ti = -0.22 \pm 0.05$, $\varepsilon^{48}Ti/^{47}Ti = -0.14 \pm 0.20$, and $\varepsilon^{50}Ti/^{47}Ti$ $-1.23 \pm 0.04$, respectively.

Further geochemical analysis of Sariçiçek sheds new light on the origin and evolution of the howardite material. The $^{207}Pb/^{206}Pb$ ages of Sariçiçek apatites at $4524 \pm 12$ Ma ($2\sigma$, 8 grains) are systematically younger by about 27 Ma compared to those of zircons at $4551.1 \pm 2.8$ Ma ($2\sigma$, 30 grains) and baddeleyites at $4558.0 \pm 8.2$ Ma ($2\sigma$, 10 grains). The latter coincide in age with a period identified earlier as the peak age of basaltic magmatism on the eucrite parent body, presumably when the crust of the parent body solidified. The possiblly younger zircon age is likely due to the progressive silica saturation in the magma where zircon formed later or metasomatic replacement of primary baddeleyite by silica-rich fluid or partial melt. The younger apatite-derived age suggests that the lower closure temperature for Pb diffusion in this phase allowed the grains to continue to equilibrate during thermal metamorphism for another ~27 Ma.

If the measured K composition is representative for the sample analysed for noble gas content, then the terrain in which Sariçiçek resided experienced at least one large impact able to reset the K-Ar chronometer ~3.6 Ga ago. Other non-anomalous HED meteorites experienced such collision(s) 4.1–3.4 Ga ago, possibly from the formation of the, partially overlapping, Rheasilvia and Veneneia impact basins.

If the measured U and Th composition is representative, stones SC12 and SC14 have U,Th-He ages of 1.8 ± 0.7 Ga and 2.6 ± 0.3 Ga, respectively, younger than the K-Ar ages. These ages are interpreted to be due to radiogenic $^{4}$He loss from smaller impacts that continued to affect the terrain where Sariçiçek was excavated.

In recent times, Sariçiçek was situated close to the terrain surface. It is rich in noble gas isotopes from solar wind implantation, with $^{20}$Ne = 1.8 nL $g^{-1}$ STP, and in siderophile elements from chondritic contamination, with an iridium abundance of 8–10 ng $g^{-1}$.

Both noble gas implantation and cosmogenic radionuclides show Sariçiçek was a meteoroid and exposed to cosmic rays during the past 22 ± 2 Ma. Thus, the Sariçiçek orbit is specifically relevant to the common HED group with cosmic ray exposure ages of 19–26 Ma. There is evidence of prior exposure for 12–14 Ma, when the meteoroid was still part of a larger body, confirming that Sariçiçek was located close to the surface at the time of the collision.

Based on the meteorite's natural thermoluminescence, the orbit did not evolve much closer to the Sun than the 1.0 AU distance at impact. Of all near-Earth objects approaching in such inclined and short orbits, ~66% are expected to originate from the 16–34º inclined Hungaria asteroid family and ~31% from lower inclined sources in the inner Main Belt that deliver through the $\nu_6$ resonance. Hungaria asteroids do not have Sariçiçek's two prominent pyroxene absorption bands at 0.93 ± 0.01 and 1.98 ± 0.02 µm wavelength, but Vesta and the 5–8º inclined Vesta asteroid family (the Vestoids) do, which are located in the inner Main Belt.

The 16.7 km Antonia impact crater on Vesta is large enough to account for the influx of 10 cm to 1 m sized HED to Earth, given the fraction that can escape the gravity of Vesta, the time it takes to evolve into a Sariçiçek-like impact orbit, and the fraction of meteoroids that impact Earth.

Antonia was formed on terrain of the same age as given by the $^{4}$He retention age of Sariçiçek. Lunar scaling for crater production to crater size-frequency distributions of its ejecta blanket show it was formed ~22 Ma ago, contemporary with the CRE age of the 22 Ma clan of HED meteorites.

Sariçiçek represents a typical 1.0 ± 0.2 m diameter sample of Antonia impact ejecta, terrain from the Rheasilvia impact basin where most V-class asteroids originated. The pre-atmospheric size of Sariçiçek was determined from cosmogenic radionuclide concentrations of $^{10}$Be, $^{36}$Cl, and $^{26}$Al and is consistent with the observed optical luminosity and infrasound energy.

This suggests that some material properties of Sariçiçek, such as the Weibull coefficient for the fracture length scale distribution, may be representative for large V-class asteroids. The manner in which Sariçiçek fragmented in Earth's atmosphere, depositing its kinetic energy at 37–27 km altitude and generating an airburst that was detected at the surface, provided ground truth for models designed to forecast the damage from future Vestoid impacts.

*Acknowledgments*–We thank N. Ergün and family in the village of Sariçiçek for donating the meteorites studied here and collecting meteorite fall coordinates. We thank E. Atalan and S. Özdemir at Bingöl University, and E. Necip Yardım and M. Çiçek at Muş Alparslan University, for facilitating our research at the campuses, and S. Pamuk at the Bingöl police headquarters. We thank A. and T. Özduman, police officers in Bingöl, for assisting with the field study. For technical assistance, we further acknowledge support from M. Fehr, Y.-J. Lai, and L. Hoffland (NASA Ames Research Center), David Mittlefehldt (NASA JSC), K. Wimmer (Ries Crater Museum), J. Sanchez (Planetary Science Institute), A. Neesemann (Free University Berlin), S. Atanasova-Vladimirova and I. Piroeva (Institute of Physical Chemistry, BAS), and B. Georgieva and V. Strijkova (Institute of Optical Materials and Technologies, BAS). This work was supported by Istanbul University (Project No. 40339 and 58261), the Scientific and Technological Research Council of Turkey (MFAG/113F035), the Swiss National Science foundation (PZ00P2_154874 and NCCR PlanetS), the Ministry of Science and Higher Education of the Russian Federation (Project # 3.1959.2017/4.6), Act 211 of the Government of the Russian Federation, contract № 02.A03.21.0006, the National Natural Science Foundation of China (41403055), the Simons Foundation (302497), the Academy of Finland (299543), the NASA Cosmochemistry Program (NNX14AM62G), the NASA Emerging Worlds Program (NNX16AD34G), and the NASA NEOO program (NNX14-AR92G).

*Editorial handling* – Dr. Akira Yamaguchi

## REFERENCES

Akram W., Schönbächler M., Bisterzo S., and Gallino R. 2015. Zirconium isotope evidence for the heterogeneous distribution of s-process materials in the solar system. *Geochimica et Cosmochimica Acta* **165**:484–500.

Amelin Y., Li C., and Naldrett, A. J. 1999. Geochronology of the Voisey's Bay intrusion, Labrador, Canada, by precise U–Pb dating of coexisting baddeleyite, zircon, and apatite. *Lithos* **47:**33-51.

Arpesella C. 1996. A low background counting facility at Laboratori Nazionali del Gran Sasso. *Applied Radiation and Isotopes* **47**:991–996.

Asphaug E., Ryan E. V., and Zuber M. T. 2002. Asteroid interiors. In: *Asteroids III.* W. F. Bottke Jr., A. Celino, P. Paolicchi, R. P. Binzel, eds., University of Arizona Press, Tucson, p. 463–484.

Beck A. W., Welten K. C., McSween H. Y. Jr. Viviano C., and Caffee M. W. 2012. Petrologic and textural diversity among the PCA 02 howardite group, one of the largest pieces of the Vestan surface. *Meteoritics and Planetary Science* **47**:947–969.

Benedix G. K., Bland P. A., Friedrich J. M., Mittlefehldt D. W., Sanborn M. E., Yin Q.-Z., Greenwood R. C., Franchi I. A., Bevan A. W. R., Towner M. C., Perotta G. C., and Mertzman S. 2017. Bunburra Rockhole: Exploring the geology of a new differentiated asteroid. *Geochemica et Cosmochimica Acta*. **208**, 145–159.

Bhandari N., Mathew K. J., Rao M. N., Herpers U., Bremer K., Vogt S., Wölfli W., Hofmann H. J., Michel R., Bodemann R., and Lange H.-J. 1993. Depth and size dependence of cosmogenic nuclide production rates in stony meteoroids. *Geochimica et Cosmochimica Acta* **57**:2361–2375.

Binzel R. P., Lupishko D. F., Martino M. D., Whiteley R. J., and Hahn G. J. 2002. Physical properties of Near-Earth Objects. In: *Asteroids III.* W. B. Bottke, A. Cellino, P. Paolicchi, and R. P. Binzel (eds.), University of Arizona Press, Tucson, AZ, p. 255–271.

Bland P. A., Spurny P., Towner M. C., Bevan A. W. R., Singleton A. T., Bottke W. F., Greenwood R. C., Chesley S. R., Shrbeny L., Borovicka J., Ceplecha Z., McCLafferty T. P., Vaughan D., Benedix G. K., Deacon G., Howard K. T., Franchi I. A., and Hough R. M. 2009. An anomalous basaltic meteorite from the innermost main belt. *Science* **325**:1525–1527.

Bogard D. D., Garrison D. H. 2003. 39Ar-40Ar ages of eucrites and thermal history of asteroid 4 Vesta. *Meteoritics and Planetary Science* **38**:669–710.

Bogard D. D. 2011. K-Ar ages of meteorites: Clues to parent-body thermal histories. *Chemie der Erde–Geochemistry* **71**:207–226.

Bonnet E., Bour O., Odling N. E., and Davy P. 1999. Scaling of fracture systems in geological media. *Reviews of Geophysics* **39**:347–383.

Borovička J., Spurny P., Brown P., Wiegert P., Kalenda P., Clark D., and Shrbeny L. 2013. The trajectory, structure and origin of the Chelyabinsk asteroidal impactor. *Nature* **503**:235–237.

Boslough M. B. E. and Crawford D. A. 2008. Low-altitude airbursts and the impact threat. *International Journal of Impact Engineering* 35:1441–1448.

Bottke W. F., Vokrouhlický D. Walsh K. J., Delbo M., Michel P., Lauretta D. S., Campins H., Connolly H. C., Scheeres D. J., and Chelsey S. R. 2015. In search of the source of asteroid (101955) Bennu: Applications of the stochastic YORP model. *Icarus* **247**:191–217.

Bouvier A., Gattaceaca J., Grossman J., and Metzler K. 2016. The Meteoritical Bulletin, No. 105. *Meteoritics & Planetary Science*1 (2017). Doi: 10.1111/maps.12944.

Bowell E., Hapke B., Domingue D., Lumme K., Peltoniemi J., and Harris A. W. 1989. Application of photometric models to asteroids. *Asteroids II*. Proceedings of the conference, Tucson, AZ, Mar. 8-11, 1988 (A90-27001 10-91), 524–556.

Brown P., Assink J. D., Astiz L., Blaauw R., Boslough M. B., Borovicka J., Brachet N., Brown D., Campbell-Brown M., Ceranna L., Cooke W., de Groot-Hedlin C., Drob D. P., Edwards W., Evers L. G., Garces M., Gill J., Hedlin M., Kingery A., Laske G., Le Pichon A., Mialle P., Moser D. E., Saffer A., Silber E., Smets P., Spalding R. E., Spurny P., Tagliaferri E., Uren D., Weryk R. J., Whitaker R., and Krzeminski Z. 2013. A 500-kiloton airburst over Chelyabinsk and an enhanced hazard from small impactors. *Nature* **503**:238–241.

Brown P., Wiegert P., Clark D., and Tagliaferri E. 2016. Orbital and physical characteristics of meter-scale impactors from airburst observations. *Icarus* **266**:96–111.

Campo Bagatin A., Aleman R. A., Benavidez P. G., and Richardson D. C. 2018. Internal structure of asteroid gravitational aggregates. *Icarus* **302**:343–359.

Cansi Y. 1995. An automatic seismic event processing for detection and location: the P.M.C.C. method. *Geophysical Research Letters* **22**:1021–1024.

Cartwright J. A., Ott U., Mittlefehldt D. W., Herrin J. S., Herrmann S., Mertzman S. A., Mertzman K. R., Peng Z. X., and Quinn J. E. 2013. The quest for regolithic howardites. Part 1: Two trends uncovered using noble gases. *Geochimica et Cosmochimica Acta* **105**:395–421.

Cartwright J. A., Ott U., and Mittlefehldt D. W. 2014. The quest for regolithic howardites. Part 2: Surface origins highlighted by noble gases. *Geochimica et Cosmochimica Acta* **140**:488–508.

Ceplecha Z. and ReVelle D. 2005. Fragmentation model of meteoroid motion, mass loss, and radiation in the atmosphere. *Meteoritics & Planetary Science* **40**:35–54.

Christie D. R. and Campus P. 2010. The IMS Infrasound Network: Design and establishment of infrasound stations. In: *Infrasound monitoring for atmospheric studies*, A. Le Pichon, E. Blanc, A. Hauchecorne (eds.) 1st ed. Dordrecht: Springer, 27–73.

Clark D. and Wiegert P. 2011. A numerical comparison with the Ceplecha analytical meteoroid orbit determination method. *Meteoritics & Planetary Science* **46**:1217–1225.

Clayton R. N. and Mayeda T. K. 1996. Oxygen isotope studies of achondrites. *Geochimica et Cosmochimica Acta* **60**:1999–2017.

Clayton R. N. and Mayeda, T. K. 1999. Oxygen isotopes studies of carbonaceous chondrites. *Geochimica et Cosmochimica Acta* **63**:2089–2104.

Cloutis E. A., Mann P., Izawa M. R. M., Nathues A., Reddy V., Hiesinger H., Le Corre L., and Palomba E. 2013. The 2.5–5.1 mm reflectance spectra of HED meteorites and their constituent minerals: Implications for Dawn. *Icarus* **255**:581–601.

Cohen B. 2013. The Vestan cataclysm: Impact-melt clasts in howardites and the bombardment history of 4 Vesta. *Meteoritics & Planetary Science* **48**:771–785.

Consolmagno G. J. and Drake M. J. 1977. Composition and evolution of eucrite parent body: Evidence from rare earth elements. *Geochimica et Cosmochimica Acta* **41**:1271–1282.

Cotto-Figueroa D., Asphaug E., Garvie L., Morris M., Rai A., Chattopadhyay A., and Chawla N. 2015. Scale-Dependent Measurements of Meteorite Strength and Fragmen- tation: Tamdakht (H5) and Allende (CV3). In AAS/Division for Planetary Sciences Meeting Abstracts, volume 47 of AAS/Division for Planetary Sciences Meeting Abstracts, page 213.20.

Cruikshank D. P., Tholen D. J., Bell J. F., Hartmann W. K., and Brown R. H. 1991. Three basaltic earth-approaching asteroids and the source of the basaltic meteorites. *Icarus* **89**:1–13.

Davidson A. and van Breemen O. 1988. Baddeleyite-zircon relationships in coronitic metagabbro, Grenville Province, Ontario: implications for geochronology. *Contributions to Mineralogy and Petrology* **100:**291–299

Davis D. R., Durda D. D., Mrzari F., Bagatin A. C., and Gil-Hutton R. 2002. Cosmic evolution of small body populations. In: *Asteroids III*. W. Bottke, A. Cellino, P. Paolicchi, R. P. Binzel (eds.), University of Arizona Press, Tucson, Az., p. 545–558.

Edwards W. N., Brown P. G., and ReVelle D. O. 2006. Estimates of meteoroid kinetic energies from observations of infrasonic airwaves. *Journal of Atmospheric and Solar-Terrestial Physics* **68**:1136–1160.

Ens T. A., Brown P. G., Edwards W. N., and Silber E.A. 2012. Infrasound production by bolides: A global statistical study. *Journal of Atmospheric and Solar-Terrestrial Physics* **80**:208–229.

Eugster O. and Michel T. 1995. Common asteroid break-up events of eucrites, diogenites, and howardites and cosmic-ray production rates for noble gases in achondrites. *Geochimica et Cosmochimica Acta* **59**, 177–199.

Everhart E. 1985. An efficient integrator that uses Gauss-Radau spacings. (A. Carus, & G. B. Valsecchi, Eds.) *Dynamics of Comets: Their Origin and Evolution*. Proceedings of IAU Colloq. **83**, held in Rome, Italy, June 11-15, 1984, pp. 185–202.

Friedrich J. M., Wang M.-S., and Lipschutz M. E. 2003 Chemical studies of L chondrites. V: Compositional patterns for 49 trace elements in 14 L4–6 and 7 LL4–6 Falls. *Geochimica et Cosmochimica Acta* **67**:2467–2479.

Fujiwara A., Cerroni P., Davis D. R., Ryan E., di Martino M., Holsapple K., and Housen K. 1989. Experiments and scaling laws for catastrophic collisions. *Asteroids II*, R. P. Binzel, T. Gehrels, M. Shapley, eds., University of Arizona Press, Tucson, p. 240–265.

Furnish M., Boslough M., Gray G., and Remo J. 1995. Dynamical properties measurements for asteroid, comet and meteorite material applicable to impact modeling and mitigation calculations. *International Journal of Impact Engineering* **17**:341–352.

Garcés M., Willis M., and Le Pichon A. 2010. Infrasonic observations of open ocean swells in the Pacific: Deciphering the song of the sea. In: *Infrasound monitoring for atmospheric studies*, Le Pichon, A., Blanc E., & Hauchecome A. (eds.), Springer: Dordrecht, pp. 231–344.

Garry W. B., Williams D. A., Yingst R. A., Mest S. C., Buczkowski D. L., Tosi F., Schäfer M., Le Corre L., Reddy V., Jaumann R., Pieters C. M., Russell C. T., and Raymond C. A., the Dawn Science Team 2014. Geologic mapping of ejecta deposits in Oppia Quadrangle, Asteroid (4) Vesta. *Icarus* **244**:104–119.

Gattacceca J., Rochett P., Gounelle M., and Van Ginneken M. 2008. Magnetic anisotropy of HED and Martian meteorites and implications for the curst of Vesta and Mars. *Earth and Planetary Science Letters* **270**:280–289.

Glavin D. P., Dworkin J. P., Aubrey A., Botta O., Doty J. H., Martins Z., and Bada J. L. 2006. Amino acid analyses of Antarctic CM2 meteorites using liquid chromatography-time of flight-mass spectrometry. *Meteoritics & Planetary Science* **41**:889–902.

Glavin D. P., Callahan M. P., Dworkin J. P., and Elsila J. E. 2010. The effects of parent body processes on amino acids in carbonaceous chondrites. *Meteoritics & Planetary Science* **45:**1948–1972.

Goldmann A., Brennecka G., Noordmann J., Weyer S., and Wadhwa M. 2015. The uranium isotopic composition of the Earth and the Solar System. *Geochimica et Cosmochimica Acta* **148**:145–158.

Granvik M., Morbidelli A., Jedicke R., Bolin B., Bottke W. F., Beshore E., Vokrouhlicky D., Delbo M., and Michel P. 2016. Super-catastrophic disruption of asteroids at small perihelion distances. *Nature* **530**:303–306.

Granvik M., Morbidelli A., Vokrouhlický D., Bottke W. F., Nesvorný D., and Jedicke R. 2017. Escape of asteroids from the main belt. *Astronomy & Astrophysics* **598**:id.A52, 13pp.

Grimberg A., Baur H., Bochsler P., Bühler F., Burnett D. S., Hays C. C., Heber V. S., Jurewicz A. J. G., and Wieler R. 2006. Solar wind neon from Genesis: Implications for the lunar noble gas record. *Science* **314**:1133–1135.

Hasan F. A., Haq M., and Sears D. W. G. 1987. Natural thermoluminescence levels in meteorites, I: 23 meteorites of known Al-26 content. *Journal of Geophysical Research* **92**:E703–E709.

Hirabayashi M., Sánchez D. P., and Scheeres D. J. 2015. Internal structure of asteroids having surface shedding due to rotational instability. *Astrophysical Journal* **808**:63–75.

Housen K. R. and Holsapple K. A. 2011. Ejecta from impact craters. *Icarus* **211**:856–875.

Ivanov B. A. and H. J. Melosh 2013. Two-dimensional numerical modeling of the Rheasilvia impact formation. *Journal of Geophysical Research: Planets* **118**:1545–1557.

Jacchia L. G., Verniani F., and Briggs R. E. 1967. An analysis of the atmospheric trajectories of 413 precisely reduced photographic meteors. *Smithsonian Contributions to Astrophysics* **10**:1–45.

Jarosewich E., Clarke R. S., and Barrows J. N. 1987. The Allende meteorite reference sample. *Smithsonian Contributions to the Earth Sciences* **27**:1–49.

Jenniskens P. 2006. *Meteor showers and their parent comets.* Cambridge University Press, Cambridge, U.K., 790 pp.

Jenniskens P., Gural P. S., Dynneson L., Grigsby B. J., Newman K. E., Borden M., Koop M., and Holman D. 2011. CAMS: Cameras for allsky meteor surveillance to establish minor meteor showrs. *Icarus* **216**:40–61.

Jenniskens P., Albers J., Tillier C. E., Edgington S. F., Longenbaugh R. S., Goodman S. J., Rudlosky S. D., Hilebrand A. R., Hanton L., Ciceri F., Nowell R., Lyytinen E., Hladiuk D., Free D., Moskovitz N., Bright L., Johnston C. O., and Stern E. 2018. Detection of meteoroid impacts by the Geostationary Lightning Mapper on the GOES-16 satellite. *Meteoritics & Planetary Science* **53**:2445–2469.

Johnson A. and Remo J. 1974. New interpretation of mechanical-properties of Gibeon meteorite. *Journal of Geophysical Research* **79**:1142–1146.

Karr C. (ed.) 1975. *Infrared and Raman Spectroscopy of Lunar and Terrestrial Minerals*. Academic Press, New York, 390 pp.

Kelley M. S. and M. J. Gaffey 2002. High-albedo asteroid 434 Hungaria: Spectrum, composition, and genetic conditions. *Meteoritics & Planetary Science* **37**:1815–1827.

Kimberley J. and Ramesh, K. T. 2011. The dynamic strength of an ordinary chondrite. *Meteoritics & Planetary Science* **46**:1653–1669.

Klima R. L., Pieters C. M., and Dyar M. D. 2008. Characterization of the 1.2 mm M1 pyroxene band: Extracting cooling history from near-IR spectra of pyroxenes and pyroxene-dominated rocks. *Meteoritics & Planetary Science* **43**:1591–1604.

Kneissl T., Schmedemann N. , Reddy V., Williams D. A., Walter S. H. G., Neesemann A., Michael G. G., Jaumann R., Krohn K., Preusker F., Roatsch T., Le Corre L., Nathues A., Hoffmann M., Schäfer M., Buczkowski D., Garry W. B., Yingst R. A., Mest S. C., Russell C. T., and Raymond C. A. 2014. Morphology and formation ages of mid-sized post-Rheasilvia craters - Geology of quadrangle Tuccia, Vesta. *Icarus* **244**:133–157.

Kohman T. P. and Bender M. L. 1967. Nuclide production by cosmic rays in meteorites and on the Moon. In: *High-Energy Nuclear Reactions in Astrophysics*, B. S. P. Shen, W. A. Benjamin (eds.), New York, N.Y. p. 169–245.

Krohn K., Jaumann R., Elbeshausen D., Kneissl T., Schmedemann N., Wagner R., Voigt J., Otto K., Matz K. D., Preusker F., Roatsch T., Stephan K., Raymond C. A., and Russell C. T. 2014. Asymmetric craters on Vesta: Impact on sloping surfaces. *Planetary & Space Science* **103**:36–56.

Le Corre L., Reddy V., Schmedemann N., Becker K. J., O'Brien D. P., Yamashita N., Peplowski P. N., Prettyman T. H., Li J.-Y., Cloutis E. A., Denevi B., Kneissl T., Palmer E., Gaskell R., Nathues A., Gaffey M. J., Garry B., Sierks H., Russell C. T., Raymond C. A., De Sanctis M. C., and Ammanito E. 2013. Olivine or impact melt: Nature of the orange material on (4) Vesta from Dawn. *Icarus* **226**:1568–1594.

Le Corre L., Reddy V., Sanchez J. A., Dunn T., Cloutis E. A., Izawa M. R. M., Mann P., and Nathues A. 2015. Exploring exogenic sources for the olivine on asteroid (4) Vesta. *Icarus* **258**:483–499.

Levison H. F. and Duncan M. J., 1994. The long-term dynamical behavior of short-period comets. *Icarus* **108**:18–36.

Leya I., Neumann S., Wieler R., and Michel R. 2001. The production of cosmogenic nuclides by galactic cosmic-ray particles for 2π exposure geometries. *Meteoritics & Planetary Science* **36**:1547–1561.

Leya I. and Masarik J. 2009. Cosmogenic nuclides in stony meteorites revisited. *Meteoritics & Planetary Science* **44**:1061–1086.

Li Q.-L., Li X.-H., Liu Y., Tang G.-Q., Yang J.-H., and Zhu W.-G. 2010. Precise U-Pb and Pb-Pb dating of phanerozoic baddeleyite by SIMS with oxygen flooding technique. *Journal of Analytical Atomic Spectrometry* **25**:1107–1113.

Lindsay F. N., Delaney J. S., Herzog G. F., Turrin B. D., Park J., and Swisher C. C. 2015. Rheasilvia provenance of the Kapoeta howardite inferred from ~1 Ga $^{40}Ar/^{39}Ar$ feldspar ages. *Earth & Planetary Science Letters* **413**:208–213.

Liu Y., Li Q.-L., Tang G.-Q., Li X.-H., and Yin Q.-Z. 2015. Towards higher precision SIMS U-Pb zircon geochronology via dynamic multi-collector analysis. *Journal of Analytical Atomic Spectrometry* **30**:979–985.

Llorca J., Casanova I., Trigo-Rodgriguez J. M., Madiedo J. M., Roszjar J., Bischoff A., Ott U., Franchi I. A., Greenwood R. C., and Laubenstein M., 2009. The Puerto Lápice eucrite. *Meteoritics & Planetary Science* **44**:159–174.

Ludwig K.R. (ed.) 2003. *User's Manual for Isoplot 3.00, a geochronological toolkit for Microsoft Excel.* Special Publication No. 4, Berkeley Geochronology Center, Berkeley, CA, 74 pp.

Macke R. J., Britt D. T., and Consolmagno G. J. 2011. Density, porosity, and magnetic susceptibility of achondritic meteorites. *Meteoritics & Planetary Science* **46**:311–326.

Marchi S., McSween H. Y., O'Brien D. P., Schenk P., De Sanctis M. C., Gaskell R., Jaumann R., Mottola S., Preusker F., Raymond C. A., Roatsch T., and Russell C. T. 2012. The violent collisional history of asteroid 4 Vesta. *Science* **336**:690–692.

McCord T. B., Adams J. B., and Johnson T. V. 1970. Asteroid Vesta: Spectral reflectivity and compositional implications. *Science* **168**:1445–1447.

McGlaun J. M., Thompson S. L., Kmetyk L. N., and Elrick M. G. 1990. CTH: A three-dimensional shock wave physics code. *International Journal of Impact Engineering* 10:351–360.

McKay D. S., Heiken G., Basu A., Blanford G., Simon S., Reedy R., French B. M., and Papike J. 1991. The lunar regolith. In Lunar sourcebook. A user’s guide to the Moon, edited by Heiken G., Vaniman D., and French B. M. Cambridge, UK: Cambridge University Press. pp. 285–356.

Meier M. M. M., Schmitz B., Alwmark C., Trappitsch R., Maden C., and Wieler R. 2014. He and Ne in individual chromite grains from the regolith breccia Ghubara (L5): Exploring the history of the L chondrite parent body regolith. *Meteoritics & Planetary Science* **49**:576–594.

Meier M. M. M., Welten K. C., Riebe M. E. I., Caffee M. W., Gritsevitch M., Maden C., and Busemann H. 2017. Park Forest (L5) and the asteroidal source of shocked L chondrites. *Meteoritics & Planetary Science* **52**:1561–1576.

Mittlefehldt D. W. 2015. Asteroid (4) Vesta I: The howardite-eucrite-diogenite (HED) clan of meteorites. *Chemie der Erde* **75**:155–183.

Moskovitz N. A., Jedicke R., Gaidos E., William M., Nesvorny D., Fevig R., and Ivezic Z. 2008. The distribution of basaltic asteroids in the Main Belt. *Icarus* **198**:77–90.

Nasdala L., Hofmeister W., Norberg N., Mattinson J. M., Corfu F., Dorr W., Kamo S. L., Kennedy A. K., Kronz A., Reiners P. W., Frei D., Kosler J., Wan Y. S., Gotze J., Hager T., Kroner A., and Valley J. W. 2008. Zircon M257 – a homogeneous natural reference material for the ion microprobe U-Pb analysis of zircon. *Geostandards and Geoanalytical Research* **32**:247–265.

National Research Council 1996. Rock fractures and fluid flow: Contemporary understanding and applications. The National Academies Press, Washington D. C., Chapter 2. p. 29–102 (https://doi.org/10.17226/2309).

Nesvorny D., Roig F., Gladman B., Lazzaro D., Carruba V., and Mothé-Diniz T. 2008. Fugitives from the Vesta family. *Icarus* **193**:85–95.

Niederer F. R., Papanastassiou D. A., and Wasserburg G. J. 1981. The isotopic composition of titanium in the Allende and Leoville meteorites. *Geochimica et Cosmochimica Acta* **45**:1017–1031.

Niihara T. 2011. Uranium-lead age of baddeleyite in shergottite Roberts Massif 04261: Implications for magmatic activity on Mars. *Journal of Geophysical Research* **116**, Article No. E12008, 12 pp.

Nishiizumi K. 2004. Preparation of $^{26}$Al AMS standards. *Nuclear Instruments and Methods in Physics Research* **B223–224**:388–392.

Nishiizumi K., Imamura M., Caffee M. W., Southon J. R., Finkel R. C., and McAninch J. 2007. Absolute calibration of $^{10}$Be AMS standards. *Nuclear Instruments and Methods in Physics Research* **B258**:403–413.

Nittler L. R., McCoy T. J., Clark P. E., Murphy M. E., Trombka J. I., and Jarosewich E. 2004. Bulk element compositions of meteorites: A guide for interpreting remote-sensing geochemical measurements of planets and asteroids. *Antarctic Meteorite Research* **17**:231–251.

O'Brien D. P., Marchi S., Morbidelli A., Bottke W. F., Schenk P. M., Russell C. T., and Raymond C. A. 2014. Constraining the cratering chronology of Vesta. *Planetary & Space Science* **103**:131–142.

Piggott A. R. 1997. Fractal relations for the diameter and trace length of disc-shaped fractures. *Journal of Geophysical Research* **102**:18122–18125.

Popova O. P., Jenniskens P., Emel'yanenko V., Kartashova A., Biryukov E., Khaibrakhmanov S., Shuvalov V., Rybnov Y., Dudorov A., Grokhovsky V. I., Badyukov D. D., Yin Q.-Z., Gural P. S., Albers J., Granvik M., Evers L. G., Kuiper J., Kharlamov V., Solovyov A., Rusakov Y. S., Korotkiy S., Serdyuk I., Korochantsev A. V., Larionov M. Y., Glazachev D., Mayer A. E., Gisler G., Gladkovsky S. V., Wimpenny J., Sanborn M. E., Yamakawa A., Verosub K., Rowland D. J., Roeske S., Botto N. W., Friedrich J. M., Zolensky M., Le L., Ross D., Ziegler K., Nakamura T., Ahn I., Lee J. I., Zhou Q., Li X.-H., Li Q.-L., Liu Y., Tang G.-Q., Hiroi T., Sears D., Weinstein I. A., Vokhmintsev A. S., Ishchenko A. V., Schmitt-Kopplin P., Hertkorn N., Nagao K., Haba M. K., Komatsu M., and Mikouchi T., and The Chelyabinsk Airburst Consortium. 2013. Chelyabinsk airburst, damage assessment, meteorite recovery, and characterization. *Science* **342**:1069–1073.

Pouchou J. L. and Pichoir F. 1984. Extension of quantitative possibilities by a new formulation of matrix effects. *Journal de Physique Colloques* **45**:17–20.

Prettyman T. H., Yamashita N., Reedy R. C., McSween H. Y., Mittlefehldt D. W., Hendricks J. S., and Toplis M. J. 2015. Concentrations of potassium and thorium within Vesta's regolith. *Icarus* **259**:39–52.

Reddy V., Nathues A., and Gaffey M. J. 2011. First fragment of asteroid 4 Vesta's mantle detected. *Icarus* **212**:175–179.

Reddy V., Gary B.L., Sanchez J.A., Takir D., Thomas C.A., Hardersen P.S., Ogmen Y., Benni P., Kaye T.G., Gregorio J., Garlitz J., Polishook D., Le Corre L., and Nathues A. 2015. The Physical Characterization of Potentially Hazardous Asteroid 2004 BL86: A Fragment of Differentiated Asteroid. *The Astrophysical Journal* **811**:65–74.

ReVelle D. O. 1997. Historical detection of atmospheric impacts by large bolides using acoustic-gravity waves. *Annals of the New York Academy of Sciences* **822**:284–302.

ReVelle D. O. 2003. *BLDM: Bolide Luminosity and Detonation Model – Software User's Manual*. Los Alamos National Laboratory, Los Alamos, NM. Manual updated on Dec. 3, 2015 by Edward Stokan, University of Western Ontario, Canada.

ReVelle D. O. 2004. Recent advances in bolide entry modeling: A bolide potpourri. *Earth, Moon, and Planets* **95**:441–476.

Robertson D. and Mathias D. 2015. Effect of Different Rock Models on Hydrocode Simulations of Asteroid Airburst, American Geophysical Union (AGU) Fall Meeting. Abstract NH11A-1891.

Ruesch O., Hiesinger H., Blewett D. T., Williams D. A., Buczkowski D., Scully J., Yingst R. A., Roatsch T., Preusker F., Jaumann R., Russell C. T., and Raymond C. A. 2014. Geologic map of the northern hemisphere of Vesta based on Dawn Framing Camera (FC) images. *Icarus* **244**:41–59.

Ruesch O., Hiesinger H., Cloutis E., Le Corre L., Kallisch J., Mann P., Markus K., Metzler K., Nathues A., and Reddy V. 2015. Near infrared spectroscopy of HED meteorites: Effects of viewing geometry and compositional variations, *Icarus* **258**:384-401.

Ruf A., Kanawati B., Hertkorn N., Yin Q.-Z., Moritz F., Harir M., Lucio M., Michalke B., Wimpenny J., Shilobreeva S., Bronsky B., Saraykin V., Gabelica Z., Gougeon R., Quirico E., Ralew S., Jakubossi T., Haack H., Gonsior M., Jenniskens P., Hinman N. W., and Schmitt-Kopplin P., 2016. Previously unknown class of metalorganic compounds revealed in meteorites. *Proceedings of the National Academy of Sciences* **114**:2819–2824.

Sanborn M. E. and Yin Q.-Z. 2014. Chromium isotopic composition of the anomalous eucrites: An additional geochemical parameter for evaluating their origin. *Lunar & Planetary Sciences Conference XLV*, abstract 2018.

Sanborn M. E., Yin Q.-Z., and Mittlefehldt D. W. 2016. The diversity of anomalous HEDs: Isotopic constraints on the connection of EET 92023, GRA 98098, and Dhofar 700 with Vesta. *Lunar & Planetarry Sciences Conference XLVII*, abstract 2256.

Sanchez P. and Scheeres D. 2014. The strength of regolith and rubble pile asteroids. *Meteoritics & Planetary Science* **49**:788–811.

Sano Y., Oyama T., Terada K., and Hidaka H. 1999. Ion microprobe U–Pb datingof apatite. *Chemical Geology* **153**:249–258.

Schmedemann N., Kneissl T., Ivanov B. A., Michael G. G., Wagner R. J., Neukum G., Ruesch O., Hieslinger H., Krohn K., Roatsch T., Preusker F., Sierks H., Jaumann R., Reddy V., Nathues A., Walter S. H., Neesemann A., Raymond C. A., and Russell C. T. 2014. The cratering record chronology and surface ages of (4) Vesta in comparison to smaller asteroids and the ages of HED meteorites. *Planetary & Space Science* **103**:103–130.

Schmitt-Kopplin P., Harir M., Kanawati B., Tziozis D., and Hertkon N., 2012. Chemical footprint of the solvent soluble extraterrestrial organic matter occluded in Soltmany ordinary chondrite. *Meteorites, Tektites, Impactites* **2:**79–92.

Schönbächler M., Rehkämper M., Lee D.-C., and Halliday A. N. 2004. Ion exchange chromatography and high precision isotopic measurements of zirconium by MC-ICP-MS. *Analyst* **129**:32–37.

Schultz L. and Franke L. 2004. Helium, neon and argon in meteorites: a data collection. *Meteoritics & Planetary Science* **39**:1889–1890.

Scott E. R. D., Greenwood R. C., Franchi I. A., and Sanders I. S. 2009. Oxygen isotopic constraints on the origin and parent bodies of eucrites, diogenites, and howardites. *Geochimica et Cosmochimica Acta* **73**:5835–5853.

Sears D. W. G., Benoit P. H., Sears H., Batchelor J. D., and Symes S. 1991. The natural thermoluminescence of meteorites: III. Lunar and basaltic meteorites. *Geochimica et Cosmochimica Acta* **55**:3167–3180.

Sharma P., Kubik P. W., Fehn U., Gove G. E., Nishiizumi K., and Elmore D. 1990. Development of $^{36}$Cl standards for AMS. *Nuclear Instruments and Methods in Physics Research* **B52**:410–415.

Sharma P., Bourgeous M., Elmore D., Granger D., Lipschutz M. E., Ma X., Miller T., Mueller K., Rickey F., Simms P., and Vogt S. 2000. PRIME lab AMS performance, upgrades and research applications. *Nuclear Instruments and Methods in Physics Research* **B172**:112–123.

Sharp Z. D. 1990. A laser-based microanalytical method for the in situ determination of oxygen isotope ratios of silicates and oxides. *Geochimica et Cosmochimica Acta* **54**:1353–1357.

Shields W. R., Murphy T. J., Catanzaro E. J., and Garner E. L. 1966. Absolute isotopic abundance ratios and the atomic weight of a reference sample of chromium. *Journal Research National Bureau of Standards* **70A**:193–197.

Spurný P., Bland P. A., Shrbeny L., Borovicka J., Ceplecha Z., Signelton A., Bevan A. W. R., Vaughan D., Towner M. C., McClafferty T. P., Toumi R., and Deacon G. 2012. The Bunburra Rockhole meteorite fall in SW Asutralia: firball trajectory, luminosity, dynamics, orbit, and impact position from photographic and photoelectric records. *Meteoritics & Planetary Science* **47**:163–185.

Steger C. 1998. An unbiased detector of curvilinear structures. *IEEE Transactions on Pattern Anaysis and Machine Intelligence* **20**:113–125.

Stöffler D., Keil K., Scott E. R. D. 1991. Shock metamorphism of ordinary chondrites. *Geochimica et Cosmochimica Acta* **55**:3845–3867.

Takeda H. and Graham A. L. 1991. Degree of equilibration of eucritic pyroxenes and thermal metamosphism of the earliest planetary crust. *Meteoritics* **26:**129–134.

Tatsumoto M., Knight R. J., and Allegre C. J. 1973. Time difference in the formation of meteorites as determined from the ratio of lead-207 to lead-206. *Science* **180:**1279–1283.

Trinquier A., Birck J.-L., and Allegre C. 2007. Widespread $^{54}$Cr heterogeneity in the inner solar system. *Astrophysical Journal* **655**:1179–1185.

Trinquier A., Elliott T., Ulfbeck D., Coath C., Krot A. N., and Bizzarro M. 2009. Origin of nucleosynthetic isotope heterogeneity in the solar protoplanetary disk. *Science* **324**:374–376.

Trotter J. A. and Eggins S. M. 2006. Chemical systematics of conodont apatite determined by laserablation ICPMS. *Chemical Geology* **233**:196–216.

Tziotis D., Hertkorn N., and Schmitt-Kopplin Ph. 2011. Kendrick-analogous network visualisation of ion cyclotron resonance Fourier Transform (FTICR) mass spectra: Improved options to assign elemental compositions and to classify organic molecular complexity. *European Journal of Mass Spectrometry* **17**:415-421.

Warren P. H., Kallemeyn G. W., Huber H., Ulff-Moller F., and Choe W. 2009. Siderophile and other geochemical constraints on mixing relationships among HED-meteoritic breccias. *Geochimica et Cosmochimica Acta* **73**:5918–5943.

Weibull W. 1951. A statistical distribution function of wide applicability. *Journal of Applied Mechanics* **18**:293–297.

Welten K. C., Lindner L., van der Borg K., Loeken T., Scherer P., and Schultz L. 1997. Cosmic-ray exposure ages of diogenites and the recent collisional history of the howardite, eucrite and diogenite parent body/bodies. *Meteoritics & Planetary Science* **32**:891–902.

Welten K. C., Nishiizumi K., Masarik J., Caffee M. W., Jull A. J. T., Klandrud S. E., and Wieler R. 2001. Cosmic-ray exposure history of two Frontier Mountain H-chondrite showers from spallation and neutron-capture products. *Meteoritics & Planetary Science* **36**:301–317.

Welten K. C., Meier M. M. M., Caffee M. W., Laubenstein M., Nishiizumi K., Wieler R., Bland P. A., Towner M. C., and Spurný P. 2012. Cosmic-ray exposure age and pre-atmospheric size of the Bunburra Rockhole achondrite. *Meteoritics & Planetary Science* **47**:186–196.

Wheeler L. F., Register P. J., and Mathias D. L. 2017. A fragment-cloud model for asteroid breakup and atmospheric energy deposition. *Icarus* **295**, 149–169.

Wieler R., Baur H., Pedroni A., Signer P., and Pellas P. 1989. Exposure history of the regolithic chondrite Fayetteville. I – Solar-gas-rich matrix. *Geochimica et Cosmochimica Acta* **53**:1441–1448.

Williams N. H. 2015. The origin of titanium isotopic anomalies within solar system material. Ph.D. thesis. The University of Manchester, Manchester, UK. 175 p.

Williams D. A., Denevi B. W., Mittlefehldt D. W., Mest S. C., Schenk P. M., Yingst R. A., Buczkowski D. L., Scully J. E. C., Garry W. B., McCord T. B., Combe J. P., Jaumann R., Pieters C. M., Nathues A., Le Corre L., Hoffmann M., Reddy V., Schäfer M., Roatsch T., Preusker F., Marchi S., Kneissl T., Schmedemann N., Neukum G., Hiesinger H., De Sanctis M. C., Ammannito E., Frigeri A., Prettyman T. H., Russell C. T., and Raymond C. A.

2014a. The geology of the Marcia quadrangle of asteroid Vesta: Assessing the effects of large, young craters. *Icarus* **244**:74–88.

Williams D. A., Jaumann R., McSween H. Y., Marchi S., Schmedemann N., Raymond C. A., and Russell C. T. 2014b. The chronostratigraphy of protoplanet Vesta. *Icarus* **244**:158–165.

Wingate M. T. D. and Compston W. 2000. Crystal orientation effects during ion microprobe U-Pb analysis of baddeleyite. *Chemical Geology* **168:**75–97.

Wolf S. F., Compton J. R., and Gagnon C. J. L. 2012. Determination of 11 major and minor elements in chondritic meteorites by inductively coupled plasma mass spectrometry. *Talanta* **100**:276–281.

Yamaguchi A., Tayor G. J., and Keil K. 1996. Global crustal metamorphism of the eucrite parent body. *Icarus* **124**, 97–112.

Yamaguchi A., Taylor G. J., and Keil K. 1997. Metamorphic history of the eucritic crust of 4 Vesta. *Journal of Geophysical Research* **102**, 13381–13386.

Yamakawa A., Yamashita K., Makishima A., and Nakamura E. 2009. Chemical separation and mass spectrometry of Cr, Fe, Ni, Zn, and Cu in terrestrial and extraterrestrial materials using thermal ionization mass spectrometry. *Analytical Chemistry* **81**:9787–9794.

Zhang J., Dauphas N., Davis A. M., and Pourmand A. 2011. A new method for MC-ICPMS measurement of titanium isotopic composition: Identification of correlated isotope anomalies in meteorites. *Journal of Analytical Atomic Spectrometry* **26**:2197–2205.

Zhang J., Dauphas N., Davis A. M., Leya I., and Fedkin A. 2012. The proto-Earth as a significant source of lunar material. *National Geoscience* **5**:251–255.

Zhou Q., Yin Q.-Z., Edward D. Y., Li X.-H., Wu F.-Y., Li Q.-L., Liu Y., and Tang G.-Q. 2013. SIMS Pb-Pb and U-Pb age determination of eucrite zircons at <5 micron scale and the first 50Ma of the thermal history of Vesta. *Geochimica et Cosmochimica Acta* **110:**152–175.

Table 1. Location of camera sites.

| Site | a | Abbr.§ | Δt (s) | FPS (Hz) | O-C (') | Lat. (ºN) ±0.00002 | Lon. (ºE) ±0.00002 | Alt. (m) ±1 | Notes |
|---|---|---|---|---|---|---|---|---|---|
| **Bingöl University:** | | | | | | | | | |
| Rectorate | 1422 | BR | –7.9 | 10 | ~120 | 38.89611 | 40.49072 | 1180 | Shadow building |
| Soccer Court | 1415 | BS | –8.0 | 25 | ~60 | 38.89815 | 40.49269 | 1155 | Shadow fence |
| Economics Faculty | B | BL | –8.1 | 25 | ~60 | 38.89785 | 40.48847 | 1171 | Shadow lamp |
| **Muş Alparslan Univ.:** | | | | | | | | | |
| Street | SE3 | MS | –5.6 | 30 | 1.6 | 38.76662 | 41.42455 | 1338 | Meteor |
| Fence 64 | 64 | M4 | –53.9 | 30 | 5.4 | 38.76820 | 41.41758 | 1333 | Meteor (reflection) |
| Fence 66 | 66 | M6 | +5.2 | 30 | -.-† | 38.76781 | 41.41852 | 1336 | Meteor |
| Lantern Post | 91 | ML | –53.9 | 30 | ~30 | 38.77265 | 41.42781 | 1294 | Shadow lantern |
| **Others:** | | | | | | | | | |
| Karliova High School | Fixed | KL | –6.5 | 30 | ~180 | 39.29470 | 41.01036 | 1828 | Shadow sign |
| Kiği Street | Dortyol | KI | –312.0 | 30 | ~480 | 39.31004 | 40.34913 | 1521 | Shadow building |

Δt, Time offset from 20:10:30 UTC (+ is system clock ahead of UTC); FPS, Frame rate; O-C = Random error in astrometry; Lat., Long., and Alt., Position of camera or shadow obstacle.

[a]Camera identifier.

[b]Abbreviation for site used in text and figures.

[c]Not calibrated.

Table 2. Meteorite mass and find locations.

| SC# | Mass (g) | Latitude (ºN) | Longitude (ºE) | Altitude (m) | Date of find | Name finder |
|---|---|---|---|---|---|---|
| 1 | 176.43 | 38.8981 | 40.5988 | 1044 | 9/10/2015 | Nezir Ergün |
| 2 | 166.77 | 38.9013 | 40.5904 | 1410 | 9/10/2015 | Nezir Ergün |
| 3 | 102.69 | 38.9010 | 40.5967 | 1044 | 9/10/2015 | Nezir Ergün |
| 4 | 16.60 | 38.9016 | 40.5991 | 1180 | 9/10/2015 | Nezir Ergün |
| 5 | 37.70 | 38.9007 | 40.5999 | 1130 | 9/10/2015 | Nezir Ergün |
| 6 | 42.23 | 38.9010 | 40.5998 | 1130 | 9/10/2015 | Nezir Ergün |
| 7 | 10.92 | 38.9007 | 40.5989 | 1410 | 9/10/2015 | Nezir Ergün |
| 8 | 21.56 | 38.9009 | 40.5997 | 1130 | 9/10/2015 | Nezir Ergün |
| 9 | 20.95 | 38.9010 | 40.6011 | 1410 | 9/10/2015 | Nezir Ergün |
| 10 | 682.12 | 38.8923 | 40.5934 | 1039 | 9/10/2015 | H. Sabri Ergün |
| 11 | 150.40 | 38.8939 | 40.5940 | 1041 | 9/10/2015 | Metin Ergün |
| 12 | 27.28 | 38.8938 | 40.6087 | 1105 | 9/18/2015 | Nezir Ergün |
| 13 | 4.22 | 38.9017 | 40.5975 | 1046 | 9/30/2015 | Huseyin Ergün |
| 14 | 19.72 | 38.9019 | 40.5975 | 1041 | 9/30/2015 | Aydin Sükrü Bengü |
| 15 | 7.58 | 38.9036 | 40.5949 | 1039 | 9/30/2015 | Nezir Ergün |
| 16 | 1.75 | 38.9034 | 40.5941 | 1039 | 9/30/2015 | Peter Jenniskens |
| 17 | 37.97 | -.- | -.- | -.- | -.- | Mehmet Ergün Sr |
| 18 | 6.90 | 38.9022 | 40.5933 | 1038 | 9/30/2015 | Nezir Ergün |
| 19 | 4.78 | 38.9027 | 40.5948 | 1039 | 9/30/2015 | Iskender Demirkol |
| 20 | 24.01 | 38.9029 | 40.5956 | 1039 | 9/30/2015 | Iskender Demirkol |
| 21 | 7.52 | 38.9024 | 40.5939 | 1038 | 9/30/2015 | Galip Akengin |
| 22 | 3.81 | 38.9021 | 40.5946 | 1036 | 9/30/2015 | Ozan Ünsalan |
| 23 | 16.62 | 38.8997 | 40.5933 | 1033 | 9/30/2015 | Ibrahim Y. Erdogan |
| 24 | 27.46 | 38.8986 | 40.5950 | 1032 | 9/30/2015 | Metin Ergün |
| 25 | 9.92 | 38.8987 | 40.5949 | 1056 | 9/30/2015 | Aydin Sükrü Bengü |
| 26 | 131.88 | 38.9024 | 40.5939 | 1038 | 9/30/2015 | Nezir Ergün |
| 27 | 2.33 | -.- | -.- | -.- | 9/30/2015 | Hüseyn Ergün |
| 28 | 301 | 38.8961 | 40.5951 | 1038 | -.- | Nezir Ergun |
| 29 | 17.72 | -.- | -.- | -.- | -.- | Nezir Ergün |
| 30 | 13.74 | -.- | -.- | -.- | -.- | Nezir Ergün |
| 31 | 6.19 | -.- | -.- | -.- | -.- | Nezir Ergün |
| 32 | 7.69 | -.- | -.- | -.- | -.- | Nezir Ergün |
| 33 | 2.32 | -.- | -.- | -.- | -.- | Nezir Ergün |
| 34 | 10.34 | 38.9004 | 40.6011 | 1043 | 9/9/2015 | Nezir Ergün |
| 35 | 2.87 | -.- | -.- | -.- | -.- | Nezir Ergün |
| 36 | 110.70 | 38.8946 | 40.5952 | -.- | 9/20/2015 | Hüseyin Ergün |
| 37 | 64.71 | 38.8934 | 40.5893 | -.- | 9/20/2015 | Hüseyin Ergün |
| 38 | 35.68 | 38.8932 | 40.5849 | -.- | 9/20/2015 | Hüseyin Ergün |
| 39 | 20.04 | -.- | -.- | -.- | -.- | Hüseyin Ergün |
| 40 | 18.42 | -.- | -.- | -.- | -.- | Hüseyin Ergün |
| 41 | 19.21 | -.- | -.- | -.- | -.- | Hüseyin Ergün |
| 42 | 18.31 | -.- | -.- | -.- | -.- | Hüseyin Ergün |
| 43 | 16.60 | -.- | -.- | -.- | -.- | Hüseyin Ergün |
| 44 | 21.72 | -.- | -.- | -.- | -.- | Hüseyin Ergün |
| 45 | 12.24 | -.- | -.- | -.- | -.- | Hüseyin Ergün |
| 46 | 4.35 | -.- | -.- | -.- | -.- | Hüseyin Ergün |
| 47 | 1.96 | -.- | -.- | -.- | -.- | Hüseyin Ergün |
| 48 | 1.50 | -.- | -.- | -.- | -.- | Hüseyin Ergün |
| 49 | 66.14 | -.- | -.- | -.- | -.- | Ibrahim Ergün |
| 50 | 26.78 | -.- | -.- | -.- | -.- | Selahattin Ergün |
| 51 | 33.42 | -.- | -.- | -.- | -.- | Selahattin Ergün |
| 52 | 208.22 | 38.9028 | 40.5943 | 1038 | -.- | Selahattin Ergün |
| 53 | 38.72 | 38.8983 | 40.5901 | 1040 | 9/19/2015 | Selahattin Ergün |
| 54 | 44.41 | 38.8982 | 40.5922 | 1040 | 9/19/2015 | Selahattin Ergün |
| 55 | 8.17 | -.- | -.- | -.- | -.- | Selahattin Ergün |
| 56 | 18.20 | -.- | -.- | -.- | -.- | Selahattin Ergün |
| 57 | 29.47 | -.- | -.- | -.- | -.- | Selahattin Ergün |
| 58 | 3.86 | -.- | -.- | -.- | -.- | Selahattin Ergün |
| 59 | 71.56 | 38.8911 | 40.5965 | -.- | 9/30/2015 | H. Sabri Ergün |
| 60 | 55.19 | 38.8915 | 40.5951 | -.- | 9/30/2015 | H. Sabri Ergün |
| 61 | 10.50 | 38.8980 | 40.5795 | -.- | 9/22/2015 | Tahir Baydas |
| 62 | 8.20 | 38.8997 | 40.5807 | -.- | 9/22/2015 | Tahir Baydas |
| 63 | 9.15 | -.- | -.- | -.- | -.- | H. Emin Ergün |
| 64 | 17.60 | -.- | -.- | -.- | -.- | H. Emin Ergün |
| 65 | 27.16 | -.- | -.- | -.- | -.- | H. Emin Ergün |
| 66 | 30.25 | -.- | -.- | -.- | -.- | H. Emin Ergün |
| 67 | 72.83 | -.- | -.- | -.- | -.- | H. Emin Ergün |
| 68 | 140.50 | 38.8884 | 40.5861 | -.- | 9/23/2015 | Metin Ergün |
| 69 | 20.40 | -.- | -.- | -.- | -.- | Metin Ergün |
| 70 | 10.35 | -.- | -.- | -.- | -.- | Metin Ergün |
| 71 | 10.30 | -.- | -.- | -.- | -.- | Metin Ergün |
| 72 | 10.28 | -.- | -.- | -.- | -.- | Metin Ergün |
| 73 | 6.16 | -.- | -.- | -.- | -.- | Metin Ergün |
| 74 | 5.12 | -.- | -.- | -.- | -.- | Metin Ergün |
| 75 | 4.98 | -.- | -.- | -.- | -.- | Metin Ergün |
| 76 | 3.35 | -.- | -.- | -.- | -.- | Metin Ergün |
| 77 | 169.12 | 38.8878 | 40.5908 | -.- | 9/30/2015 | Ismail Ergün |
| 78 | 71.65 | 38.8904 | 40.5898 | -.- | 9/30/2015 | Ismail Ergün |
| 79 | 32.66 | -.- | -.- | -.- | -.- | Ismail Ergün |
| 80 | 27.53 | -.- | -.- | -.- | -.- | Ismail Ergün |
| 81 | 24.20 | -.- | -.- | -.- | -.- | Ismail Ergün |
| 82 | 17.61 | -.- | -.- | -.- | -.- | Ismail Ergün |
| 83 | 15.94 | -.- | -.- | -.- | -.- | Ismail Ergün |
| 84 | 15.42 | -.- | -.- | -.- | -.- | Ismail Ergün |
| 85 | 12.73 | -.- | -.- | -.- | -.- | Ismail Ergün |
| 86 | 12.14 | -.- | -.- | -.- | -.- | Ismail Ergün |
| 87 | 11.45 | -.- | -.- | -.- | -.- | Ismail Ergün |
| 88 | 11.29 | -.- | -.- | -.- | -.- | Ismail Ergün |
| 89 | 10.41 | -.- | -.- | -.- | -.- | Ismail Ergün |
| 90 | 10.13 | -.- | -.- | -.- | -.- | Ismail Ergün |
| 91 | 8.60 | -.- | -.- | -.- | -.- | Ismail Ergün |
| 92 | 6.11 | -.- | -.- | -.- | -.- | Ismail Ergün |
| 93 | 6.03 | -.- | -.- | -.- | -.- | Ismail Ergün |
| 94 | 5.40 | -.- | -.- | -.- | -.- | Ismail Ergün |
| 95 | 5.39 | -.- | -.- | -.- | -.- | Ismail Ergün |
| 96 | 5.12 | -.- | -.- | -.- | -.- | Ismail Ergün |
| 97 | 4.82 | -.- | -.- | -.- | -.- | Ismail Ergün |
| 98 | 4.64 | -.- | -.- | -.- | -.- | Ismail Ergün |
| 99 | 4.56 | -.- | -.- | -.- | -.- | Ismail Ergün |
| 100 | 4.44 | -.- | -.- | -.- | -.- | Ismail Ergün |
| 101 | 4.41 | -.- | -.- | -.- | -.- | Ismail Ergün |
| 102 | 3.92 | -.- | -.- | -.- | -.- | Ismail Ergün |
| 103 | 3.36 | -.- | -.- | -.- | -.- | Ismail Ergün |
| 104 | 2.18 | -.- | -.- | -.- | -.- | Ismail Ergün |
| 105 | 2.17 | -.- | -.- | -.- | -.- | Ismail Ergün |
| 106 | 1.42 | -.- | -.- | -.- | -.- | Ismail Ergün |
| 107 | 1.28 | -.- | -.- | -.- | -.- | Ismail Ergün |
| 108 | 0.86 | -.- | -.- | -.- | -.- | Ismail Ergün |
| 109 | 0.79 | -.- | -.- | -.- | -.- | Ismail Ergün |
| 110 | 306.48 | 38.8974 | 40.6000 | 1040 | -.- | Kazim Sazak |
| 111 | 39.61 | -.- | -.- | -.- | -.- | Kazim Sazak |
| 112 | 17.23 | -.- | -.- | -.- | -.- | Kazim Sazak |
| 113 | 14.89 | -.- | -.- | -.- | -.- | Kazim Sazak |
| 114 | 8.39 | -.- | -.- | -.- | -.- | Kazim Sazak |
| 115 | 2.73 | -.- | -.- | -.- | -.- | Kazim Sazak |
| 116 | 40.91 | -.- | -.- | -.- | -.- | Idris Ergün |
| 117 | 40.97 | -.- | -.- | -.- | -.- | Idris Ergün |
| 118 | 31.19 | -.- | -.- | -.- | -.- | Idris Ergün |
| 119 | 19.33 | -.- | -.- | -.- | -.- | Idris Ergün |
| 120 | 14.56 | -.- | -.- | -.- | -.- | Idris Ergün |
| 121 | 5.61 | -.- | -.- | -.- | -.- | Idris Ergün |
| 122 | 10.59 | -.- | -.- | -.- | -.- | Idris Ergün |
| 123 | 4.02 | -.- | -.- | -.- | -.- | Idris Ergün |
| 124 | 4.30 | -.- | -.- | -.- | -.- | Mehmet Ergün |
| 125 | 4.30 | -.- | -.- | -.- | -.- | Mehmet Ergün |
| 126 | 4.30 | -.- | -.- | -.- | -.- | Mehmet Ergün |
| 127 | 5.50 | -.- | -.- | -.- | -.- | Mehmet Ergün |
| 128 | 5.40 | -.- | -.- | -.- | -.- | Mehmet Ergün |
| 129 | 10.40 | -.- | -.- | -.- | -.- | Mehmet Ergün |
| 130 | 10.50 | -.- | -.- | -.- | -.- | Mehmet Ergün |
| 131 | 15.00 | -.- | -.- | -.- | -.- | Mehmet Ergün |
| 132 | 25.50 | -.- | -.- | -.- | -.- | Mehmet Ergün |
| 133 | 70.15 | -.- | -.- | -.- | -.- | Mehmet Ergün |
| 134 | 50.87 | -.- | -.- | -.- | -.- | Dilan Gencay |
| 135 | 29.58 | -.- | -.- | -.- | -.- | Dilan Gencay |
| 136 | 16.24 | -.- | -.- | -.- | -.- | Dilan Gencay |
| 137 | 12.17 | -.- | -.- | -.- | -.- | Dilan Gencay |
| 138 | 91.41 | 38.8984 | 40.5987 | 1044 | 9/14/2015 | Resit Sazak |
| 139 | 39.19 | 38.9001 | 40.5988 | -.- | 9/17/2015 | Resit Sazak |
| 140 | 27.52 | 38.9007 | 40.5820 | 1044 | 9/29/2015 | Resit Sazak |
| 141 | 25.00 | 38.8985 | 40.5796 | -.- | 9/29/2015 | Resit Sazak |
| 142 | 23.30 | -.- | -.- | -.- | -.- | Resit Sazak |
| 143 | 17.01 | -.- | -.- | -.- | -.- | Resit Sazak |
| 144 | 15.48 | -.- | -.- | -.- | -.- | Resit Sazak |
| 145 | 9.05 | -.- | -.- | -.- | -.- | Resit Sazak |
| 146 | 7.68 | -.- | -.- | -.- | -.- | Resit Sazak |
| 147 | 7.54 | -.- | -.- | -.- | -.- | Resit Sazak |
| 148 | 8.22 | -.- | -.- | -.- | -.- | Resit Sazak |
| 149 | 7.36 | -.- | -.- | -.- | -.- | Resit Sazak |
| 150 | 5.33 | -.- | -.- | -.- | -.- | Resit Sazak |
| 151 | 5.05 | -.- | -.- | -.- | -.- | Resit Sazak |
| 152 | 5.14 | -.- | -.- | -.- | -.- | Resit Sazak |
| 153 | 7.70 | -.- | -.- | -.- | -.- | Resit Sazak |
| 154 | 2.04 | -.- | -.- | -.- | -.- | Resit Sazak |
| 155 | 1.60 | -.- | -.- | -.- | -.- | Resit Sazak |
| 156 | 1.06 | -.- | -.- | -.- | -.- | Resit Sazak |
| 157 | 0.91 | -.- | -.- | -.- | -.- | Resit Sazak |
| 158 | 10 | -.- | -.- | -.- | -.- | Mehmet A. Görken |
| 159 | 250 | 38.8938 | 40.6038 | 1077 | -.- | Mehmet A. Görken |
| 160 | 20.92 | -.- | -.- | -.- | -.- | Muhittin Ergün |
| 161 | 5.79 | -.- | -.- | -.- | -.- | Muhittin Ergün |
| 162 | 1.58 | -.- | -.- | -.- | -.- | Muhittin Ergün |
| 163 | 2.33 | -.- | -.- | -.- | -.- | Muhittin Ergün |
| 164 | 8.01 | -.- | -.- | -.- | -.- | Muhittin Ergün |
| 165 | 1.35 | -.- | -.- | -.- | -.- | Muhittin Ergün |
| 166 | 0.76 | -.- | -.- | -.- | -.- | Muhittin Ergün |
| 167 | 36.79 | -.- | -.- | -.- | -.- | Muhittin Ergün |
| 168 | 7.72 | -.- | -.- | -.- | -.- | Muhittin Ergün |
| 169 | 35.59 | -.- | -.- | -.- | -.- | Muhittin Ergün |
| 170 | 18.98 | -.- | -.- | -.- | -.- | Muhittin Ergün |
| 171 | 9.42 | -.- | -.- | -.- | -.- | Muhittin Ergün |
| 172 | 14.78 | -.- | -.- | -.- | -.- | Muhittin Ergün |
| 173 | 4.93 | -.- | -.- | -.- | -.- | Muhittin Ergün |
| 174 | 17.80 | -.- | -.- | -.- | -.- | Muhittin Ergün |
| 175 | 5.83 | -.- | -.- | -.- | -.- | Muhittin Ergün |
| 176 | 9.08 | -.- | -.- | -.- | -.- | Muhittin Ergün |
| 177 | 4.01 | -.- | -.- | -.- | -.- | Muhittin Ergün |
| 178 | 4.50 | -.- | -.- | -.- | -.- | Mehmet Ergün |
| 179 | 47.29 | 38.9018 | 40.5919 | 1034 | 10/3/2015 | Ozan Ünsalan |
| 180 | 4.00 | 38.8947 | 40.6055 | 1078 | 10/3/2015 | Oner Ergün |
| 181 | 17.64 | 38.9012 | 40.5907 | 1034 | 10/3/2015 | Fahrettin Baydas |
| 182 | 7.60 | 38.9022 | 40.5857 | 1035 | 10/3/2015 | Ersin Kaygisiz |
| 183 | 23.98 | 38.9056 | 40.6000 | 1045 | 10/3/2015 | Fahrettin Baydas |
| 184 | 9.21 | 38.8945 | 40.6053 | 1084 | 10/3/2015 | Burak Ergün |
| 185 | 1.77 | 38.8951 | 40.6056 | 1072 | 10/3/2015 | Ismail Ergün |
| 186 | 0.64 | 38.8991 | 40.6032 | 1043 | 10/3/2015 | Isa Cicek |
| 187 | 18.25 | -.- | -.- | -.- | -.- | Mehmet Ergün Sr. |
| 188 | 0.65 | -.- | -.- | -.- | -.- | Nezir Ergün |
| 189 | 4.73 | 38.8971 | 40.6135 | 1087 | 9/9/2015 | Iskender Demirkol |
| 190 | 3.42 | 38.8931 | 40.6084 | 1106 | 9/9/2015 | Iskender Demirkol |
| 191 | 6.81 | 38.8921 | 40.6016 | 1080 | 9/9/2015 | Ibrahim Y. Erdogan |
| 192 | 5.35 | 38.8943 | 40.6049 | 1079 | 9/9/2015 | Ibrahim Y. Erdogan |
| 193 | 2.12 | 38.8939 | 40.6140 | 1096 | 9/9/2015 | Aydin Sukru Bengu |
| 194 | 34.50 | 38.8982 | 40.5939 | 1040 | 10/14/2015 | Ibrahim Y. Erdogan |
| 195 | 25 | 38.8999 | 40.5922 | 1040 | 10/13/2015 | Burak Ergün |
| 196 | 100 | 38.9025 | 40.5952 | 1045 | 10/10/2015 | Metin Ergün |
| 197 | 1 | 38.9023 | 40.5953 | 1045 | 10/10/2015 | Metin Ergün |
| 198 | 15 | 38.9034 | 40.5954 | 1044 | 10/10/2015 | Metin Ergün |
| 199 | 10 | 38.9034 | 40.5954 | 1044 | 10/10/2015 | Metin Ergün |
| 200 | 5 | 38.9034 | 40.5970 | 1046 | 10/10/2015 | Metin Ergün |
| 201 | 110 | 38.9035 | 40.5971 | 1046 | 10/5/2015 | Mehmet Ergün |
| 202 | 70 | 38.8894 | 40.5953 | 1049 | 10/2/2015 | Muhammet T.Ergün |
| 203 | 35 | 38.8921 | 40.6123 | 1115 | -.- | Selahattin Ergün |
| 204 | 40 | 38.8920 | 40.6123 | 1114 | -.- | Selahattin Ergün |
| 205 | 45 | 38.8919 | 40.6126 | 1112 | -.- | Selahattin Ergün |
| 206 | 50 | 38.9083 | 40.6072 | 1052 | -.- | Selahattin Ergün |
| 207 | 250 | 38.9088 | 40.6065 | 1052 | -.- | Selahattin Ergün |
| 208 | 2.2 | 38.8994 | 40.6007 | 1046 | 10/11/2015 | Hasan Ergün |
| 209 | 1.1 | 38.8993 | 40.6015 | 1048 | 10/11/2015 | Burak Ergün |
| 210 | 1.3 | 38.8997 | 40.6018 | 1048 | 10/11/2015 | Hasan Ergün |
| 211 | 9.4 | 38.9000 | 40.6021 | 1048 | 10/11/2015 | Hasan Ergün |
| 212 | 6.5 | 38.9007 | 40.6027 | 1048 | 10/11/2015 | Hasan Ergün |
| 213 | 20.3 | 38.8990 | 40.6004 | 1045 | 10/15/2015 | Ismail Ergün |
| 214 | 28.1 | 38.8964 | 40.6016 | 1046 | 10/16/2015 | Muhammet Sazak |
| 215 | 120.6 | 38.9016 | 40.6026 | 1048 | 10/16/2015 | Onur Ergün |
| 216 | 22.4 | 38.9016 | 40.6026 | 1048 | 10/16/2015 | Yasir Ergün |
| 217 | 2.1 | 38.9016 | 40.6026 | 1048 | 10/17/2015 | Furkan Ergün |
| 218 | 2.75 | 38.9016 | 40.6026 | 1048 | 10/17/2015 | Ibrahim Y. Erdogan |
| 219 | 1.12 | 38.9016 | 40.6026 | 1048 | 10/17/2015 | Iskender Demirkol |
| 220 | 217 | 38.9014 | 40.6067 | 1049 | 10/13/2015 | Hatice Bullukara |
| 221 | 4.3 | 38.8991 | 40.6063 | 1049 | 10/3/2015 | Resit Sazak |
| 222 | 7.1 | 38.8982 | 40.6079 | 1051 | 10/3/2015 | Firat Gorken |
| 223 | 65.7 | 38.8984 | 40.6039 | 1050 | 10/3/2015 | Metin Ergün |
| 224 | 26.2 | 38.8965 | 40.5941 | 1040 | 10/7/2015 | Eli Emci |
| 225 | 6.2 | 38.8962 | 40.5974 | 1043 | 10/10/2015 | Zelihan Sazak |
| 226 | 5.1 | 38.8965 | 40.5979 | 1043 | 10/10/2015 | Zelihan Sazak |
| 227 | 7.3 | 38.8964 | 40.5985 | 1043 | 10/10/2015 | Zelihan Sazak |
| 228 | 16.1 | 38.8968 | 40.5989 | 1043 | 10/10/2015 | Yasir Ergün |
| 229 | 225.8 | 38.9067 | 40.6081 | 1052 | 10/16/2015 | Firat Ergün |
| 230 | 12.72 | 38.9066 | 40.6090 | 1053 | 11/2/2016 | Ibrahim Y. Erdogan |
| 231 | 1250 | 38.8787 | 40.6131 | 1102 | 10/31/2015 | Zeki Ozel |
| 232 | 1470 | 38.8816 | 40.6156 | 1144 | 11/1/2015 | Hasan Beldek |
| 233 | 410 | 38.8872 | 40.6021 | 1092 | 10/30/2015 | Mucahit Emci |
| 234 | 420 | 38.8857 | 40.6046 | 1101 | 10/30/2015 | Celal Ergormus |
| 235 | 2.34 | 38.9114 | 40.5865 | 1050 | 11/2/2015 | Aydin Ozdemir |
| 236 | 4.17 | 38.9113 | 40.5861 | 1050 | 11/2/2015 | Iskender Demirkol |
| 237 | 3.66 | 38.9113 | 40.5856 | 1050 | 11/2/2015 | Ibrahim Y. Erdogan |
| 238 | 12.17 | -.- | -.- | -.- | -.- | Mesut Gencay |
| 239 | 16.29 | -.- | -.- | -.- | -.- | Mesut Gencay |
| 240 | 29.58 | -.- | -.- | -.- | -.- | Mesut Gencay |
| 241 | 50.87 | -.- | -.- | -.- | -.- | Mesut Gencay |
| 242 | 230.30 | 38.9024 | 40.6213 | 1098 | 11/1/2015 | Ferit Karaoba |
| 243 | 115.76 | 38.9023 | 40.6189 | 1083 | 11/1/2015 | Ferit Karaoba |
| 244 | 205.62 | 38.8916 | 40.5931 | 1037 | 10/10/2015 | Ferit Karaoba |
| 245 | 5.1 | 38.9166 | 40.5832 | 1095 | 11/12/2015 | Mustafa Ramiz |
| 246 | 15.2 | 38.9129 | 40.5825 | 1085 | 11/1/2015 | Turan Morkoyun |
| 247 | 3.4 | 38.9131 | 40.5864 | 1080 | 11/3/2015 | Murat Ergün |
| 248 | 1054.2 | 38.8821 | 40.6107 | 1090 | 11/8/2015 | Abdullah Ercan |
| 249 | 80.3 | 38.9056 | 40.6123 | 1058 | 11/10/2015 | Yaşar Belgin |
| 250 | 8.7 | 38.9133 | 40.5858 | 1090 | 11/2/2015 | Nihat Buluş |
| 251 | 2.78 | 38.9152 | 40.5730 | 1122 | 11/9/2015 | Iskender Demirkol |
| 252 | 18.2 | 38.9120 | 40.5818 | 1075 | 11/12/2015 | Kazım Temiz |
| 253 | 1.2 | 38.9126 | 40.5660 | 1088 | 11/1/2015 | Kazım Temiz |
| 254 | 5.4 | 38.9130 | 40.5725 | 1094 | 11/11/2015 | Kadir Temiz |
| 255 | 2.34 | 38.9154 | 40.5710 | 1119 | 11/9/2015 | Ibrahim Y. Erdogan |
| 256 | 12.4 | 38.9074 | 40.5696 | 1051 | 11/2/2015 | Menderes Atlı |
| 257 | 8.3 | 38.9107 | 40.5773 | 1061 | 11/3/2015 | Menderes Atlı |
| 258 | 5.1 | 38.9098 | 40.5680 | 1083 | 11/8/2015 | Menderes Atlı |
| 259 | 1.0 | 38.9209 | 40.5789 | 1163 | 11/9/2015 | Menderes Atlı |
| 260 | 1.1 | 38.9201 | 40.5672 | 1210 | 11/4/2015 | Menderes Atlı |
| 261 | 2.6 | 38.9157 | 40.5779 | 1105 | 11/7/2015 | Ahmet Aras |
| 262 | 1.7 | 38.9170 | 40.5684 | 1124 | 11/2/2015 | Veysel Aras |
| 263 | 3.5 | 38.9112 | 40.5696 | 1086 | 11/6/2015 | Veysel Aras |
| 264 | 13.18 | 38.9142 | 40.5880 | 1085 | 11/5/2015 | Veysel Elaltuntas |
| 265 | 2.59 | 38.9143 | 40.5696 | 1098 | 11/5/2015 | Veysel Elaltuntas |
| 266 | 1.2 | 38.9164 | 40.5649 | 1152 | 11/10/2015 | Mustafa Kişeçok |
| 267 | 1.5 | 38.9154 | 40.5690 | 1107 | 11/11/2015 | Mustafa Kişeçok |
| 268 | 1.9 | 38.9205 | 40.5836 | 1177 | 11/1/2015 | Halil Gürmen |
| 269 | 2.4 | 38.9110 | 40.5660 | 1082 | 11/8/2015 | Halil Gürmen |
| 270 | 4.5 | 38.9108 | 40.5701 | 1083 | 11/8/2015 | Halil Gürmen |
| 271 | 3.7 | 38.9117 | 40.5714 | 1081 | 11/5/2015 | Hasan Kondu |
| 272 | 4.5 | 38.9128 | 40.5746 | 1079 | 11/5/2015 | Hasan Kondu |
| 273 | 2.6 | 38.9124 | 40.5690 | 1091 | 11/4/2015 | Fethi Kondu |
| 274 | 7.8 | 38.9166 | 40.5901 | 1106 | 11/5/2015 | Fethi Kondu |
| 275 | 6.5 | 38.9170 | 40.5919 | 1104 | 11/12/2015 | Fethi Kondu |
| 276 | 9.2 | 38.9049 | 40.5705 | 1045 | 11/12/2015 | Uğur Korkucu |
| 277 | 7.6 | 38.9064 | 40.5713 | 1048 | 11/12/2015 | Uğur Korkucu |
| 278 | 2.9 | 38.9171 | 40.5798 | 1105 | 11/6/2015 | Gülden Koçuk |
| 279 | 5.6 | 38.9135 | 40.5782 | 1072 | 11/7/2015 | Gülden Koçuk |
| 280 | 52.1 | 38.9120 | 40.5839 | 1063 | 11/9/2015 | Muaz Korkutata |
| 281 | 7.8 | 38.9088 | 40.5729 | 1060 | 11/2/2015 | Metin Korkutata |
| 282 | 2.8 | 38.9184 | 40.5843 | 1119 | 11/5/2015 | Serap Erbil |
| 283 | 3.7 | 38.9137 | 40.5818 | 1089 | 11/6/2015 | Serap Erbil |
| 284 | 2.7 | 38.9188 | 40.5877 | 1110 | 11/9/2015 | Remziye Alaçay |
| 285 | 8.4 | 38.9086 | 40.5779 | 1052 | 11/9/2015 | Remziye Alaçay |
| 286 | 5.3 | 38.9145 | 40.5840 | 1109 | 11/2/2015 | Aysel Gümren |
| 287 | 8.2 | 38.9057 | 40.5756 | 1049 | 11/6/2015 | Ayşe Alcı |
| 288 | 1.1 | 38.9188 | 40.5781 | 1133 | 11/4/2015 | Faruk Korkucu |
| 289 | 1.9 | 38.9193 | 40.5822 | 1159 | 11/4/2015 | Faruk Korkucu |
| 290 | 2.3 | 38.9173 | 40.5827 | 1103 | 11/11/2015 | Osman Temiz |
| 291 | 8.6 | 38.9132 | 40.5893 | 1057 | 11/4/2015 | Ali Atlı |
| 292 | 9.4 | 38.9148 | 40.5932 | 1066 | 11/11/2015 | Ali Atlı |
| 293 | 12.3 | 38.9160 | 40.5950 | 1065 | 11/3/2015 | Birol Kişeçok |
| 294 | 20.6 | 38.9106 | 40.6047 | 1059 | 11/5/2015 | Kadri Kondu |
| 295 | 14.7 | 38.9010 | 40.5759 | 1034 | 11/5/2015 | Kadri Kondu |
| 296 | 24.3 | 38.9080 | 40.6082 | 1052 | 11/7/2015 | Mustafa Koçuk |
| 297 | 28.1 | 38.9044 | 40.6078 | 1052 | 11/8/2015 | Mustafa Koçuk |
| 298 | 11.8 | 38.9098 | 40.5962 | 1046 | 11/7/2015 | Recep Korkutata |
| 299 | 9.7 | 38.9050 | 40.5929 | 1042 | 11/1/2015 | Fatma Aras |
| 300 | 15.8 | 38.9143 | 40.5960 | 1056 | 11/8/2015 | Fatma Aras |
| 301 | 30.4 | 38.8982 | 40.6110 | 1061 | 11/8/2015 | Meral Alcı |
| 302 | 165.2 | 38.8878 | 40.5835 | 1031 | 11/5/2015 | Ismail Çetin |
| 303 | 225.08 | -.- | -.- | -.- | 9/7/2015 | Abdullah Çurman |
| 304 | 6.0 | 38.8873 | 40.5937 | 1056 | 10/5/2015 | Ibrahim Arifoğlu |
| 305 | 4.50 | 38.8872 | 40.5933 | 1056 | 10/5/2015 | Ibrahim Arifoğlu |
| 307 | 245.7 | 38.8843 | 40.5934 | 1061 | 12/2/2015 | Bedri Alakuş |
| 308 | 10.0 | -.- | -.- | -.- | 10/4/2015 | Vedat Serttağ, Gülcan Serttağ |
| 309 | 13.83 | -.- | -.- | -.- | 10/4/2015 | Selahattin Ergün |
| 310 | 15.56 | -.- | -.- | -.- | 10/4/2015 | Selahattin Ergün |
| 311 | 37.58 | -.- | -.- | -.- | 10/4/2015 | Selahattin Ergün |
| 312 | 115.2 | 38.8968 | 40.5818 | 1034 | 12/4/2015 | Mustafa Bilmen |
| 313 | 4.20 | 38.9185 | 40.5946 | 1109 | 12/4/2015 | Ayşe Baytimur |
| 314 | 1.60 | 38.9209 | 40.5957 | 1159 | 12/5/2015 | Ayşe Baytimur |
| 315 | 2.40 | 38.9186 | 40.5939 | 1117 | 12/7/2015 | Sevinç Öge |
| 316 | 1.50 | 38.9262 | 40.5950 | 1256 | 12/7/2015 | Sevinç Öge |
| 317 | 7.10 | 38.9163 | 40.5989 | 1063 | 12/8/2015 | Gülşen Koçuk |
| 318 | 4.90 | 38.9068 | 40.5847 | 1040 | 12/11/2015 | Meral Alcı |
| 319 | 2.18 | 38.9234 | 40.5948 | 1215 | 12/12/2015 | İskender Demirkol |
| 320 | 1.74 | 38.9246 | 40.5934 | 1235 | 12/12/2015 | Ibrahim Y. Erdogan |
| 321 | 1.70 | 38.9214 | 40.5917 | 1149 | 12/13/2015 | Zeliha Kulaş |
| 322 | 1.80 | 38.9201 | 40.5890 | 1138 | 12/13/2015 | Zeliha Kulaş |
| 323 | 4.70 | 38.9086 | 40.5834 | 1045 | 12/13/2015 | Menderes Atlı |
| 324 | 5.50 | 38.9097 | 40.5845 | 1047 | 12/13/2015 | Menderes Atlı |
| 325 | 11.30 | -.- | -.- | -.- | 11/17/2015 | Yunus Taşçi |
| 326 | 12.73 | -.- | -.- | -.- | 11/17/2015 | Yunus Taşçi |
| 327 | 3.34 | -.- | -.- | -.- | 11/17/2015 | Yunus Taşçi |
| 328 | 12.15 | -.- | -.- | -.- | 11/17/2015 | Yunus Taşçi |
| 329 | 4.37 | -.- | -.- | -.- | 11/17/2015 | Yunus Taşçi |
| 330 | 8.59 | -.- | -.- | -.- | 11/17/2015 | Yunus Taşçi |
| 331 | 11.45 | -.- | -.- | -.- | 11/17/2015 | Yunus Taşçi |
| 332 | 15.95 | -.- | -.- | -.- | 11/17/2015 | Yunus Taşçi |
| 333 | 220.0 | -.- | -.- | -.- | 11/18/2015 | Uğur Ataoğlu |
| 334 | 150.0 | -.- | -.- | -.- | 11/18/2015 | Uğur Ataoğlu |
| 335 | 46 | -.- | -.- | -.- | -.- | Yunus Taşçi |
| 336 | 30 | -.- | -.- | -.- | -.- | Yunus Taşçi |
| 337 | 10 | -.- | -.- | -.- | -.- | Ahmet Becerikli |
| 338 | 8 | -.- | -.- | -.- | -.- | Ahmet Becerikli |
| 339 | 2 | -.- | -.- | -.- | -.- | Ahmet Becerikli |
| 340 | 2 | -.- | -.- | -.- | -.- | Ahmet Becerikli |
| 341 | 2 | -.- | -.- | -.- | -.- | Ahmet Becerikli |
| 342 | 3 | -.- | -.- | -.- | -.- | Ahmet Becerikli |
| 343 | 210 | -.- | -.- | -.- | -.- | Onur Ergün |

Note: SC306 was removed from the list because it is likely the same meteorite as SC10; † Location of meteorite.

Table 3. Atmospheric trajectory and pre-impact orbit of the Sariçiçek meteoroid. All angles are for equinox J2000.

| Trajectory | Apparent | Geocentric: | Orbit | Schiaparelli[a] | Dynamic[b] |
|---|---|---|---|---|---|
| Date | 2015-09-22 | 2015-09-22 | Epoch (TD) | 2015-09-022 | 2015-07-04 |
| Time at start (UT) | 20:10:26.92 | 20:10:26.92 | Time at Start (UT) | 20:10:26.92 | 20:11:36 |
| Right Ascension (º) | 276.5 ± 1.4 | 264.8 ± 13.4 | Solar Longitude (º) | 159.8392 ± 0.0001 | 159.849 ± 0.004 |
| Declination (º) | 59.7 ± 0.8 | 59.4 ± 4.8 | Perihelion Distance (AU) | 1.009 ± 0.012 | 1.0086 ± 0.0004 |
| Entry Speed (km $s^{-1}$) | 17.1 ± 0.8 | 13.1 ± 1.1 | Semi-major Axis (AU) | 1.44 ± 0.17 | 1.454 ± 0.083 |
| Ecliptic Longitude (º) | – | 249.4 ± 47.6 | Eccentricity | 0.301 ± 0.071 | 0.304 ± 0.039 |
| Ecliptic Latitude (º) | – | 82.5 ± 3.0 | Inclination (º, J2000) | 22.6 ± 1.8 | 22.6 ± 1.6 |
| Begin Altitude (km) | 58.4[c] | – | Argument of Perihelion (º) | 182.9 ± 17.0 | 182.8 ± 1.6 |
| Latitude (ºN) | 39.1163±0.009 | – | Node (º) | 159.832 ± 0.003 | 159.849 ± 0.004 |
| Longitude (ºE) | 40.3687±0.009 | – | Longitude of Perihelion (º) | 342.7 ± 17.0 | 324.6 ± 2.7 |
| Fragmentation Altitude (km) | 36.5 ± 1.0 | – | True Anomaly (º) | – | –63.1 ± 1.4 |
| Disruption Altitude (km) | 27.4 ± 1.4 | – | Mean Anomaly (º) | – | 324.6 ± 2.7 |
| End Altitude (km) | 21.3 ± 0.5 | – | Heliocentric Speed (km $s^{-1}$) | 33.8 ± 1.1 | – |
| Latitude (ºN) | 38.9268±0.009 | – | Tisserand Parameter $T_J$ | 4.53 ± 0.37 | 4.52 ± 0.20 |
| Longitude (ºE) | 40.5683±0.009 | – | Aphelion Distane (AU) | 1.88 ± 0.26 | 1.898 ± 0.166 |
| Azimuth (º, N) | 320.7 ± 1.0 | – | Date at Perihelion | – | 2015-09-05 ± 1.42 |
| Zenith angle (º) | 36.6 ± 0.8 | – | Time at Perihelion | – | 09:50:26 |

[a]Jenniskens et al. (2011).

[b]Orbit calculated as in Clark and Wiegert (2011). An uncertainty of 0.5 km in beginning altitude is assumed. See text for ascending node discrepancy

[c]First creating detectable shadows at ~60.2 km altitude.

Table 4. Compression strength of meteorites.

| SC# | P (MPa) | Measurement |
|---|---|---|
| Cube: | | |
| 12 | 79.1 ± 0.3 | New Mexico Tech |
| 14 | (broke) | New Mexico Tech |
| Intact stone[a] | | |
| 12 | 380 ± 40 | NASA Ames |
| 14 | 7 ± 5 | NASA Ames |
| 50 | ≥45 | Univ. Istanbul |
| 54 | 66 ± 23 | Univ. Istanbul |
| 57 | 168 ± 110 | Univ. Istanbul |
| 239 | 71 ± 19 | Univ. Istanbul |

[a]Uncertainties derive from uncertainty in area.

Table 5. Sariçiçek bulk elemental composition. Compilation of data by UC Davis (UCD) and Fordham University.

| Element | Z | Units | SC12 (UCD) | SC14 (Fordham) | Element | Z | Units | SC12 (UCD) | SC14 (Fordham) |
|---|---|---|---|---|---|---|---|---|---|
| Li | 3 | μg g$^{-1}$ | 5.68 | 5.7 | Cd | 48 | μg g$^{-1}$ | 0.020 | -.- |
| Be | 4 | μg g$^{-1}$ | 0.174 | -.- | Sn | 50 | μg g$^{-1}$ | 0.119 | -.- |
| Na | 11 | wt% | 0.255 | 0.20 | Te | 52 | μg g$^{-1}$ | 0.052 | -.- |
| Mg | 12 | wt% | 9.94 | 9.9 | Cs | 55 | μg g$^{-1}$ | 0.007 | 0.007 |
| Al | 13 | wt% | 3.77 | 5.23 | Ba | 56 | μg g$^{-1}$ | 10.1 | 12 |
| Si | 14 | wt% | – | – | La | 57 | μg g$^{-1}$ | 1.37 | 1.79 |
| P | 15 | wt% | – | – | Ce | 58 | μg g$^{-1}$ | 4.06 | 4.91 |
| Cl | 17 | wt% | – | – | Pr | 59 | μg g$^{-1}$ | 0.623 | 0.75 |
| K | 19 | wt% | 0.0248 | 0.022 | Nd | 60 | μg g$^{-1}$ | 2.95 | 3.56 |
| Ca | 20 | wt% | 5.34 | 6.28 | Sm | 62 | μg g$^{-1}$ | 1.02 | 1.23 |
| Sc | 21 | μg g$^{-1}$ | 24.3 | 25.1 | Eu | 63 | μg g$^{-1}$ | 0.331 | 0.41 |
| Ti | 22 | μg g$^{-1}$ | 2630 | – | Gd | 64 | μg g$^{-1}$ | 1.34 | 1.62 |
| V | 23 | μg g$^{-1}$ | 102 | 88.7 | Tb | 65 | μg g$^{-1}$ | 0.260 | 0.32 |
| Cr | 24 | wt% | 0.756 | – | Dy | 66 | μg g$^{-1}$ | 1.43 | 1.68 |
| Mn | 25 | μg g$^{-1}$ | 4638 | 4560 | Ho | 67 | μg g$^{-1}$ | 0.349 | 0.42 |
| Fe | 26 | wt% | 14.4 | 14.6 | Er | 68 | μg g$^{-1}$ | 0.998 | 1.21 |
| Co | 27 | μg g$^{-1}$ | 42.5 | 25.0 | Tm | 69 | μg g$^{-1}$ | 0.158 | 0.19 |
| Ni | 28 | μg g$^{-1}$ | 530 | 150 | Yb | 70 | μg g$^{-1}$ | 0.939 | 1.14 |
| Cu | 29 | μg g$^{-1}$ | 3.32 | 3.4 | Lu | 71 | μg g$^{-1}$ | 0.171 | 0.20 |
| Zn | 30 | μg g$^{-1}$ | 2.09 | 2.1 | Hf | 72 | μg g$^{-1}$ | 0.848 | 0.80 |
| Ga | 31 | μg g$^{-1}$ | 0.887 | 1.7 | Ta | 73 | μg g$^{-1}$ | 0.074 | 0.06 |
| Ge | 32 | μg g$^{-1}$ | 11.6 | – | W | 74 | μg g$^{-1}$ | 0.197 | – |
| As | 33 | μg g$^{-1}$ | 13.6 | – | Re | 75 | μg g$^{-1}$ | – | – |
| Se | 34 | μg g$^{-1}$ | 114 | – | Os | 76 | μg g$^{-1}$ | 0.010 | – |
| Rb | 37 | μg g$^{-1}$ | 0.209 | 0.20 | Ir | 77 | μg g$^{-1}$ | 0.008 | 0.01 |
| Sr | 38 | μg g$^{-1}$ | 39.3 | 43 | Pt | 78 | μg g$^{-1}$ | 0.011 | 0.018 |
| Y | 39 | μg g$^{-1}$ | 9.96 | 13 | Au | 79 | μg g$^{-1}$ | 0.033 | – |
| Zr | 40 | μg g$^{-1}$ | 40.9 | 34 | Tl | 81 | μg g$^{-1}$ | 0.002 | – |
| Nb | 41 | μg g$^{-1}$ | 1.48 | – | Pb | 82 | μg g$^{-1}$ | 0.196 | – |
| Mo | 42 | μg g$^{-1}$ | 0.977 | – | Th | 90 | μg g$^{-1}$ | 0.181 | 0.23 |
| Ru | 44 | μg g$^{-1}$ | 0.036 | 0.02 | U | 92 | μg g$^{-1}$ | 0.034 | 0.050 |
| Ag | 47 | μg g$^{-1}$ | 0.338 | – | | | | | |

Table 6. The oxygen isotope data for Sariçiçek. Stable isotope results are given in ‰ V-SMOW. The delta values are linearized, using a slope of 0.528. Averages of the bulk rock measurements do not include the mono-mineralic feldspar analyses.

| SC12 | mg | $\delta^{17}O'$ | $\delta^{18}O'$ | $\Delta^{17}O'$ | SC14 | mg | $\delta^{17}O'$ | $\delta^{18}O'$ | $\Delta^{17}O'$ |
|---|---|---|---|---|---|---|---|---|---|
| Bulk | 1.8 | 1.627 | 3.666 | –0.309 | Bulk | 1.1 | 1.589 | 3.554 | –0.288 |
| ″ | 2.1 | 1.519 | 3.450 | –0.303 | ″ | 1.1 | 1.542 | 3.505 | –0.309 |
| ″ | 2.1 | 1.437 | 3.453 | –0.386 | ″ | 2.2 | 1.602 | 3.627 | –0.313 |
| ″ | 1.1 | 1.632 | 3.638 | –0.289 | ″ | 1.9 | 1.543 | 3.543 | –0.328 |
| ″ | 1.5 | 1.571 | 3.502 | –0.278 | ″ | 1.7 | 1.487 | 3.498 | –0.360 |
| – | – | – | – | – | ″ | 1.4 | 1.706 | 3.700 | –0.248 |
| – | – | – | – | – | ″ | 1.5 | 1.631 | 3.662 | –0.303 |
| – | – | – | – | – | ″[a] | 1.6 | 1.621 | 3.633 | –0.297 |
| – | – | – | – | – | ″[a] | 1.8 | 1.651 | 3.688 | –0.296 |
| – | – | – | – | – | ″[a] | 1.5 | 1.589 | 3.609 | –0.317 |
| – | – | – | – | – | ″ feldspar | 2.3 | 1.783 | 3.981 | –0.319 |
| – | – | – | – | – | ″ feldspar[a] | 1.8 | 1.714 | 3.856 | –0.322 |
| SC 12 avg. | | 1.557 | 3.542 | -0.313 | SC 14 avg. | | 1.596 | 3.602 | –0.306 |
| SD (σ) | | 0.081 | 0.103 | 0.043 | SD (σ) | | 0.062 | 0.073 | 0.029 |

[a]Portion of SC14 that was not acid-treated. SD, standard deviation.

Table 7. Cr isotopic composition.

| | ε$^{53}$Cr (±2SE) | ε$^{54}$Cr (±2SE) |
|---|---|---|
| Sariçiçek SC12 | 0.15 ± 0.04 | –0.69 ± 0.10 |
| Bholghati | 0.23 ± 0.04 | –0.63 ± 0.10 |

Table 8. Ti isotopic composition.

| Howardites | ε$^{46}$Ti/$^{47}$Ti | ε$^{48}$Ti/$^{47}$Ti | ε$^{50}$Ti/$^{47}$Ti | Eucrites | ε$^{46}$Ti/$^{47}$Ti | ε$^{48}$Ti/$^{47}$Ti | ε$^{50}$Ti/$^{47}$Ti |
|---|---|---|---|---|---|---|---|
| Sariçiçek (HR) | –0.24 | -0.05 | –1.06 | Béréba (MR)[a] | –0.09 | 0.04 | –1.19 |
| "(HR) | –0.03 | -0.05 | –1.34 | Béréba (MR)[a] | –0.32 | 0.01 | –1.22 |
| "(HR) | –0.14 | 0.06 | –1.22 | Juvinas (MR)[b] | –0.43 | –0.01 | –1.20 |
| "(HR) | –0.16 | 0.01 | –1.36 | Juvinas (MR)[b] | –0.28 | –0.20 | –1.20 |
| "(MR) | –0.27 | 0.01 | –1.25 | Pasamonte (HR)[b] | –0.11 | –0.02 | –1.24 |
| "(MR) | –0.20 | 0.16 | –1.17 | | | | |
| "(MR) | –0.28 | 0.05 | –1.37 | | | | |
| "(MR) | –0.24 | 0.15 | –1.36 | | | | |
| "(MR, UCD) | –0.17 | -0.05 | –1.20 | | | | |
| Average | –0.19 | 0.03 | –1.26 | Average | –0.24 | –0.03 | –1.21 |
| 2SD | 0.16 | 0.16 | 0.21 | 2SD | 0.29 | 0.19 | 0.04 |
| 2SE | 0.05 | 0.05 | 0.07 | 2SE | 0.13 | 0.08 | 0.02 |
| Kapoeta[d] | –0.22 | -0.14 | –1.23 | Ave. Eucrites | –0.27 | 0.03 | –1.28 |
| 2SE (*N* = 16) | 0.05 | 0.20 | 0.04 | 2SE (*N* = 6)[c] | 0.09 | 0.04 | 0.02 |
| | | | | Ave. Eucrites | –0.23 | –0.01 | –1.25 |
| | | | | 2SE (*N* = 4)[d] | 0.04 | 0.04 | 0.06 |

A single sample aliquot was analysed several times. N, number of measurements; HR, high resolution analyses, MR, medium resolution analyses, SD:, standard deviation, SE, standard error (SD divided by square root of number of measurements); UCD, samples measured at UC Davis.

[a]Solution from *Akram et al. (*2015), reanalyzed in this study

[b]Solution from Williams (2015), reanalyzed in this study

[c]Data from Williams (2015)

[d]Data from Zhang et al. (2012).

Table 9. Summary of the amino acid abundances in parts-per-billion (ppb) in the free (nonhydrolyzed) and total (6M HCl-hydrolyzed) hot-water extracts of SC12 and SC14, and soil and a small pebble collected from the SC16 and SC14 meteorite recovery locations, respectively.

| | **Sariçiçek Meteorite (SC14)** | | **Sariçiçek Meteorite (SC12)** | | **Soil near SC16** | | **Pebble near SC14** | |
|---|---|---|---|---|---|---|---|---|
| **Amino acid** | Free | Total | Free | Total | Free | Total | Free | Total |
| D-aspartic acid | 1.5 ± 0.1 | 34.9 ± 0.4 | < 0.1 | 1.2 ± 0.2 | 16.9 ± 4.6 | 358.9 ± 27.0 | 36.5 ± 2.6 | 697.3 ± 53.6 |
| L-aspartic acid | 13.6 ± 0.6 | 98.8 ± 1.0 | < 0.1 | 4.9 ± 0.6 | 470.6 ± 121.5 | 1,840.2 ± 164.8 | 135.1 ± 5.0 | 1,271.9 ± 111.2 |
| L-glutamic acid | 5.1 ± 0.3 | 141.5 ± 1.1 | < 0.1 | 6.1 ± 0.3 | 841.3 ± 228.4 | 3,087.0 ± 555.0 | 276.4 ± 23.5 | 2,067.1 ± 133.0 |
| D-glutamic acid | 0.5 ± 0.0 | 17.7 ± 0.4 | < 0.1 | 1.1 ± 0.1 | 115.9 ± 21.6 | 621.7 ± 102.9 | 23.7 ± 1.0 | 604.5 ± 52.6 |
| D-serine | 1.6 ± 0.1 | 13.2 ± 1.9 | < 0.1 | 0.6 ± 0.1 | < 0.1 | 296.9 ± 6.7 | 29.7 ± 1.8 | 135.2 ± 13.7 |
| L-serine | 6.7 ± 0.3 | 79.7 ± 12.4 | < 0.1 | 3.8 ± 1.1 | 156.7 ± 40.7 | 929.7 ± 52.3 | 84.3 ± 5.4 | 229.6 ± 20.1 |
| **C2 amino acid** | | | | | | | | |
| glycine | 15.9 ± 0.6 | 164.2 ± 24.5 | < 0.1 | 4.4 ± 0.3 | 539.2 ± 132.6 | 1,378.1 ± 86.9 | 645.9 ± 32.9 | 2,671.4 ± 60.6 |
| **C3 amino acids** | | | | | | | | |
| β-alanine | 1.1 ± 0.1 | 3.7 ± 0.4 | < 0.1 | < 0.1 | 54.2 ± 15.0 | 36.2 ± 2.6 | 12.4 ± 1.0 | ≤ 0.1 |
| D-alanine | 2.3 ± 0.1 | 6.5 ± 0.3 | < 0.1 | 0.2 ± 0.1 | 63.0 ± 10.9 | 160.3 ± 4.3 | 33.6 ± 4.5 | 281.6 ± 16.0 |
| L-alanine | 5.0 ± 0.4 | 55.5 ± 0.7 | < 0.1 | 5.0 ± 0.6 | 220.3 ± 33.1 | 681.0 ± 24.6 | 112.3 ± 12.8 | 869.7 ± 33.1 |
| **C4 amino acids** | | | | | | | | |
| γ-amino-*n*-butyric acid | 2.2 ± 0.3 | 7.4 ± 0.6 | < 0.1 | 2.1 ± 0.7 | 1,100.3 ± 198.2 | 2,135.5 ± 96.5 | 67.1 ± 12.1 | 133.4 ± 20.9 |
| D-β-amino-*n*-butyric acid | < 0.1 | 0.2 ± 0.0 | < 0.1 | 0.2 ± 0.1 | 2.1 ± 0.4 | < 18.8 | 5.6 ± 0.2 | 4.3 ± 0.8 |
| L-β-amino-*n*-butyric acid | < 0.1 | 0.2 ± 0.0 | < 0.1 | < 0.1 | 2.1 ± 0.4 | < 59.8 | 5.9 ± 0.2 | ≤ 0.1 |
| α-aminoisobutyric acid | < 0.1 | 0.2 ± 0.1 | < 0.1 | 1.9 ± 0.4 | 2.6 ± 0.5 | 50 ± 3.1 | 1.7 ± 0.2 | ≤ 0.1 |
| D,L-α-amino-*n*-butyric acid | < 0.1 | 2.5 ± 0.1 | < 0.1 | 0.3 ± 5.4 | 2.5 ± 3.0 | 58.6 ± 4.5 | < 0.1 | 20.7 ± 3.1 |
| **C5 amino acids** | | | | | | | | |
| D-isovaline | < 0.1 | < 0.1 | < 0.1 | < 0.1 | < 0.1 | < 0.1 | < 0.1 | ≤ 0.1 |
| L-isovaline | < 0.1 | < 0.1 | < 0.1 | < 0.1 | < 0.1 | < 0.1 | < 0.1 | ≤ 0.1 |
| L-valine | 2.2 ± 0.0 | 27.0 ± 2.4 | < 0.1 | 7.0 ± 0.0 | 95.0 ± 12.5 | 681.1 ± 33.7 | 55.2 ± 4.4 | 1,172.8 ± 144.1 |
| D-valine | < 0.1 | 0.5 ± 0.0 | < 0.1 | < 0.1 | < 0.1 | 182.1 ± 10.3 | 1.6 ± 0.2 | 34.5 ± 6.1 |
| **C6 amino acid** | | | | | | | | |
| ε-amino-*n*-caproic acid | 0.1 ± 0.0 | 12.5 ± 1.5 | < 0.1 | 9.6 ± 0.7 | 0.5 ± 0.1 | < 0.1 | 0.6 ± 0.4 | 4.9 ± 0.7 |
| Total (ppb) | 57 | 666 | < 0.1 | 45 | 3,683 | 12,497 | 1,529 | 10,199 |

[a]Extracts were analyzed by OPA/NAC derivatization and UPLC separation with UV fluorescence detection and TOF-MS detection at NASA Goddard Space Flight Center. The uncertainties are based on the standard deviation of the average value of three separate measurements.

Table 10. Concentrations of major elements (measured by ICP-OES) and cosmogenic radionuclides (measured by AMS) in two fragments of the Sariçiçek howardite. Concentrations of Si and O are estimated from other elements, as in Beck *et al.* (2012) and Welten *et al.* (2012).

| | **SC12** | **SC14** |
|---|---|---|
| Mass (mg) | 52.0 | 58.5 |
| Element | | |
| O[a](wt%) | 43.0 | 43.0 |
| Mg (wt%) | 8.79 | 7.55 |
| Al (wt%) | 4.29 | 5.41 |
| Si[a] (wt%) | 23.9 | 23.6 |
| K (ppm) | 250 | 250 |
| Ca (wt%) | 4.64 | 5.64 |
| Ti (wt%) | 0.23 | 0.26 |
| Mn (wt%) | 0.41 | 0.39 |
| Fe (wt%) | 14.0 | 13.4 |
| Co (ppm) | 30 | 17 |
| Ni (ppm) | 320 | 130 |
| *Radionuclides* | | |
| $^{10}$Be (dpm kg$^{-1}$) | 22.9 ± 0.2 | 24.9 ± 0. 2 |
| $^{26}$Al (dpm kg$^{-1}$) | 86.7 ± 1.4 | 97.3 ± 1.6 |
| $^{36}$Cl (dpm kg$^{-1}$) | 13.8 ± 0.2 | 17.5 ± 0.3 |

[a]Estimated from presence of other minerals.

Table 11. Left: Massic activities (corrected to the time of fall) of cosmogenic radionuclides in the 131.88 g specimen SC26 measured by non-destructive gamma-ray spectroscopy. Errors include a 1σ uncertainty of ~10% in the detector efficiency calibration. Right: Concentration of primordial radionuclides in SC26. Errors include a 1s uncertainty of ~10% in the detector efficiency calibration.

| Nuclide | Half-life | Sariçiçek (SC26) (dpm $kg^{-1}$) |
|---|---|---|
| $^{52}Mn$ | 5.59 d | – |
| $^{48}V$ | 15.97 d | < 20 |
| $^{51}Cr$ | 27.7 d | < 56 |
| $^{7}Be$ | 53.1 d | 52 ± 9 |
| $^{58}Co$ | 70.9 d | < 2.6 |
| $^{56}Co$ | 77.3 d | < 5.2 |
| $^{46}Sc$ | 83.8 d | 5 ± 1 |
| $^{57}Co$ | 271.8 d | < 3.1 |
| $^{54}Mn$ | 312.3 d | 55.2 ± 5.6 |
| $^{22}Na$ | 2.60 y | 72.9 ± 5.0 |
| $^{60}Co$ | 5.27 y | < 2.1 |
| $^{44}Ti$ | 60 y | < 2.8 |
| $^{26}Al$ | $7.17x10^5$ y | 79.0 ± 6.1 |

| Nuclide | Sariçiçek (SC26) |
|---|---|
| U | 57.0 ± 3.1 ng $g^{-1}$ |
| Th | 203 ± 10 ng $g^{-1}$ |
| K | 200 ± 20 µg $g^{-1}$ |

Table 12. Noble gas He, Ne, and Ar results.

| Sample | Sariçiçek-SC12-Z1 | Sariçiçek-SC12-Z2 | Total (SC12) | Sariçiçek-SC14-Z2.1 | Sariçiçek-SC14-Z2.2 | Total (SC14) |
|---|---|---|---|---|---|---|
| Mass (mg) | 44.6 | 92.0 | 136.6 | 32.5 | 19.7 | 52.2 |
| $^{3}He/^{4}He_{meas}$ | 0.00331 | 0.00362 | 0.00352 | 0.00533 | 0.00502 | 0.00521 |
| $^{4}He_{meas}$ | 13100 | 8640 | 10100 | 777 | 860 | 808 |
| $^{20}Ne/^{22}Ne_{meas}$ | 8.87 | 7.74 | 8.11 | 6.08 | 6.60 | 6.29 |
| $^{21}Ne/^{22}Ne_{meas}$ | 0.317 | 0.399 | 0.372 | 0.487 | 0.446 | 0.470 |
| $^{20}Ne_{meas}$ | 249 | 153 | 184 | 111 | 134 | 120 |
| $^{20}Ne_{fSW}$ | 241 | 146 | 177 | 103 | 126 | 112 |
| $^{36}Ar/^{38}Ar_{meas}$ | 2.13 | 1.93 | 2.00 | 1.47 | 1.65 | 1.53 |
| $^{40}Ar/^{36}Ar_{meas}$ | 84.3 | 106 | 98.9 | 143 | 129 | 137 |
| $^{36}Ar_{meas}$ | 12.3 | 9.45 | 10.4 | 7.30 | 8.03 | 7.58 |
| $^{36}Ar_{SW}$ | 9.73 | 7.14 | 7.99 | 4.62 | 5.54 | 4.97 |
| $^{3}He_{cos}$ | 39.1 | 28.7 | 32.1 | 39.9 | 41.3 | 40.4 |
| $^{3}He_{cos, 4\pi}$ | – | – | 17.2 | -.- | -.- | 32.8 |
| $^{21}Ne_{cos}$ | 8.29 | 7.52 | 7.77 | 8.62 | 8.77 | 8.68 |
| $^{21}Ne_{cos, 4\pi}$ | – | – | 6.33 | – | – | 7.98 |
| $^{22}Ne/^{21}Ne_{cos}$ | – | – | 1.11 | – | – | 1.12 |
| $^{38}Ar_{cos}$ | 3.96 | 3.56 | 3.69 | 4.12 | 3.83 | 4.01 |
| $^{38}Ar_{cos, 4\pi}$ | – | – | 2.50 | – | – | ? |
| $^{4}He_{rad}$ | – | – | 1800±800 | – | – | 4000±800 |
| $^{40}Ar_{rad}$ | 1037 | 1002 | 1027 | 1044 | 1036 | 1038 |

All concentrations are given in units of $10^{-8}$ $cm^3$ STP $g^{-″}$ (= 0.01 nL $g^{-1}$ STP or $4.46\times10^{-13}$ mol $g^{-1}$). Uncertainties in concentrations are about 3%, uncertainties in ratios are about 0.5%.

Table 13. Collisional history of Sariçiçek from noble gas data.

| Sample | PR or Concentr. | SC12- Z1 | SC12- Z2 | SC12 Corrected | PR or Concentr. | SC14-Z2.1 | SC14-Z2.2 | SC14 Corrected |
|---|---|---|---|---|---|---|---|---|
| $^{3}He_{cos}$ CRE age | 1.65* | (24) | (17) | 10 | 1.65* | (24) | (25) | 18 |
| $^{21}Ne_{cos}$ CRE age | 0.281* | (31) | (28) | 22 | 0.290* | (30) | (30) | 23 |
| $^{38}Ar_{cos}$ CRE age | 0.126* | (31) | (28) | 20 | 0.134* | (31) | (29) | –[a] |
| U,Th-He age | U = 0.034<br>Th = 0.181 | – | – | 1800 ± 700 | U = 0.050<br>Th = 0.230 | – | – | 2600 ± 300 |
| K-Ar age | K = 248 | – | – | 3500 ± 70 | K = 220 | – | – | 3700 ± 100 |

All ages given in Ma (million years). Nominal ages are given in parentheses, while the regolith exposure corrected ages are given without parentheses. "PR or concentration" = U, Th, K concentration in ppm or the production rate (*) in $10^{-8}$ $cm^3$ STP $g^{-1}$ Ma calculated using the formulas given by Eugster and Michel (1995) and the elemental composition of Table 5. Cosmogenic radionuclide data suggest that production rates in SC14 are about 8%, 19%, and 17% higher for He, Ne, Ar than in SC12 (Leya and Masarik 2009); therefore, the given rates have been multiplied by an additional correction factor to determine the exposure ages given in the last column. [a]For SC14, the 4π $^{38}$Ar-age could not be determined as the "regolith trend line" for Ar is negative.

Table 14. SIMS U-Pb isotopic data of zircon from Sariçiçek.

| Spot | U ppm | Th ppm | Th/ U | $^{207}Pb^*/^{206}Pb^a$ | ±1σ (%) | $^{207}Pb^*/^{235}U$ | ±1σ (%) | $^{206}Pb^*/^{238}U$ | ±1σ (%) | $t_{207/206}$ (Ma) | ±1σ | $t_{207/235}$ (Ma) | ±1σ | $t_{206/238}$ (Ma) | ±1σ |
|---|---|---|---|---|---|---|---|---|---|---|---|---|---|---|---|
| 2 | 39 | 26 | 0.66 | 0.6159 | 0.52 | 77.5 | 5.0 | 0.913 | 5.00 | 4545 | 8 | 4430 | 52 | 4181 | 156 |
| 3 | 71 | 18 | 0.26 | 0.6155 | 0.37 | 91.8 | 3.0 | 1.082 | 2.95 | 4544 | 5 | 4600 | 30 | 4727 | 100 |
| 4-1 | 20 | 9 | 0.45 | 0.6116 | 0.71 | 84.9 | 3.2 | 1.007 | 3.16 | 4535 | 10 | 4521 | 33 | 4491 | 103 |
| 4-2-1 | 94 | 17 | 0.18 | 0.6201 | 0.46 | 91.8 | 2.4 | 1.075 | 2.32 | 4555 | 7 | 4600 | 24 | 4705 | 78 |
| 4-2-2 | 108 | 38 | 0.35 | 0.6179 | 0.38 | 79.6 | 2.8 | 0.935 | 2.73 | 4550 | 6 | 4457 | 28 | 4256 | 86 |
| 4-3 | 19 | 8 | 0.40 | 0.6186 | 0.91 | 95.1 | 5.6 | 1.116 | 5.51 | 4551 | 13 | 4636 | 58 | 4832 | 190 |
| 4-4 | 29 | 8 | 0.27 | 0.6165 | 0.61 | 89.8 | 2.6 | 1.057 | 2.56 | 4546 | 9 | 4578 | 27 | 4648 | 85 |
| 5 | 15 | 4 | 0.28 | 0.6141 | 0.35 | 90.5 | 3.4 | 1.069 | 3.38 | 4541 | 5 | 4586 | 35 | 4688 | 114 |
| 8-1 | 15 | 4 | 0.27 | 0.6175 | 0.68 | 85.6 | 4.1 | 1.006 | 4.05 | 4549 | 10 | 4530 | 42 | 4489 | 132 |
| 8-2 | 17 | 4 | 0.24 | 0.6230 | 0.75 | 87.1 | 4.9 | 1.014 | 4.83 | 4562 | 11 | 4547 | 50 | 4514 | 159 |
| 8-3 | 17 | 4 | 0.22 | 0.6174 | 0.82 | 83.4 | 5.4 | 0.981 | 5.31 | 4549 | 12 | 4504 | 55 | 4406 | 172 |
| 9 | 161 | 105 | 0.65 | 0.6203 | 0.39 | 83.7 | 2.2 | 0.979 | 2.17 | 4556 | 6 | 4507 | 22 | 4400 | 69 |
| 10 | 88 | 8 | 0.10 | 0.6213 | 0.37 | 86.5 | 1.6 | 1.011 | 1.55 | 4558 | 5 | 4541 | 16 | 4502 | 50 |
| 11 | 28 | 3 | 0.11 | 0.6208 | 0.45 | 94.2 | 3.1 | 1.101 | 3.03 | 4557 | 7 | 4626 | 31 | 4786 | 103 |
| 14 | 22 | 23 | 1.07 | 0.6104 | 0.61 | 96.3 | 4.3 | 1.145 | 4.27 | 4532 | 9 | 4648 | 44 | 4920 | 149 |
| 15 | 40 | 22 | 0.55 | 0.6146 | 0.42 | 85.1 | 4.1 | 1.005 | 4.06 | 4542 | 6 | 4524 | 42 | 4484 | 133 |
| 19-1 | 40 | 6 | 0.14 | 0.6138 | 0.53 | 82.8 | 5.0 | 0.979 | 4.94 | 4540 | 8 | 4497 | 51 | 4400 | 159 |
| 19-2 | 79 | 8 | 0.11 | 0.6183 | 0.31 | 86.5 | 2.4 | 1.015 | 2.41 | 4551 | 4 | 4540 | 25 | 4516 | 79 |
| 20 | 24 | 18 | 0.76 | 0.6189 | 0.45 | 89.1 | 5.9 | 1.045 | 5.92 | 4552 | 7 | 4571 | 61 | 4613 | 198 |
| 21-1 | 193 | 127 | 0.66 | 0.6232 | 0.35 | 88.7 | 2.7 | 1.033 | 2.66 | 4562 | 5 | 4565 | 27 | 4573 | 88 |
| 21-2 | 68 | 24 | 0.35 | 0.6177 | 0.50 | 89.8 | 2.4 | 1.055 | 2.36 | 4549 | 7 | 4578 | 25 | 4644 | 79 |
| 22-1 | 53 | 6 | 0.11 | 0.6190 | 0.42 | 83.1 | 3.4 | 0.974 | 3.34 | 4552 | 6 | 4500 | 34 | 4385 | 107 |
| 22-2 | 49 | 3 | 0.06 | 0.6191 | 0.48 | 83.9 | 3.0 | 0.984 | 2.95 | 4553 | 7 | 4510 | 30 | 4416 | 95 |
| 22-3 | 23 | 1 | 0.05 | 0.6155 | 0.61 | 81.9 | 4.0 | 0.966 | 4.00 | 4544 | 9 | 4486 | 41 | 4357 | 128 |
| 23 | 4 | 1 | 0.17 | 0.6086 | 1.05 | 92.2 | 6.5 | 1.099 | 6.39 | 4528 | 15 | 4604 | 67 | 4780 | 219 |
| 24 | 17 | 2 | 0.10 | 0.6219 | 0.76 | 83.6 | 5.0 | 0.976 | 4.90 | 4559 | 11 | 4507 | 51 | 4390 | 158 |
| 25 | 41 | 3 | 0.08 | 0.6214 | 0.39 | 90.6 | 3.8 | 1.058 | 3.83 | 4558 | 6 | 4587 | 39 | 4654 | 128 |
| 26 | 26 | 20 | 0.75 | 0.6143 | 0.44 | 90.2 | 3.4 | 1.066 | 3.35 | 4541 | 6 | 4583 | 35 | 4677 | 113 |
| 27 | 22 | 6 | 0.27 | 0.6134 | 0.47 | 115.4 | 9.3 | 1.365 | 9.28 | 4539 | 7 | 4830 | 98 | 5549 | 355 |
| 28 | 16 | 5 | 0.32 | 0.6210 | 0.52 | 86.9 | 4.8 | 1.015 | 4.79 | 4557 | 8 | 4545 | 50 | 4518 | 158 |
| 29-1 | 36 | 4 | 0.12 | 0.6238 | 0.52 | 89.8 | 4.1 | 1.045 | 4.03 | 4564 | 8 | 4578 | 42 | 4610 | 134 |
| 29-2 | 39 | 5 | 0.14 | 0.6218 | 0.55 | 81.1 | 3.3 | 0.947 | 3.21 | 4559 | 8 | 4476 | 33 | 4294 | 101 |
| 29-3 | 35 | 5 | 0.15 | 0.6241 | 0.54 | 81.0 | 3.7 | 0.942 | 3.64 | 4564 | 8 | 4475 | 38 | 4279 | 115 |
| 30 | 37 | 3 | 0.07 | 0.6203 | 0.58 | 91.1 | 3.2 | 1.066 | 3.18 | 4556 | 8 | 4593 | 33 | 4678 | 107 |

[a]Denotes radiogenic, using the CDT Pb as common-lead compositions $^{206}Pb/^{204}Pb = 9.307$, $^{207}Pb/^{206}Pb = 1.09861$ from Tatsumoto et al. (1973).

Table 15. SIMS U-Pb isotopic data of baddeleyite from Sariçiçek.

| Spot | U ppm | Th ppm | Th/U | $^{207}Pb^*/^{206}Pb^a$ | ±1σ (%) | $^{207}Pb^*/^{235}U$ | ±1σ (%) | $^{206}Pb^*/^{238}U$ | ±1σ (%) | $t_{207/206}$ (Ma) | ±1σ | $t_{207/235}$ (Ma) | ±1σ | $t_{206/238}$ (Ma) | ±1σ |
|---|---|---|---|---|---|---|---|---|---|---|---|---|---|---|---|
| 1-1 | 13 | 1 | 0.07 | 0.62 | 0.63 | 91.8 | 4.9 | 1.068 | 4.90 | 4563 | 9 | 4600 | 51 | 4685 | 165 |
| 1-2 | 19 | 2 | 0.08 | 0.61 | 0.42 | 82.5 | 7.7 | 0.979 | 7.65 | 4534 | 6 | 4492 | 80 | 4399 | 249 |
| 1-3 | 18 | 0 | 0.02 | 0.62 | 0.57 | 90.3 | 5.7 | 1.057 | 5.64 | 4554 | 8 | 4583 | 59 | 4649 | 190 |
| 1-4 | 17 | 1 | 0.06 | 0.62 | 0.57 | 88.7 | 2.2 | 1.030 | 2.12 | 4566 | 8 | 4565 | 22 | 4565 | 70 |
| 2 | 31 | 1 | 0.04 | 0.62 | 0.37 | 92.3 | 3.9 | 1.076 | 3.91 | 4561 | 5 | 4606 | 40 | 4710 | 132 |
| 6 | 31 | 5 | 0.17 | 0.62 | 0.60 | 96.3 | 6.1 | 1.126 | 6.03 | 4555 | 9 | 4648 | 63 | 4863 | 209 |
| 10 | 17 | 7 | 0.40 | 0.62 | 0.94 | 79.8 | 20.1 | 0.932 | 20.07 | 4557 | 14 | 4460 | 225 | 4247 | 657 |
| 16 | 36 | 19 | 0.55 | 0.62 | 1.24 | 104.3 | 8.2 | 1.217 | 8.14 | 4560 | 18 | 4729 | 86 | 5132 | 295 |
| 30-1 | 19 | 3 | 0.15 | 0.62 | 0.71 | 101.5 | 6.3 | 1.179 | 6.25 | 4566 | 10 | 4701 | 65 | 5022 | 222 |
| 30-2 | 57 | 10 | 0.17 | 0.63 | 0.33 | 101.0 | 5.4 | 1.171 | 5.38 | 4567 | 5 | 4696 | 56 | 4999 | 190 |

[a]Denotes radiogenic, using the CDT Pb as common-lead compositions $^{206}Pb/^{204}Pb = 9.307$, $^{207}Pb/^{206}Pb = 1.09861$ from Tatsumoto et al. (1973).

Table 16. SIMS U-Pb isotopic data of apatite from Sariçiçek.

| Spot | U ppm | Th ppm | Th /U | $^{206}Pb/^{204}Pb$ | $f_{206}$ (%) | $^{207}Pb^*/^{206}Pb^{a}$ | ±1σ (%) | $^{207}Pb^*/^{235}U$ | ±1σ (%) | $^{206}Pb^*/^{238}U$ | ±1σ (%) | $t_{207/206}$ (Ma) | ±1σ | $t_{207/235}$ (Ma) | ±1σ | $t_{206/238}$ (Ma) | ±1σ |
|---|---|---|---|---|---|---|---|---|---|---|---|---|---|---|---|---|---|
| 1-1 | 27 | 36 | 1.33 | 20505 | 0.09 | 0.6034 | 0.55 | 81.2 | 1.8 | 0.977 | 1.75 | 4515 | 8 | 4477 | 19 | 4393 | 56 |
| 1-2 | 38 | 45 | 1.19 | 33328 | 0.06 | 0.5999 | 0.38 | 83.2 | 2.2 | 1.007 | 2.17 | 4507 | 6 | 4502 | 22 | 4490 | 71 |
| 1-3 | 36 | 41 | 1.15 | 27890 | 0.07 | 0.5760 | 0.49 | 80.3 | 2.6 | 1.012 | 2.53 | 4448 | 7 | 4466 | 26 | 4508 | 83 |
| 3 | 10 | 5 | 0.51 | 2834 | 0.66 | 0.6081 | 0.79 | 81.7 | 5.5 | 0.976 | 5.42 | 4526 | 11 | 4484 | 57 | 4389 | 175 |
| 8-1 | 54 | 79 | 1.47 | 31969 | 0.06 | 0.6124 | 0.32 | 89.5 | 2.1 | 1.060 | 2.03 | 4537 | 5 | 4574 | 21 | 4659 | 68 |
| 8-2 | 19 | 34 | 1.81 | 11267 | 0.17 | 0.6123 | 0.37 | 80.2 | 3.8 | 0.951 | 3.83 | 4537 | 5 | 4465 | 39 | 4309 | 121 |
| 8-3 | 33 | 69 | 2.11 | 7692 | 0.24 | 0.6061 | 0.35 | 110.2 | 8.5 | 1.320 | 8.48 | 4522 | 5 | 4784 | 89 | 5424 | 319 |
| 8-4 | 22 | 43 | 1.93 | 914 | 2.05 | 0.5983 | 0.55 | 78.9 | 6.5 | 0.957 | 6.52 | 4503 | 8 | 4448 | 68 | 4328 | 209 |
| 8-5 | 34 | 52 | 1.52 | 1375 | 1.36 | 0.6210 | 1.29 | 94.6 | 9.3 | 1.106 | 9.20 | 4557 | 19 | 4631 | 98 | 4801 | 319 |

[a]Denotes radiogenic, using the CDT Pb as common-lead compositions $^{206}Pb/^{204}Pb = 9.307$, $^{207}Pb/^{206}Pb = 1.09861$ from Tatsumoto et al. (1973).

Table 17. Induced and natural thermoluminescence data for Sariçiçek.

| | Induced TL Sensitivity (Dhajala = 1000) | Natural TL (Equivalent Dose in krad) 250 ºC | 375 ºC |
|---|---|---|---|
| Sariçiçek SC12A | 391 ± 30 | 10.8 ± 2.1 | 17.5 ± 3.5 |
| Sariçiçek SC12B | 252 ± 42 | 8.49 ± 1.7 | 17.8 ± 3.6 |
| Mean | 322 ± 36 | 9.7 ± 1.9 | 17.7 ± 3.6 |

Table 18. Visible spectroscopy. Bands I and II center positions in Near-IR absorbance and the derived Fs and Wo contents (Ruesch et al. 2015). "BI" is the Band I, "BII" the Band II.

| Sample Number | BI center (µm) | BII center (µm) | Fs(BI) % | Fs(BII) % | Wo(BI) % | Wo(BII) % |
|---|---|---|---|---|---|---|
| SC51 | 0.93 | 2.00 | 32 ± 4 | 46 ± 4 | 6 ± 2 | 13 ± 2 |
| SC55 | 0.93 | 1.97 | 32 ± 4 | 39 ± 4 | 6 ± 2 | 10 ± 2 |
| SC239 | 0.93 | 1.97 | 32 ± 4 | 39 ± 4 | 6 ± 2 | 10 ± 2 |
| SC327 | 0.93 | 1.98 | 32 ± 4 | 41 ± 4 | 6 ± 2 | 11 ± 2 |

Table 19. Observed meteoroid fragmentation altitudes and corresponding dynamical pressure ($P_{dyn}$), and the corresponding strength parameter and ablation coefficient derived from a one-dimensional triggered progressive fragmentation model (TPFM), assuming Sariçiçek had a diameter of 1.0 m and Chelyabinsk 19.6 m.

| | **Sariçiçek** (17.1 km.s$^{-1}$) | | | | **Chelyabinsk** (19.16 km.s$^{-1}$) | | | |
|---|---|---|---|---|---|---|---|---|
| | **Observation** | | **TPFM** | | **Observation** | | **TPFM** | |
| Key observational points on the flight trajectory | Altitude (km) | Pressure $2 \times P_{dyn}$ (MPa) | Strength parameter (MPa) | Ablation coefficient (s$^2$.km$^{-2}$) | Altitude (km) | Pressure $2 \times P_{dyn}$ (MPa) | Strength parameter (MPa) | Ablation coefficient (s$^2$.km$^{-2}$) |
| Onset of emission | >60 | >0.12 | 1 | 0.021 | 94 | 0.001 | n/a | n/a |
| First flare[a] | 36.4 | 1.99 | 20 | 0.026 | 50.5[a] | 0.36 | 0.47 | 0.041 |
| First flare[b] | | | | | 42.7[b] | 1.0 | 1.17 | 0.044 |
| Final disruption | 27.4 | 7.98 | 20 | 0.015 | 29.7 | 7.10 | 7.32 | 0.034 |

For Chelyabinsk: [a]Popova et al. (2013); [b]Brown et al. (2013).

Fig. 1. Example video frames from security camera footage (left) and calibration images (right) used to measure the meteor trajectory and velocity. A map provides the relative location of each site. A) Bing€ol University rectorate building with the arrows pointing to the tracked feature. B) Bing€ol University soccer court fence, with arrows pointing to the shadow of a horizontal post. C) The lamp fixture in front of the Bing€ol University campus' faculty of economy cast a shadow on the plaza. D) Direct imaging of the meteor by the Muş Alparslan University camera #SE3, with dots marking the position of the meteor in subsequent frames, transposed to the star-background calibration image. E) Muş Alparslan University camera #64, with dots marking the position of the meteor. F) Lamp fixtures of the lamp post in view of Muş Alparslan University camera #91 (inset) cast two shadows that were tracked on the street. G) In Karliova, shadows were cast by a traffic sign at the center of the frame. H) In Kiği, the roof of a house cast a shadow on the street. (Color figure can be viewed at wileyonlinelibrary.com.)

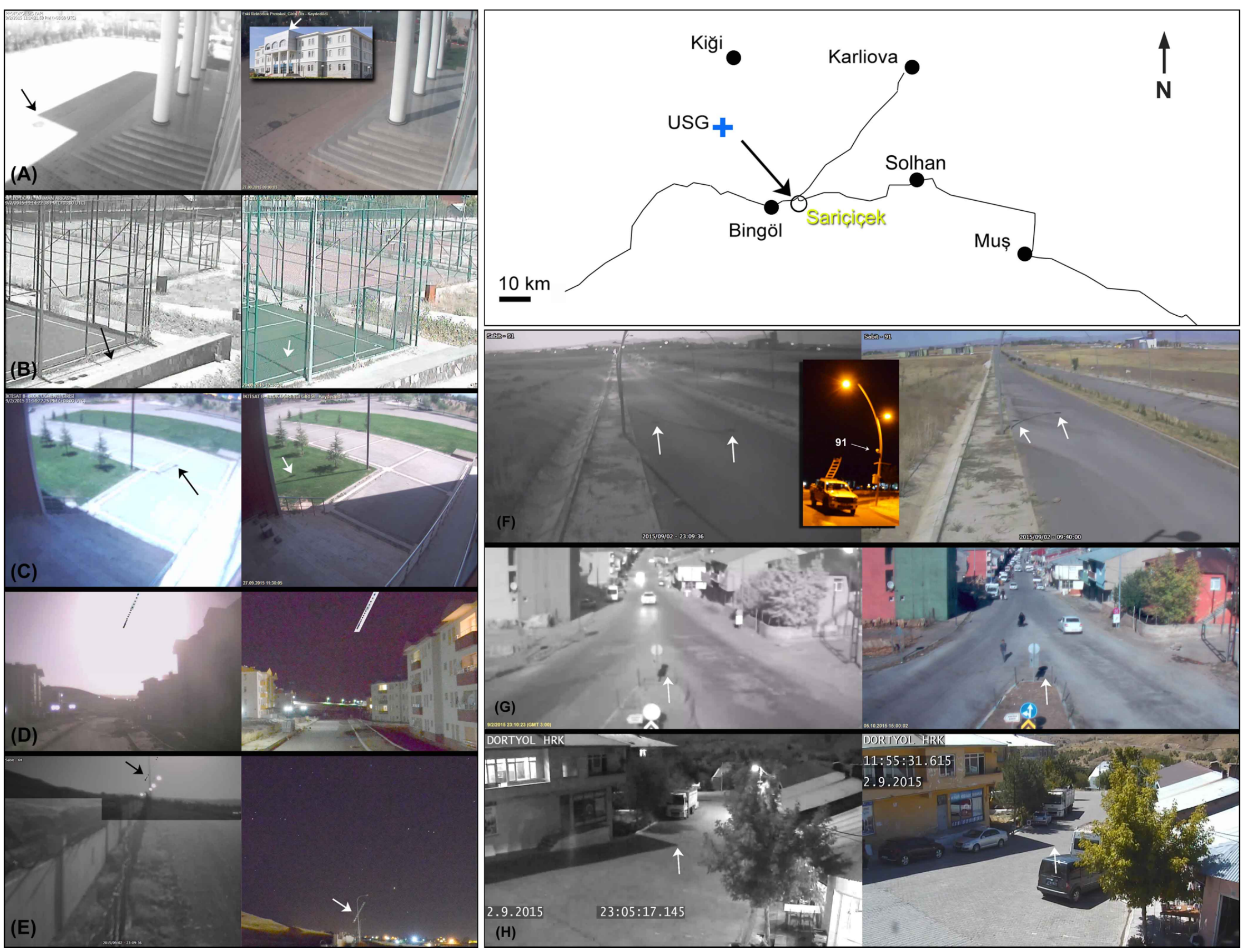

Fig. 2. Optical photographs of sample SC12 (left, with volume 8.4 $cm^3$) and SC14 (right, with volume 6.1 $cm^3$). From top to bottom: As found (minus a broken off tip for SC12), after crushing (on about same scale as above), and the subsamples SC12-b and SC14-a1. (Color figure can be viewed at wileyonlinelibrary.com.)

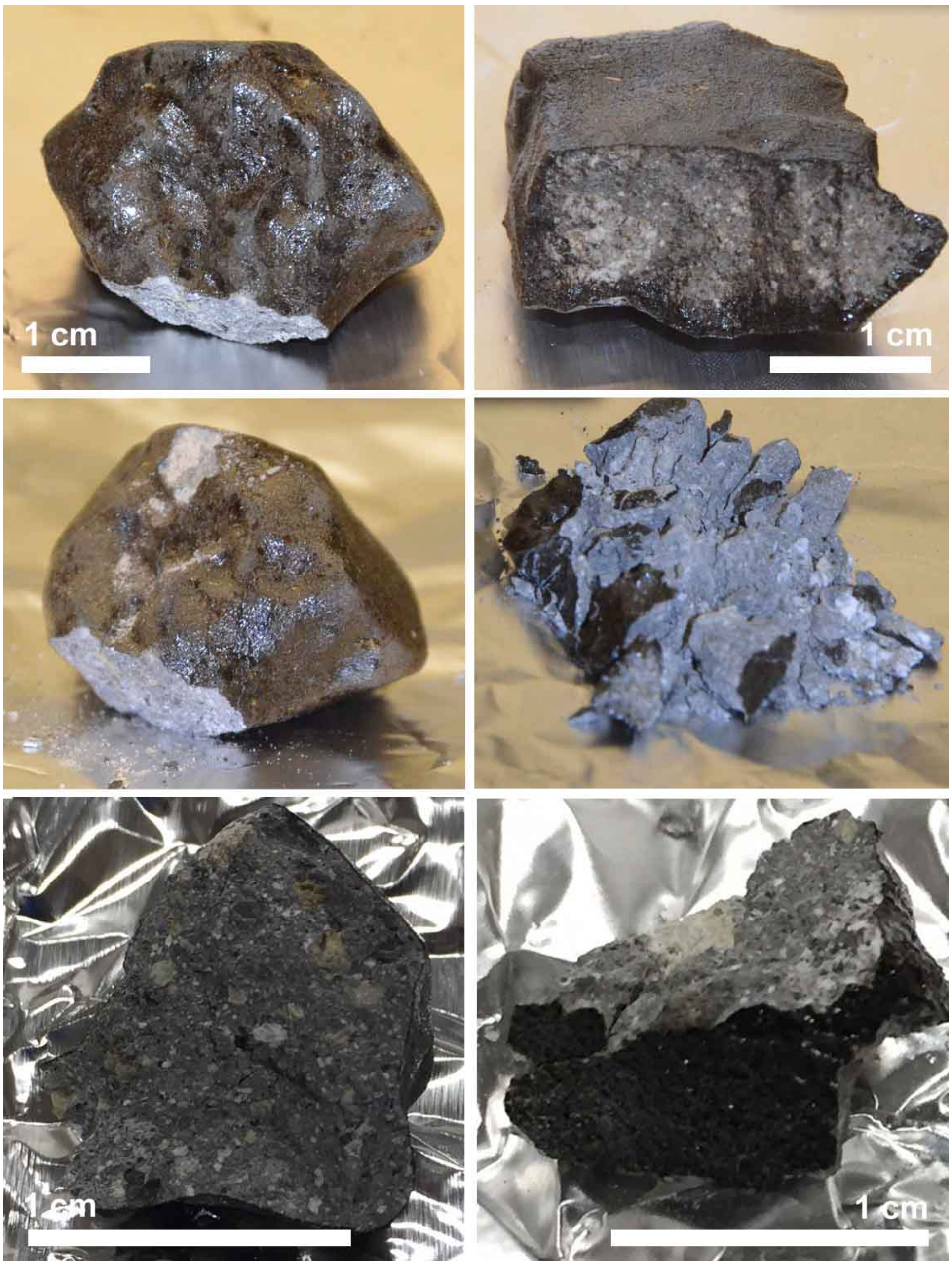

Fig. 3. Altitude and velocity (the latter only for the brighter part of the light curve) time-dependencies for the individual station directional vectors plotted on the vertical plane through the mean trajectory solution. The best-fit Jacchia-type velocity profile is shown as a gray solid line. Constant-velocity fits to sections of the trajectory are shown as open symbols, with error bar giving the range of time considered. USG marks the reported United States Government satellite data (https://cneos.jpl.nasa.gov/fireballs).

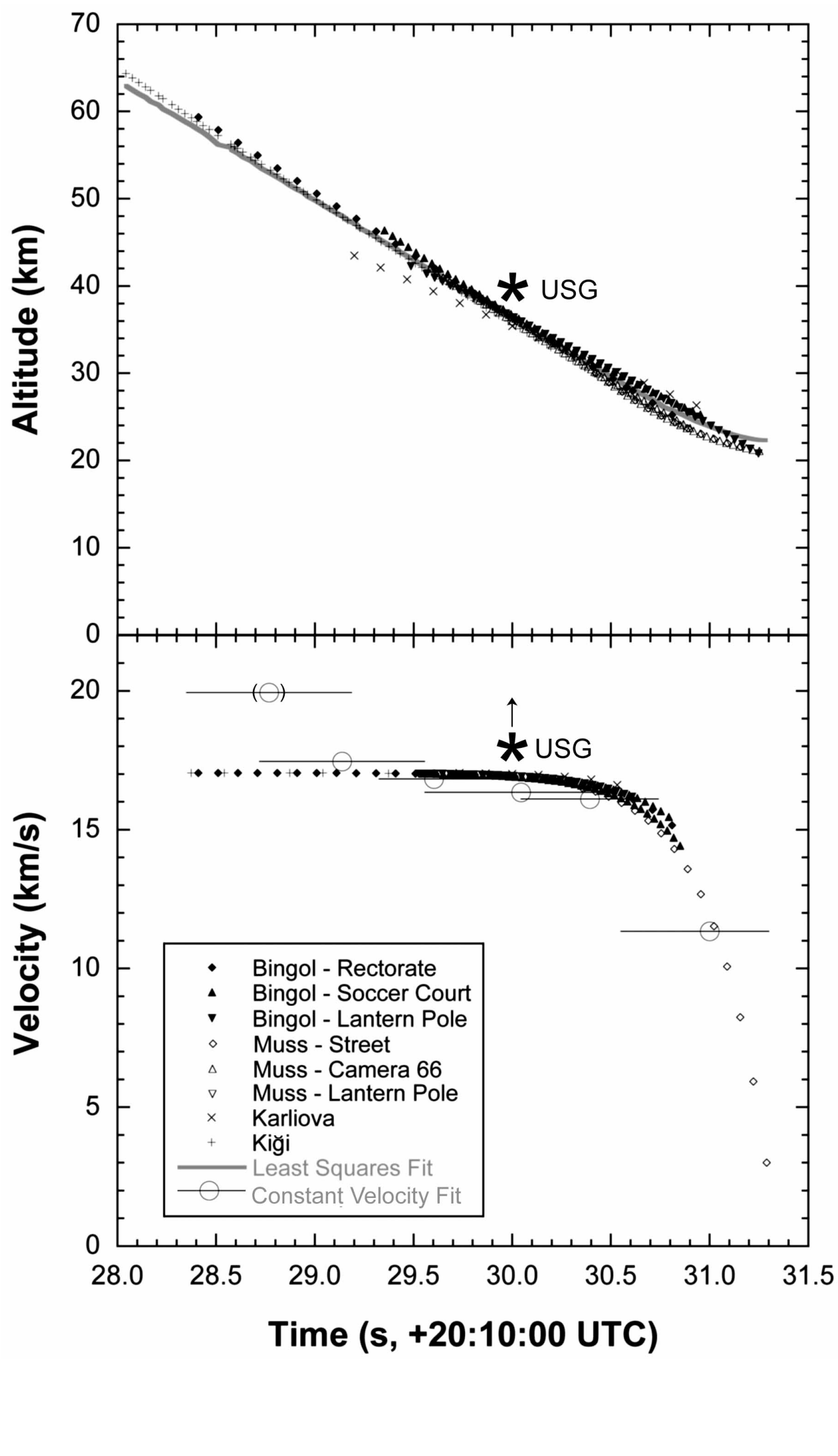

Fig. 4. Light curve based on the analyzed video records as a function of time (right) and altitude (left). The time scale is normalized to the satellite reported time of 20:10:30 UT at peak brightness. The brightness is normalized to a constant range of 100 km. Flux density is computed based on calibration to the Moon. TPFM model light curves (ReVelle 2003, 2004) pertain to meteoroid sizes of 0.8–1.2 m, 9% porosity, and strength modifier factor of 20 from an initial value of 0.05 MPa**.**

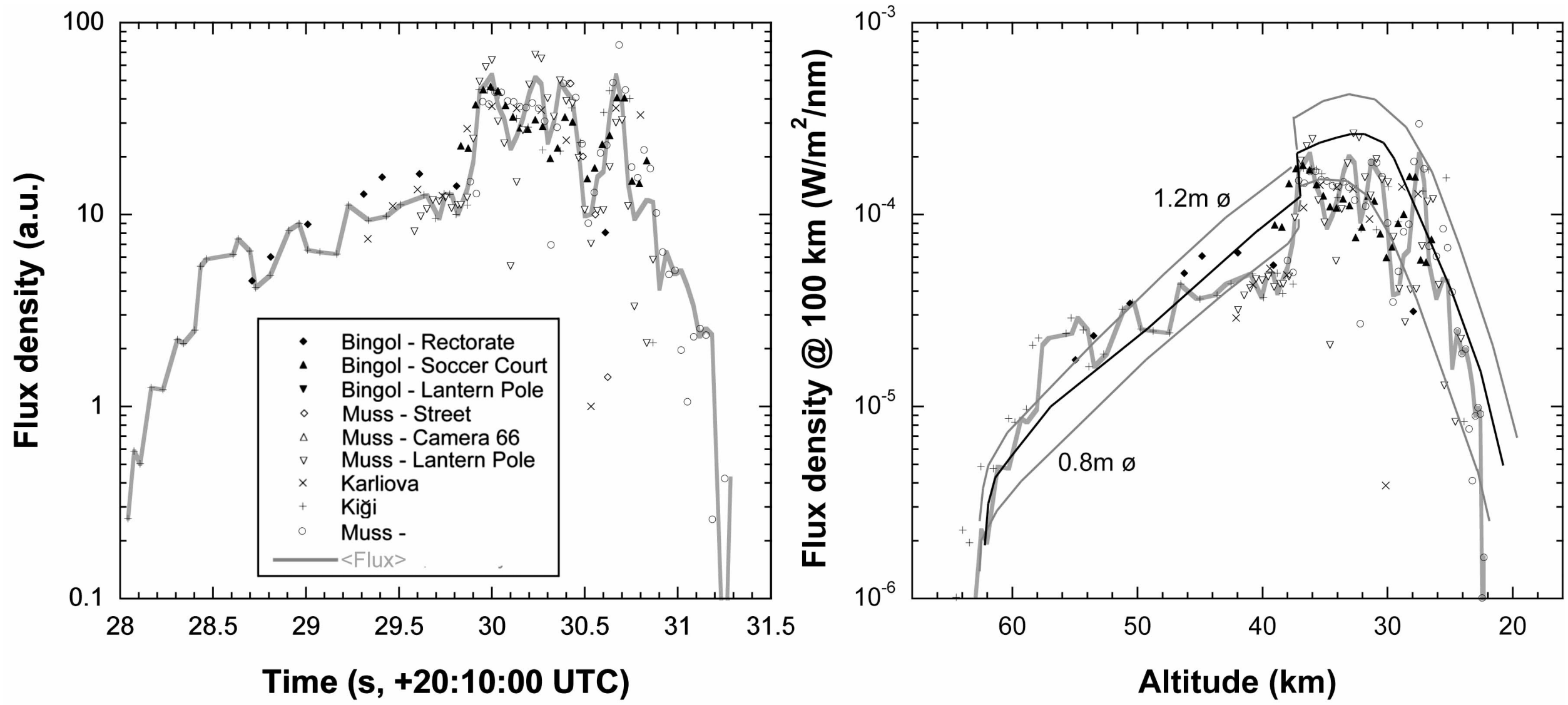

Fig. 5. Meteorite strewn field relative to the ground-projected meteor trajectory (solid line) near Sariçiçek (+), compared to modeled fall in ambient wind. Observed pattern is consistent with falling from final flare at 27.4 km (squares), not from 36.5 km (diamonds). Location of the largest sample SC232 is marked. Also shown is the area covered in the grid search during the field study. (Color figure can be viewed at wileyonlinelibrary.com.)

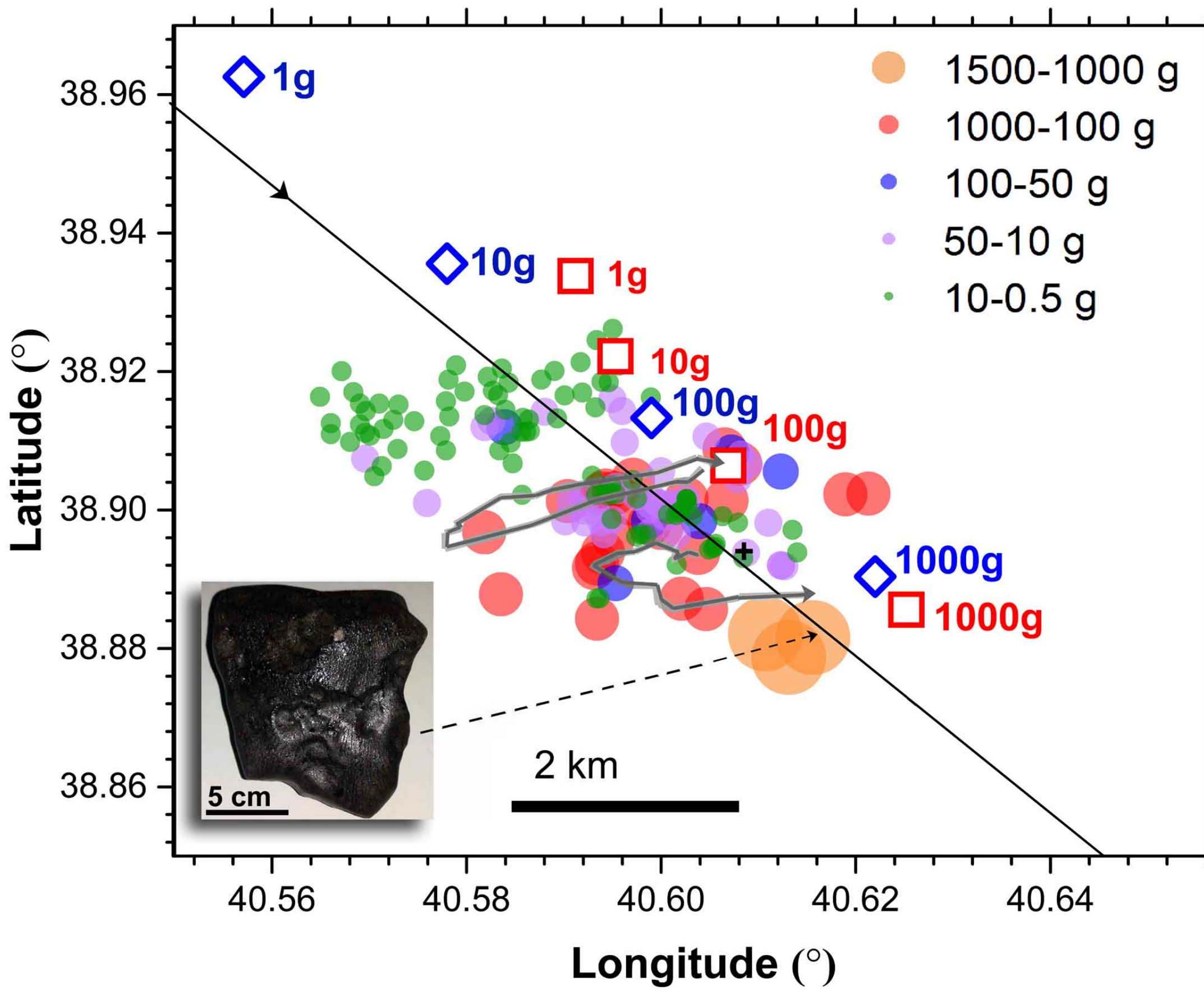

Fig. 6. Seismic signals detected at stations BNGB (Bing€ol) and SLHN (Solhan), left, and 1213 (Adakli) and 1212 (Yedisu) to the right. The significant part of airwave energy is marked by the gray area on the components of N–S (N), E–W (E), and Vertical (Z). The signal amplitudes are count and normalized common, the horizontal axis is time (20h12m00s UTC onwards). Results from airburst modeling are displayed above and below the observed data.

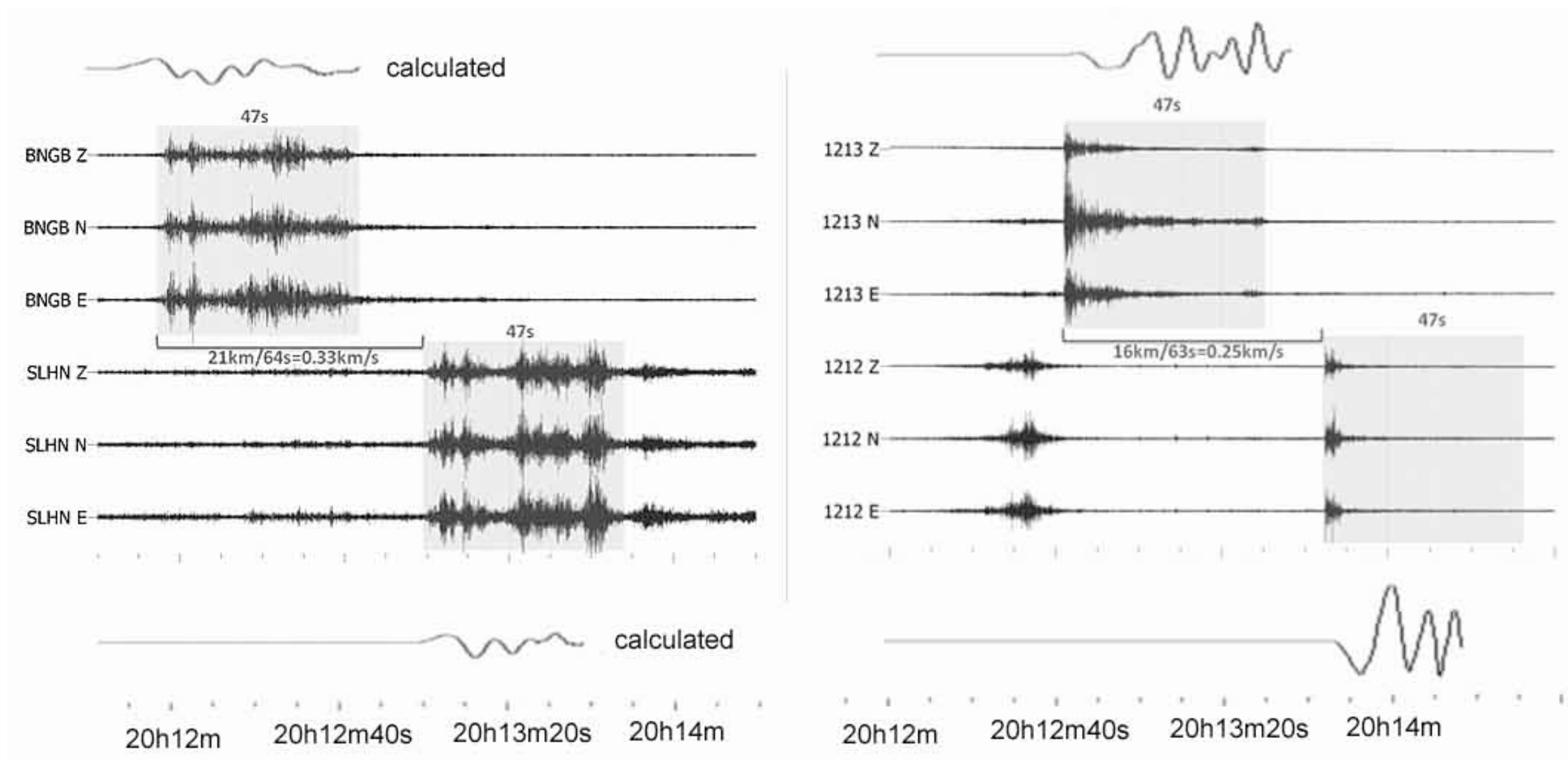

**Fig. 7.** Representative microCT images of Sariçiçek fragments SC12b (left) and SC14 (right), with sub-samples a1 (top) and a2 (bottom).

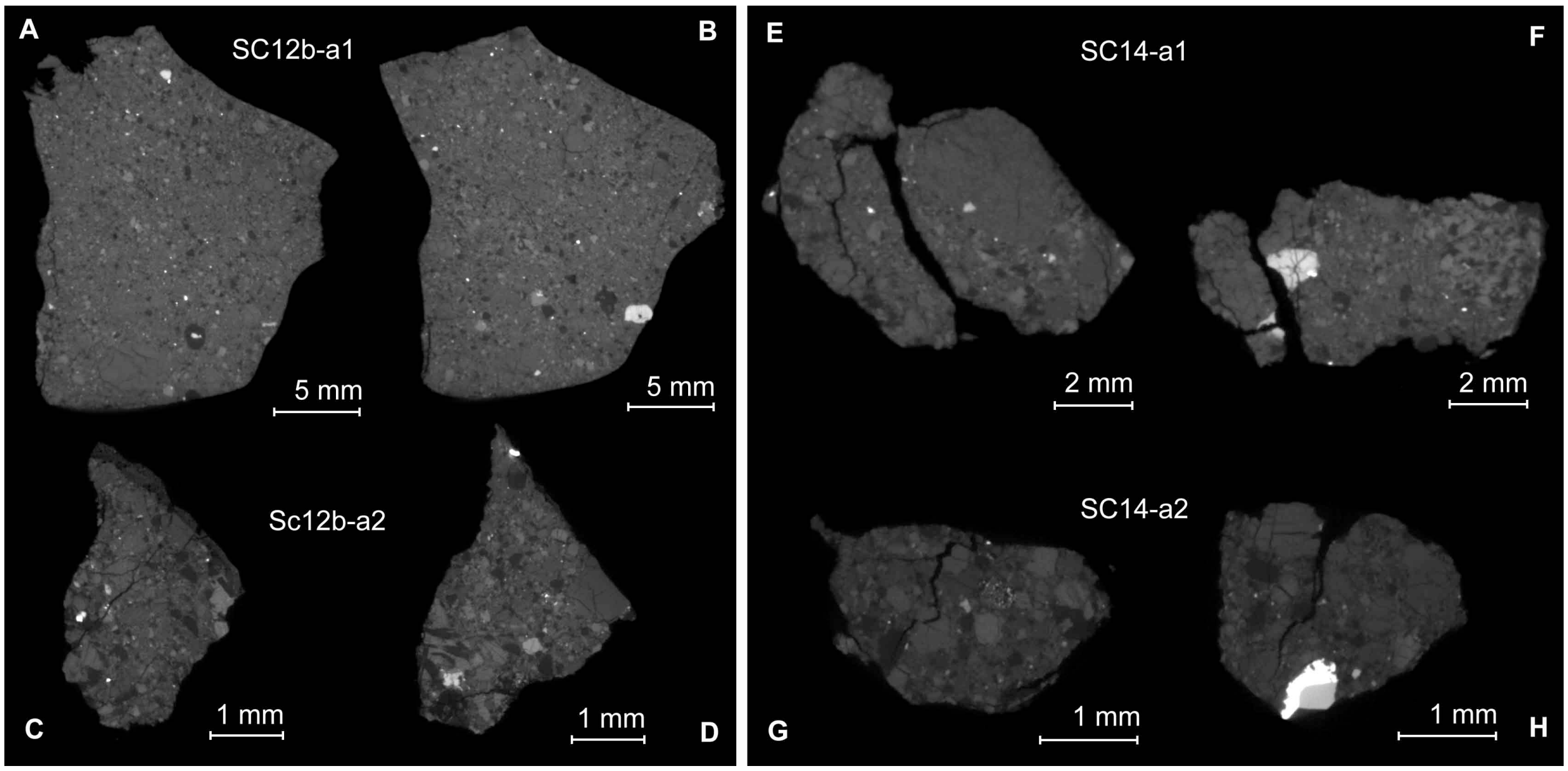

**Fig. 8.** Distribution of flaw trace length for Sariçiçek sample SC12b-a2 (A) and sample SC14-a2 (B). The gray line is based on the relationship between trace density and trace length, with a slope providing a value for α. The black line displays the same relationship, but with the value of α = 0.166 for ordinary chondrites. (Color figure can be viewed at wileyonlinelibrary.com.)

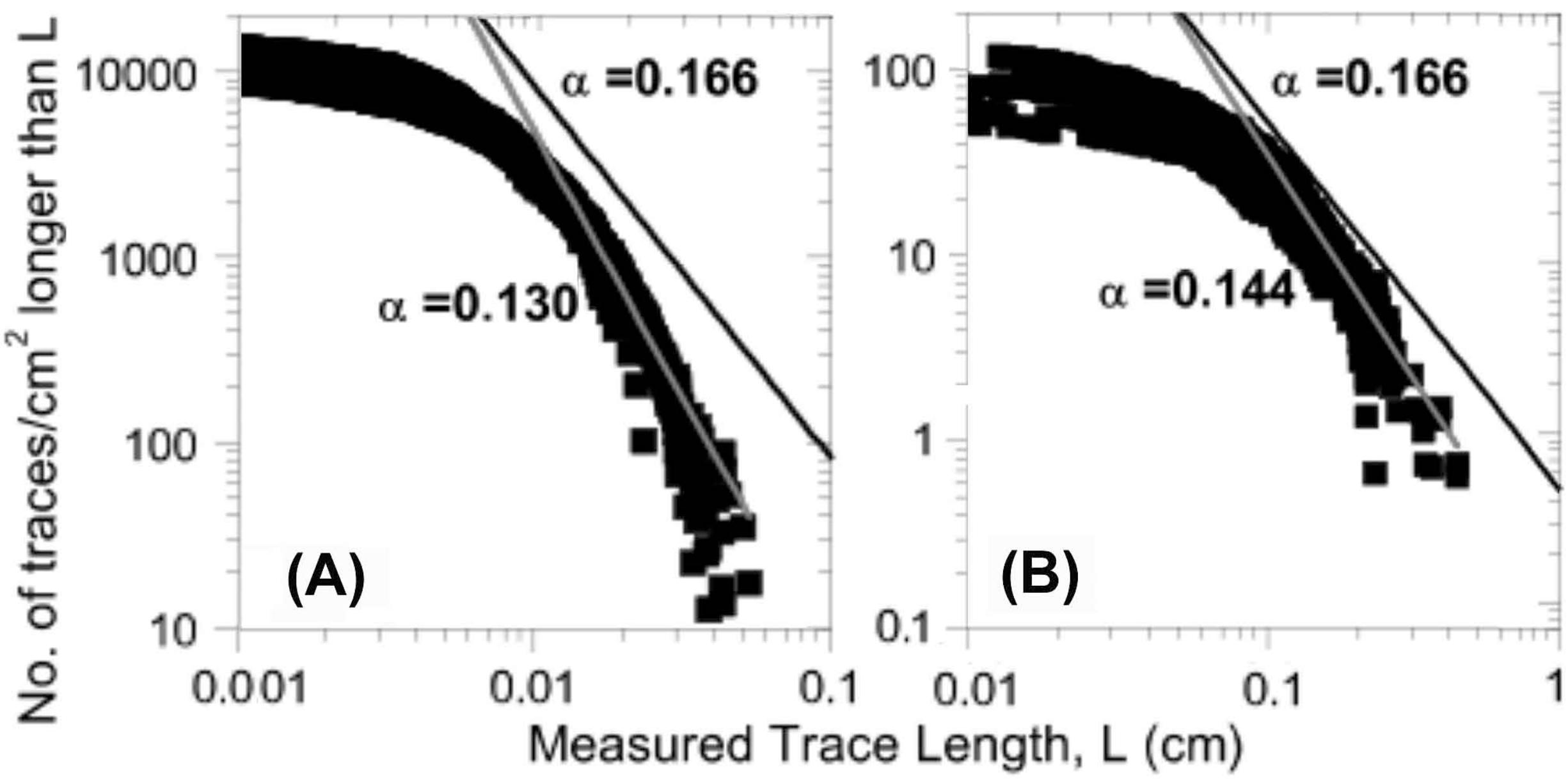

**Fig. 9.** Stereographs (lower hemisphere) displaying the orientation of the longest axis for FeNi grains (top) and FeS grains (bottom) in meteorite SC12b-a1.

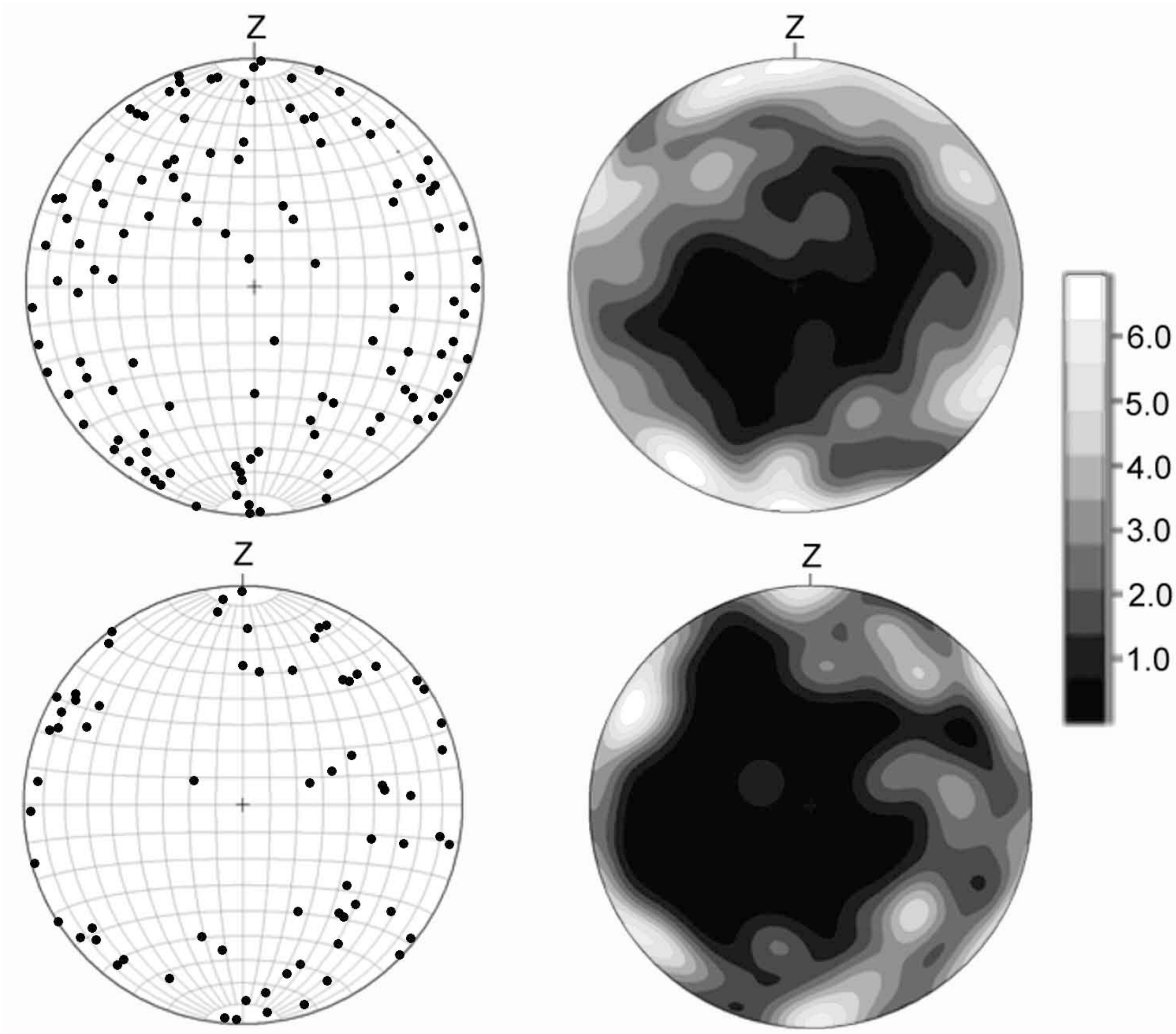

**Fig. 10.** Backscattering Electron Microscopy mosaic of SC12a. This BSE image shows various crystals and clasts.

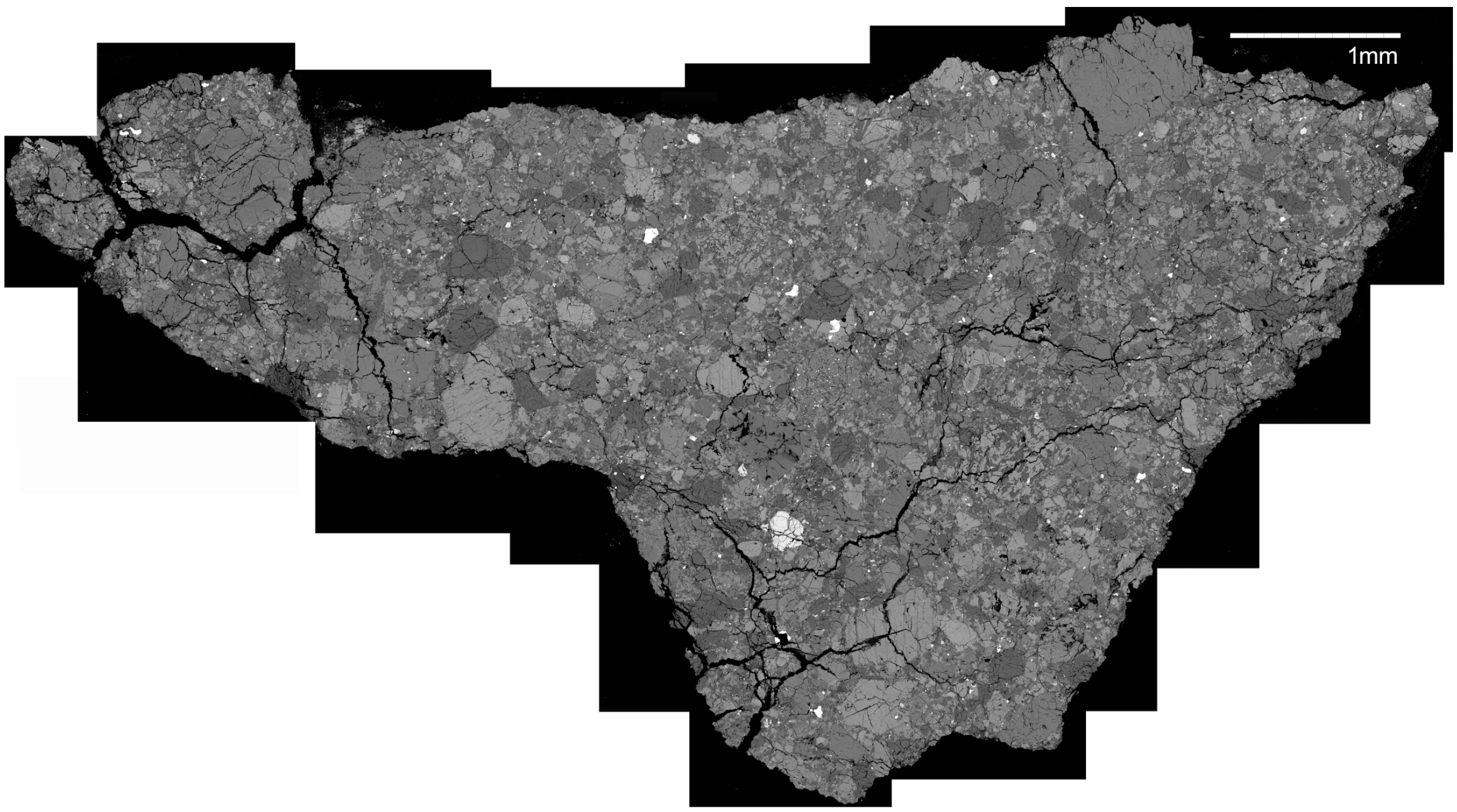

Fig. 11. Backscattering electron microscopy mosaic of SC14a. These thin sections show more diogenite components (large crystals). Detailed image 30 shows a clast consisting mainly of plagioclase (plag), silica, and troilite; image 29 shows a last consisting of plagioclase and high-Ca pyroxene (HCpx); image 31 shows a clast consisting of plagioclase, silica, and troilite, adjacent to a grain of high- and low-Ca pyroxene (LCpx); and image 35 shows a pyroxene crystal with exsolved high- and low-Ca pyroxene. Next to image 35 is a higher magnification view showing s–ubmicron-sized chromite crystals, mainly within low-Ca pyroxene.

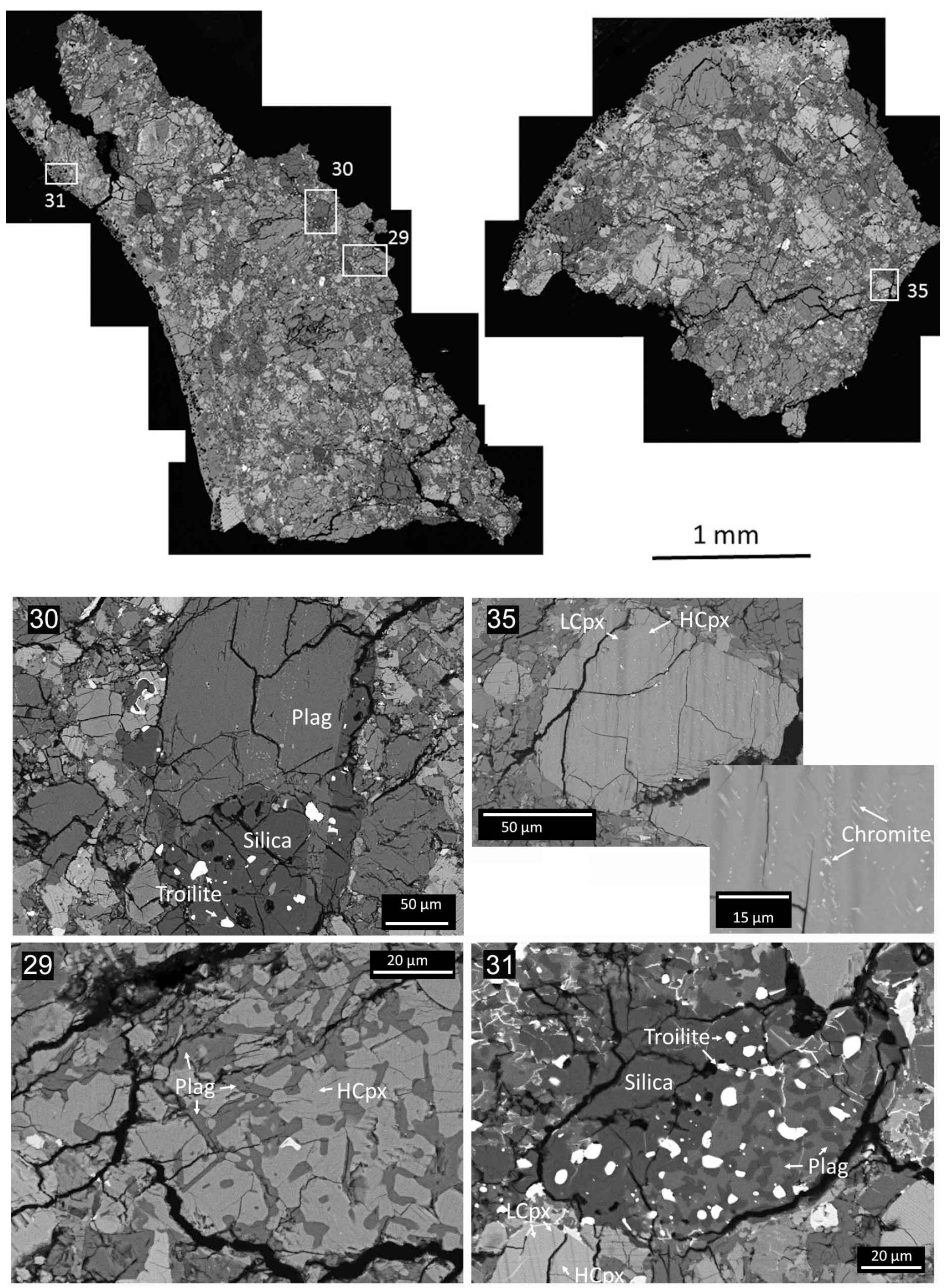

**Fig. 12. (A)** Compilation of EMPA analysis (dots) of pyroxenes plotted onto the quadritlateral Enstatite (En)–Ferrosilite (Fs)–Hedenbergite (Hd)–Diopside (Di) diagram. **(B)** Classification based on bulk analysis by EMPA, compared to data by Warren et al. (2009).

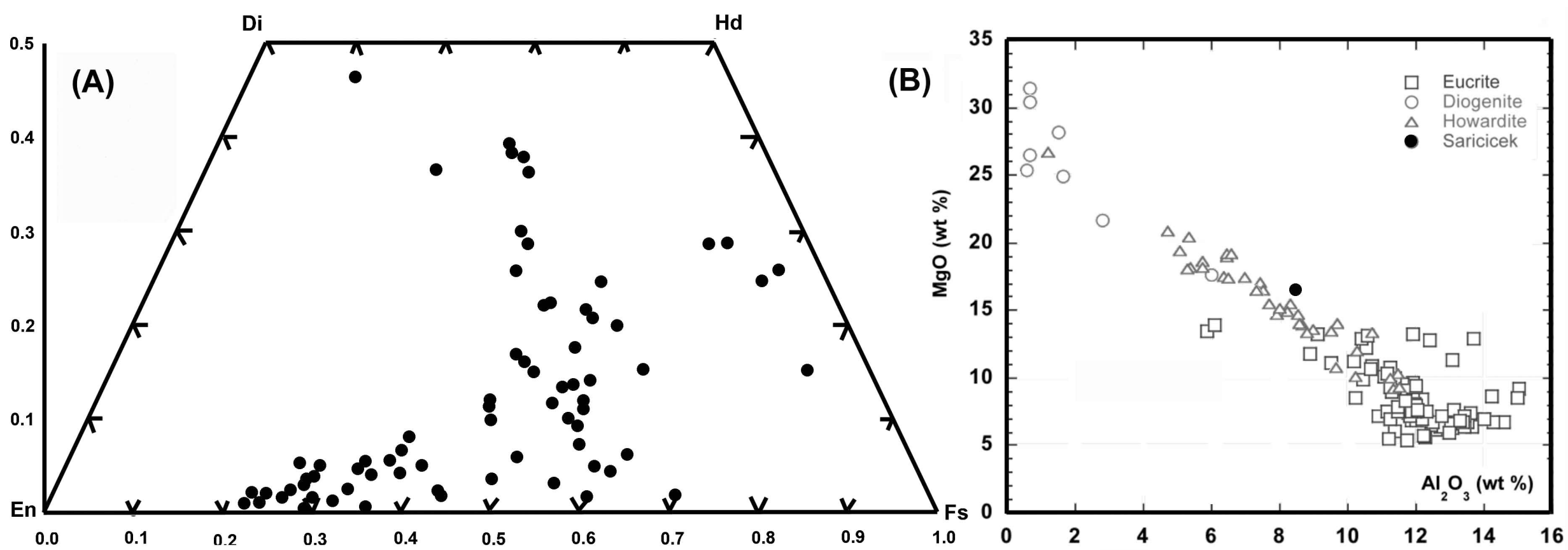

**Fig. 13.** Sariçiçek's noble gasses and Ir content place it among rare regolithic howardites. Error bars are smaller than the symbol shown. (Color figure can be viewed at wileyonlinelibrary.com.)

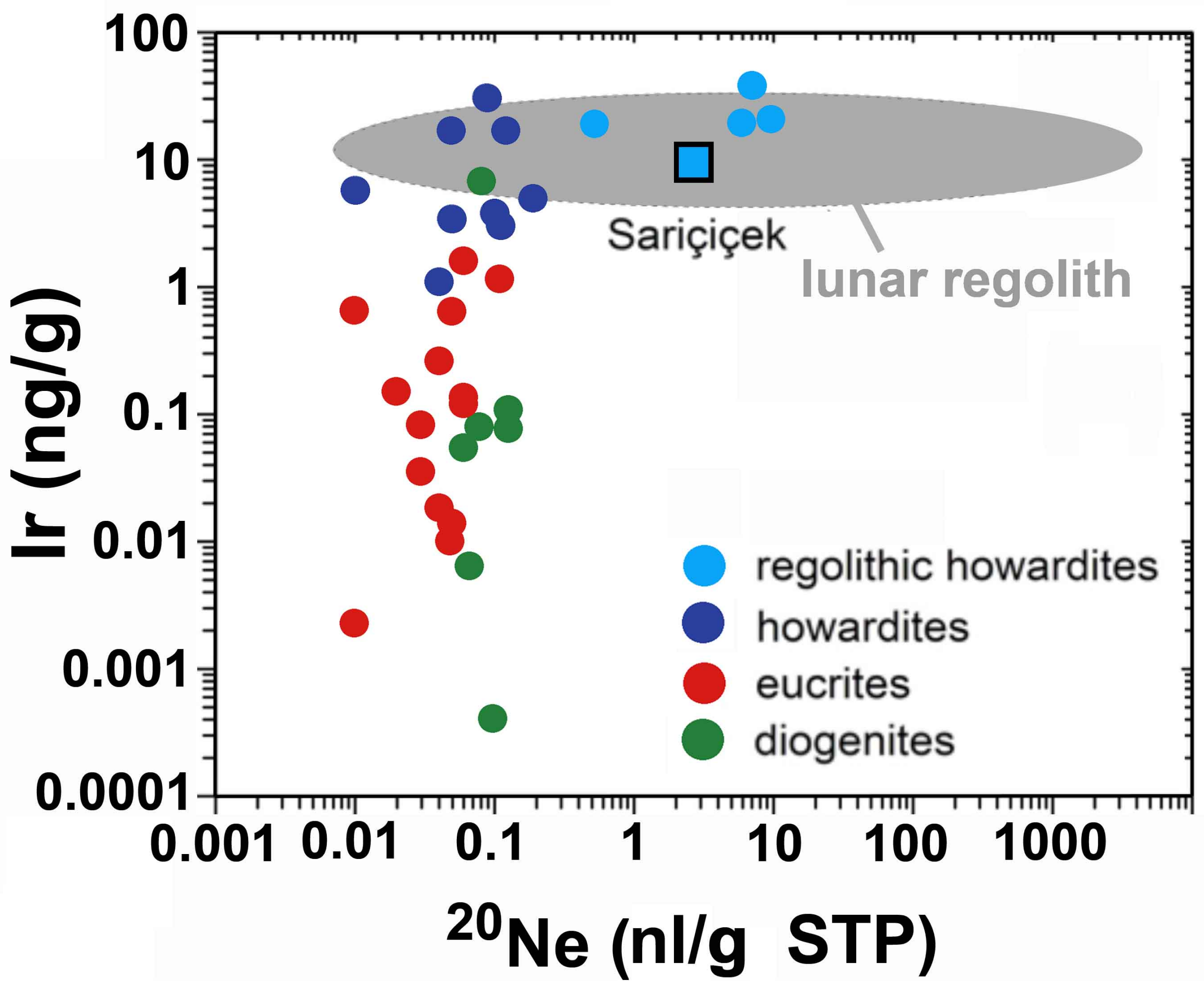

**Fig. 14.** Oxygen and chromium isotopes identify this material as HED. (Color figure can be viewed at wileyonlinelibrary.com.)

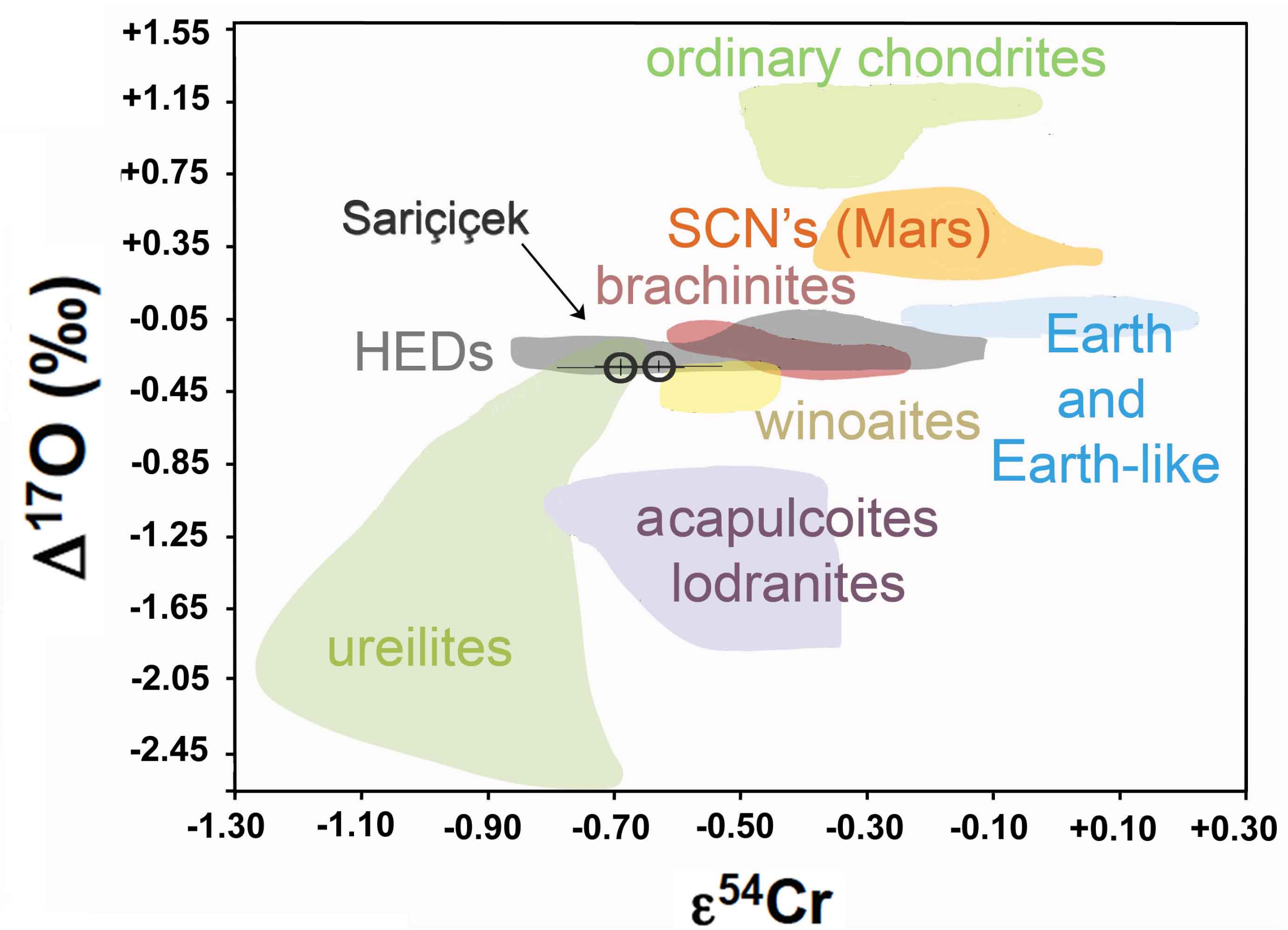

**Fig. 15.** Raman spectrum of a dark inclusion in a larger grey region of the inner face of the probed meteorite sample SC239. The D and G band of carbonaceous matter are marked. (Color figure can be viewed at wileyonlinelibrary.com.)

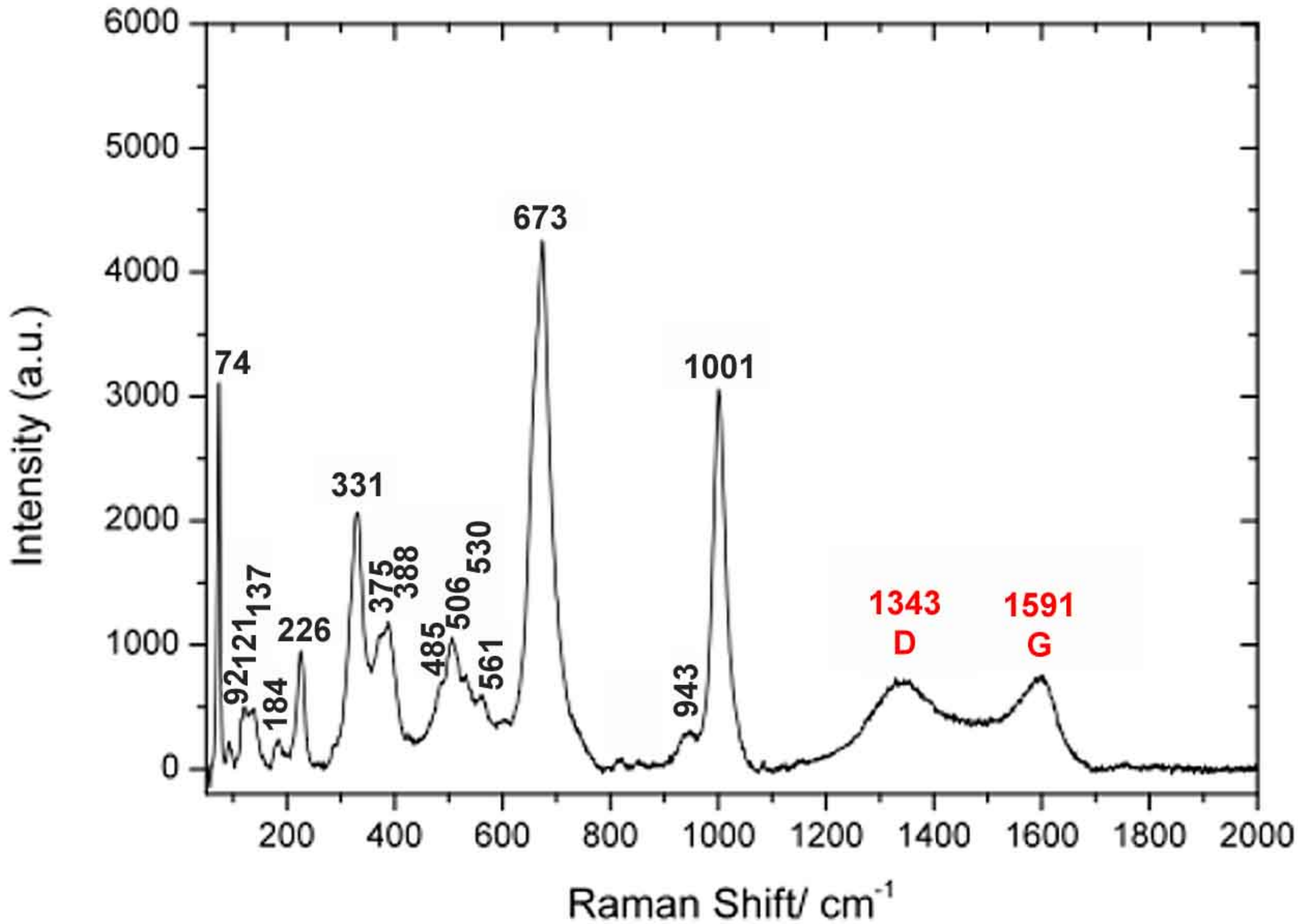

**Fig. 16.** The inferred depth of three samples inside the meteoroid based on the measured cosmogenic $^{26}Al$, in comparison with model calculations of the expected $^{26}Al$ in howardite meteoroids of different diameter (Welten et al. 2012). (Color figure can be viewed at wileyonlinelibrary.com.)

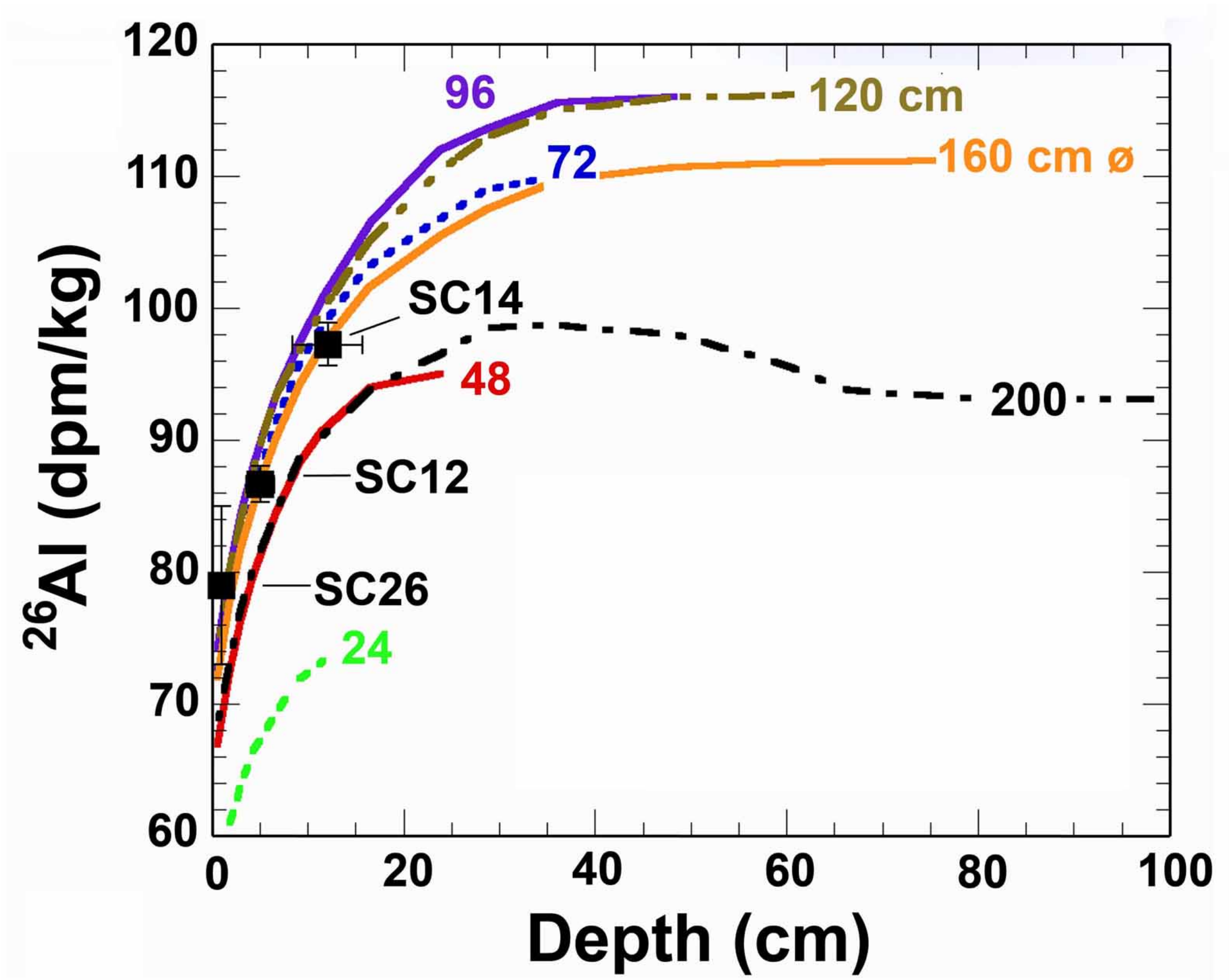

**Fig. 17.** Ne three isotope diagram. Results are compared to data for other HED meteorites compiled by Schultz & Franke (2004) and Cartwright et al. (2013, 2014). The endmember isotopic composition of solar wind (SW), fractionated solar wind (fSW) and Earth's atmosphere (Air) are shown as crosses. Trend lines for admixtures with cosmogenic isotope compositions (cosmogenic) are shown as dashed lines. (Color figure can be viewed at wileyonlinelibrary.com.)

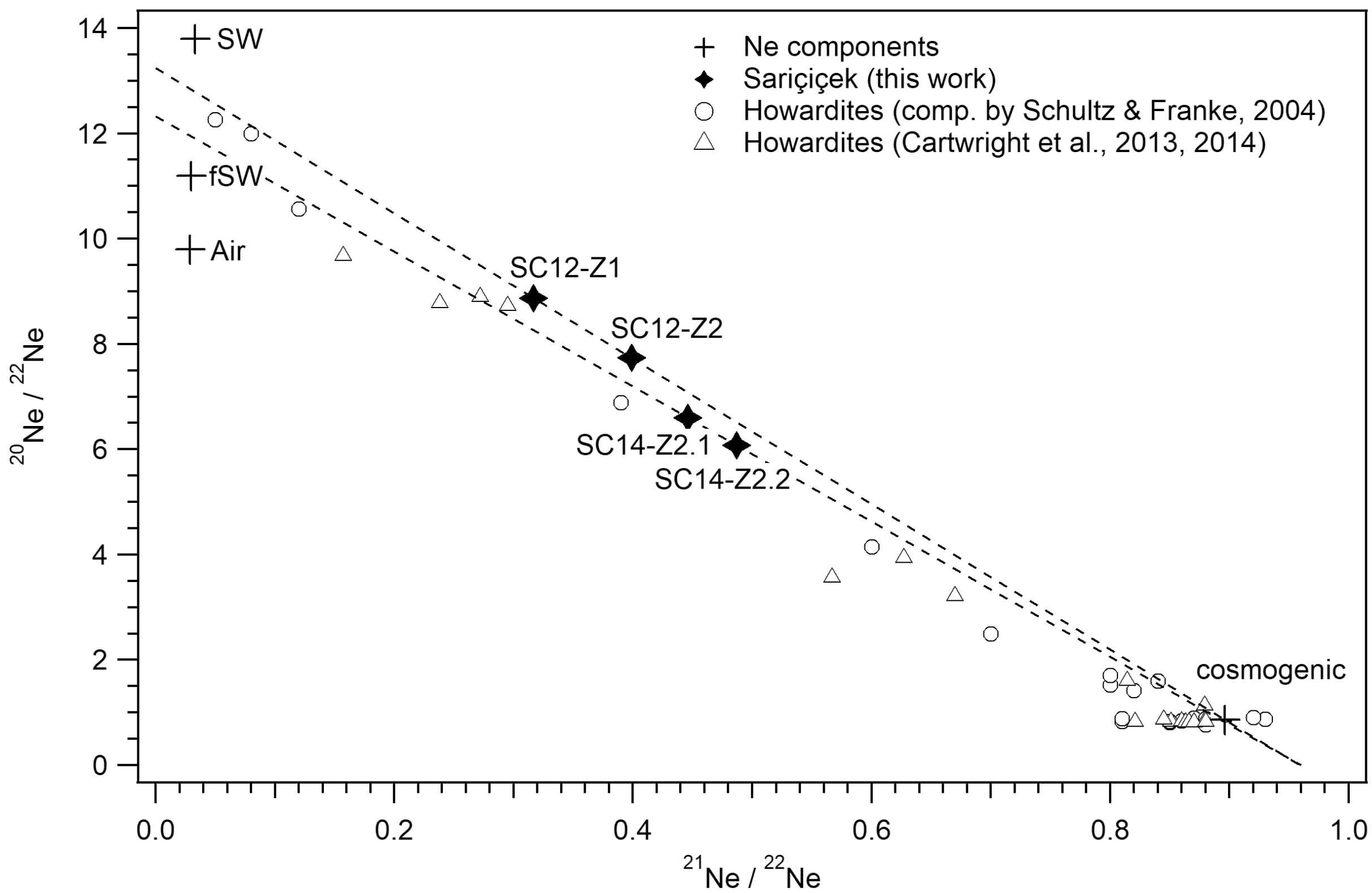

Fig. 18. Non-cosmogenic $^{4}$He vs. non-cosmogenic $^{20}$Ne. Extrapolation of non-cosmogenic $^{4}$He to $^{20}$Nenon-cos = 0 allows to infer the presence of radiogenic $^{4}$He. Both the measured and non-cosmogenic $^{4}$He concentrations are shown and confirm that the correction for cosmogenic $^{4}$He (assuming all $^{3}$He to be cosmogenic) does not significantly affect the extrapolation.

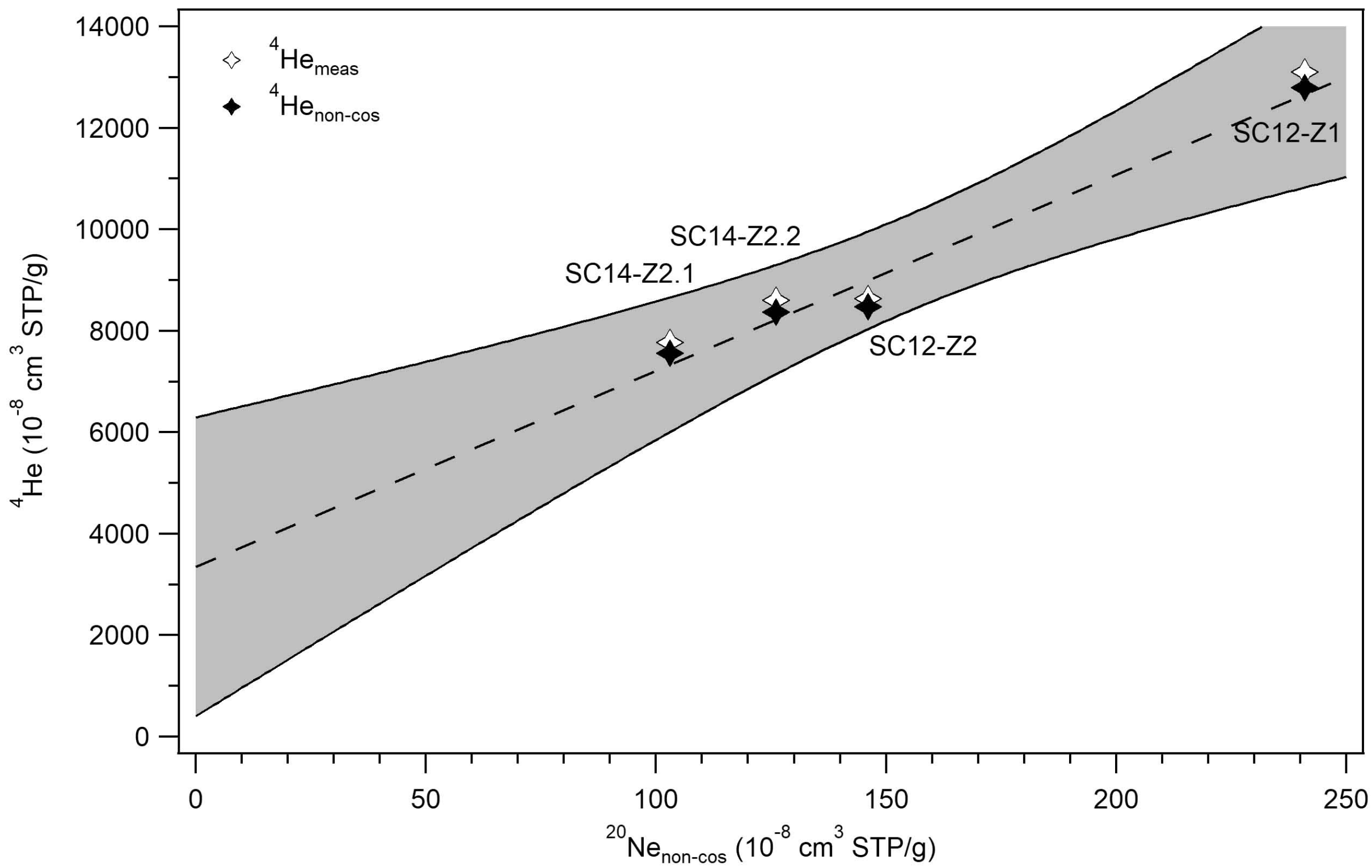

**Fig. 19.** Regolith trend lines for Ne in Sariçiçek and Kapoeta. Cosmogenic $^{21}Ne$ vs. solar wind $^{20}Ne$ for all Sariçiçek and Kapoeta samples (the latter from the data compiled by Schultz and Franke 2004). The implied regolith trend lines can be translated into solar wind pickup rates plus a 4π cosmic-ray exposure age (see main text). For Sariçiçek, this correlation is based only on four data points (the SC14 samples have been corrected for shielding using radionuclides, measured values given as open symbols).

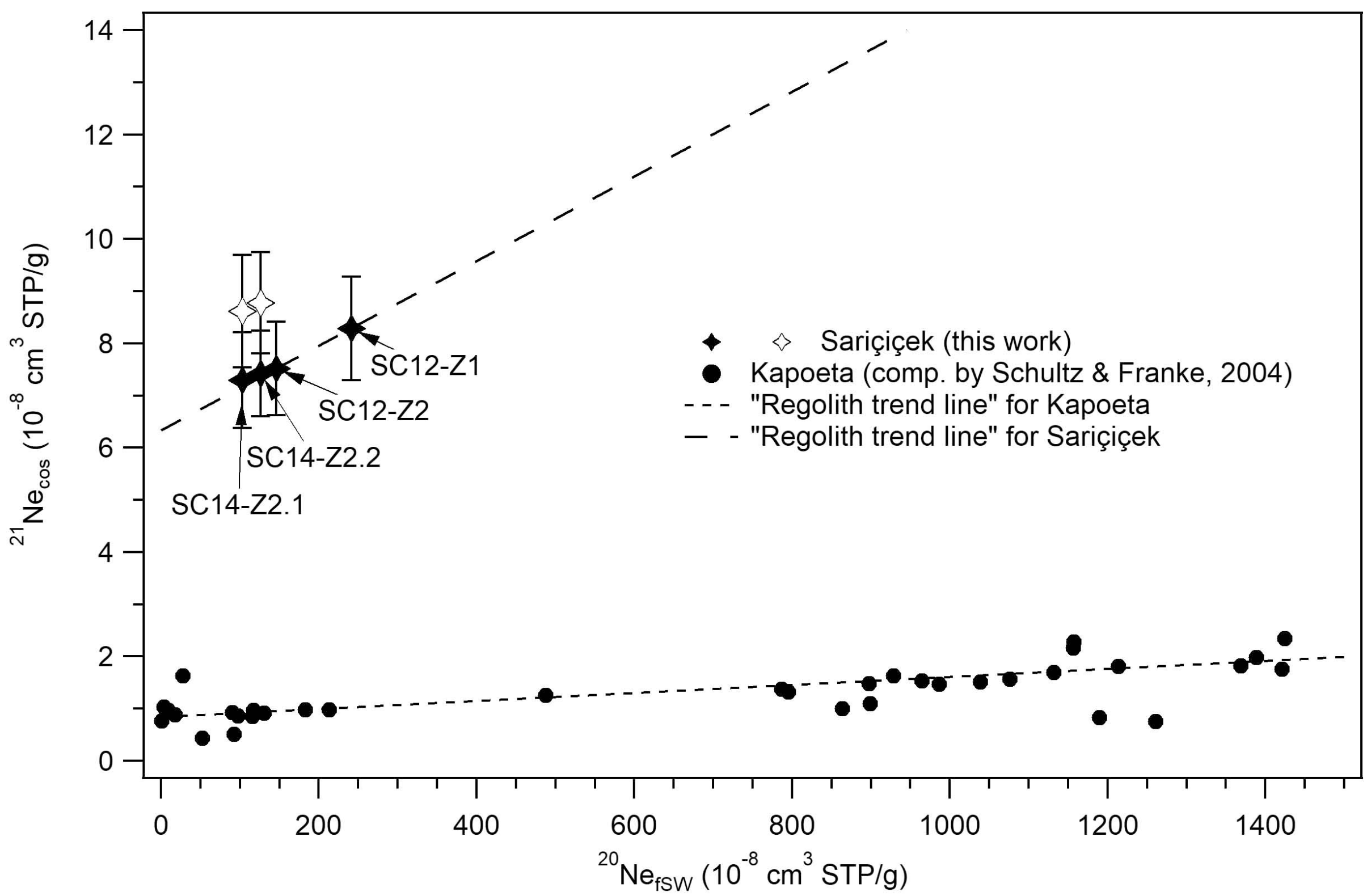

**Fig. 20a.** Zircon (Zr) grains in Sariçiçek in cathodoluminescence (CL, left column) and backscattered electron (BSE, right column) images. The zircon grain number corresponds to the U-Pb data reported in Table 14.

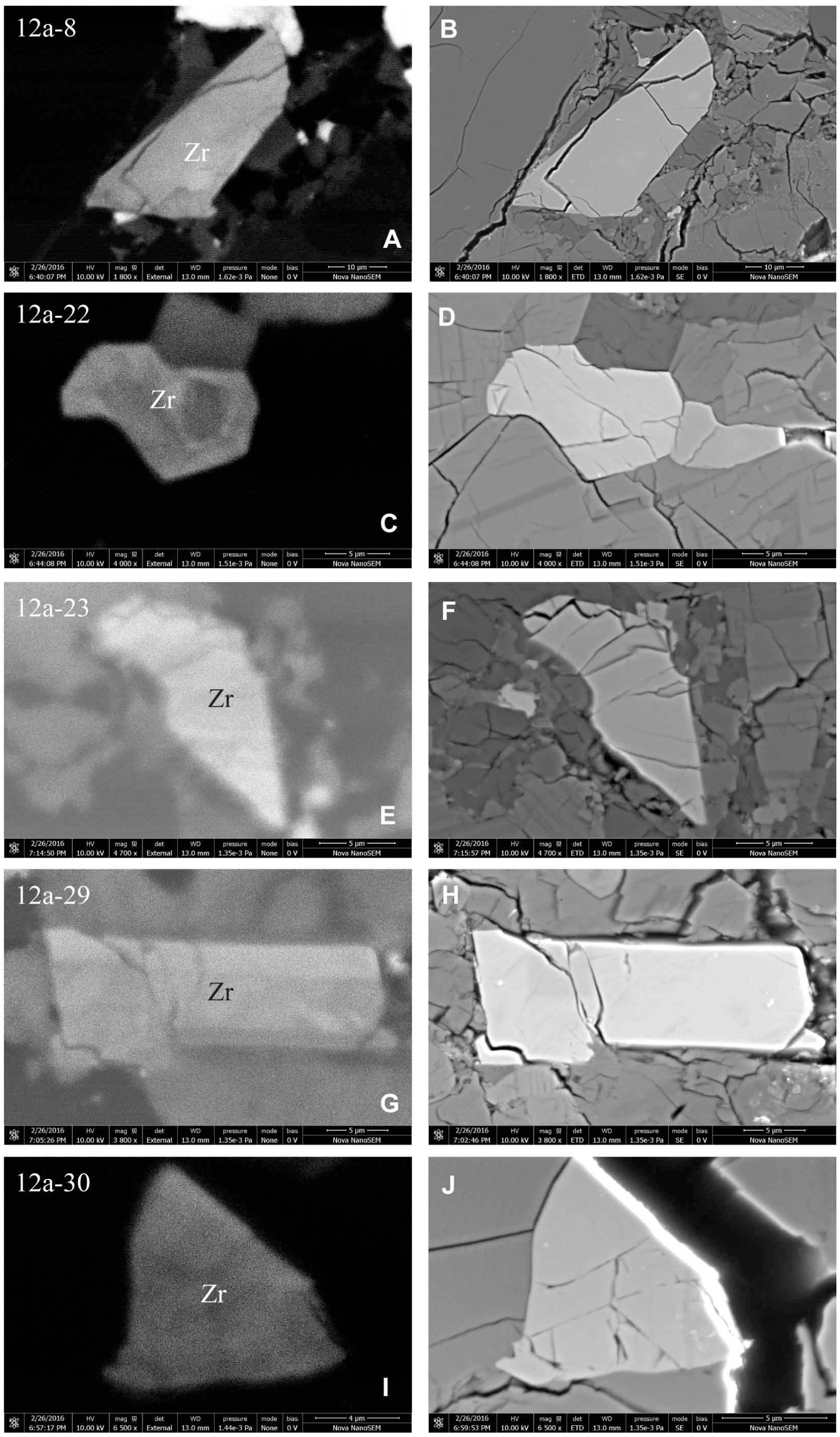

**Fig. 20b.** As Fig. 20a, for baddeleyite grains in Sariçiçek. The baddeleyite grain number corresponds to the U-Pb data reported in Table 15.

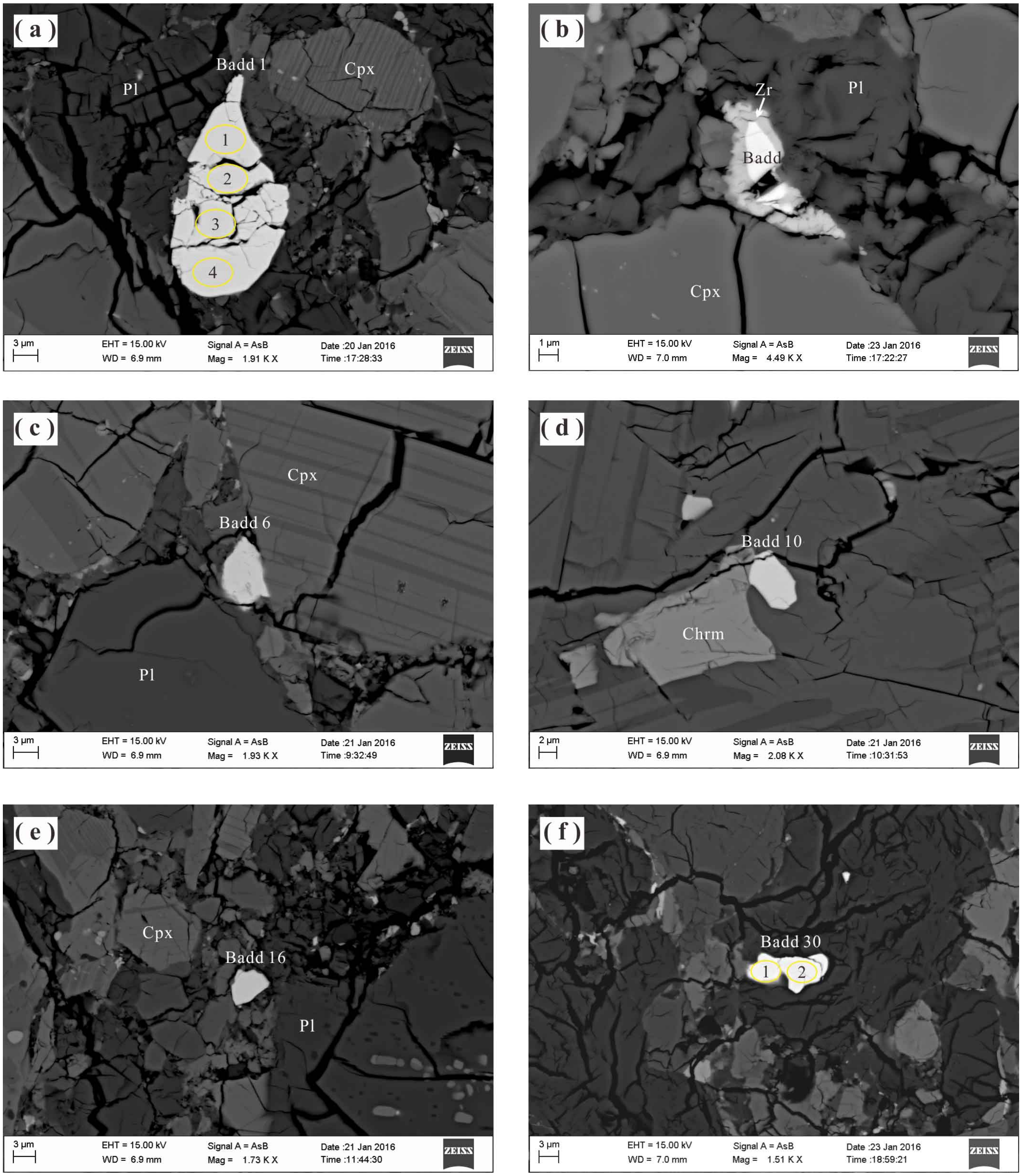

**Fig. 21.** Backscattered electron (BSE) images for all measured apatite grains (Ap) in Sariçiçek, containing also high-Ca pyroxene (cpx), low-Ca pyroxene (opx), plagioclase (Pl), chromite (Chrm), ilmenite (Ilm), troilite (FeS), silica ($SiO_2$), baddeleyite (Badd), and merrillite (Merr). Circles represent the U-Pb analysis positions for the apatite grains, while grain numbers correspond to the U-Pb data reported in Table 16.

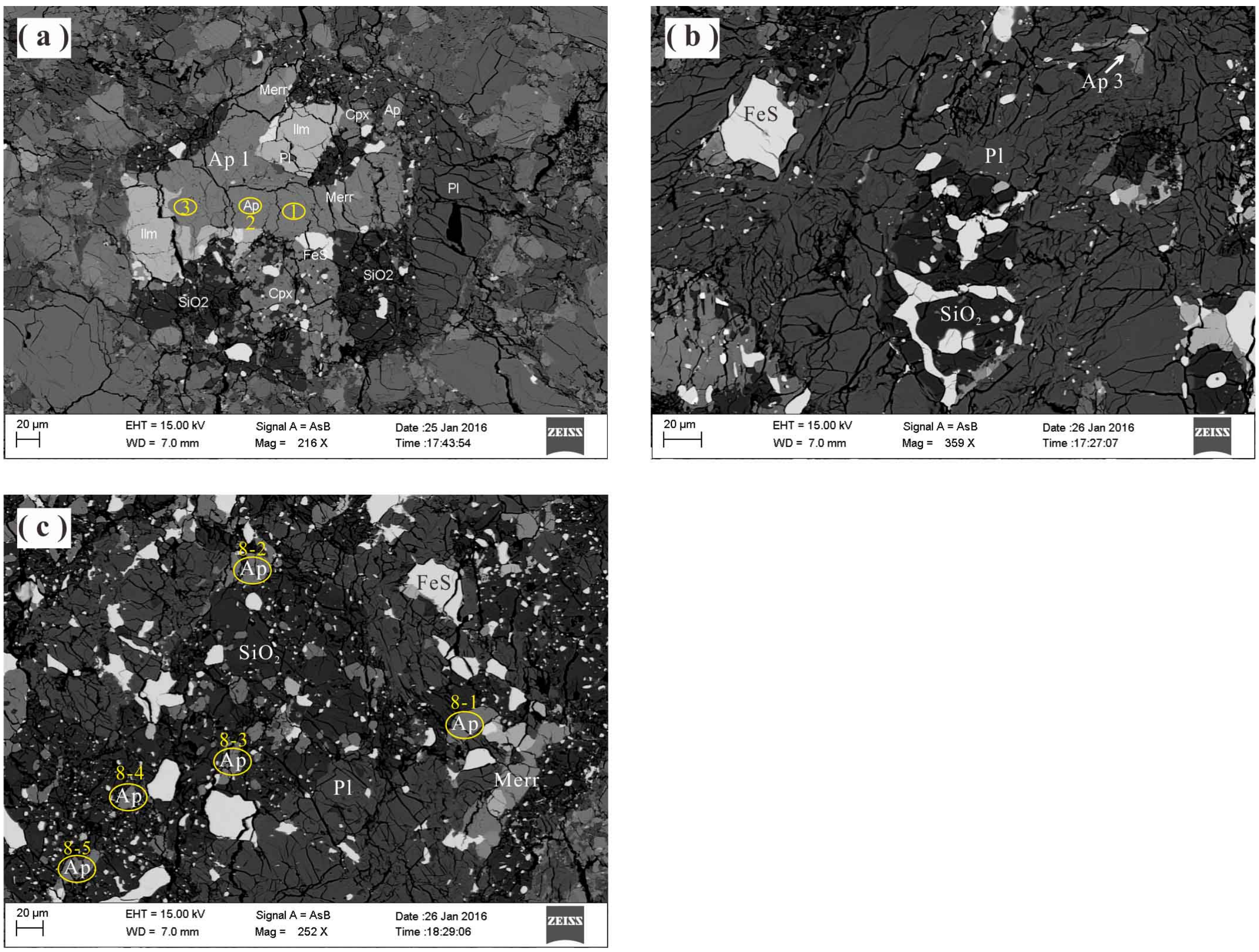

**Fig. 22.** U-Pb concordia diagram (left column) and $^{207}Pb/^{206}Pb$ age (right column) for zircons, baddeleyite and apatite in Sariçiçek. (a-b) Thirty-four data points of zircon grains intercept the concordia at 4551.2 ± 2.5 Ma and give a weighted average $^{207}Pb/^{206}Pb$ age of 4552.0 ± 2.8 Ma. (c-d) Ten data points for baddeleyite give the intercept age of 4554.4 ± 8.8 Ma and $^{207}Pb/^{206}Pb$ age of 4558.9 ± 8.2 Ma. (e-f) Eight data points for apatite intercepts the concordia at 4529 ± 32 Ma with $^{207}Pb/^{206}Pb$ age of 4525 ± 12 Ma. Data-point and age uncertainties are 2σ.

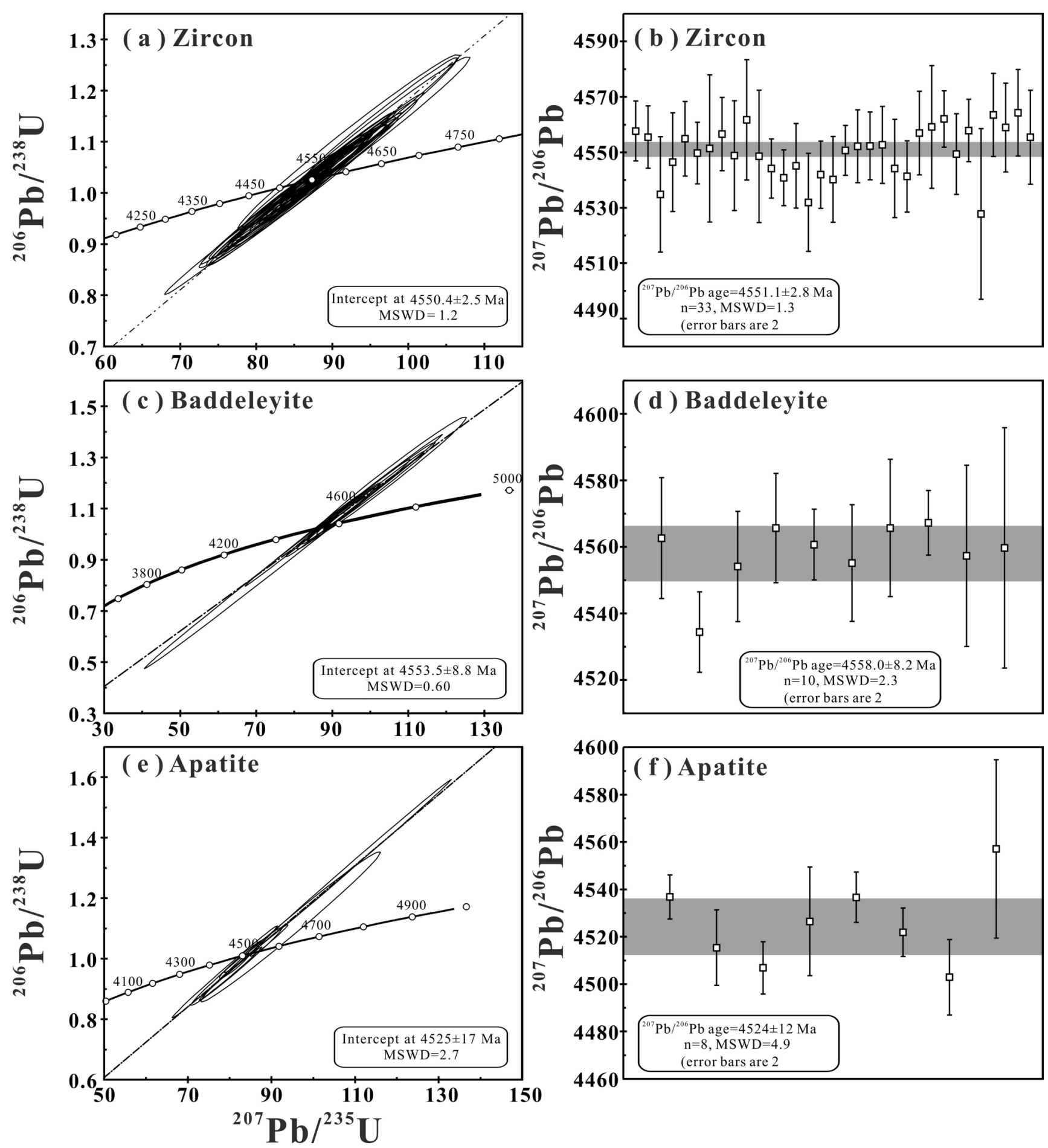

**Fig. 23.** Reflectance spectra of Sariçiçek. (A) Optical and near-IR wavelength range. From top to bottom, lines are results for a fine powder <25 micron, coarse powder < 125 micron, reflectance from the solid surface, and that of very coarse grains 125–500 micron; (B) Same for Mid-IR wavelength range; Band center (C) and band area ratio (D) of the measured reflectance spectra are compared to those of other eucrites and diogenites, as well as to that of the mean surface reflectance spectrum of Vesta itself (Reddy et al. 2015).

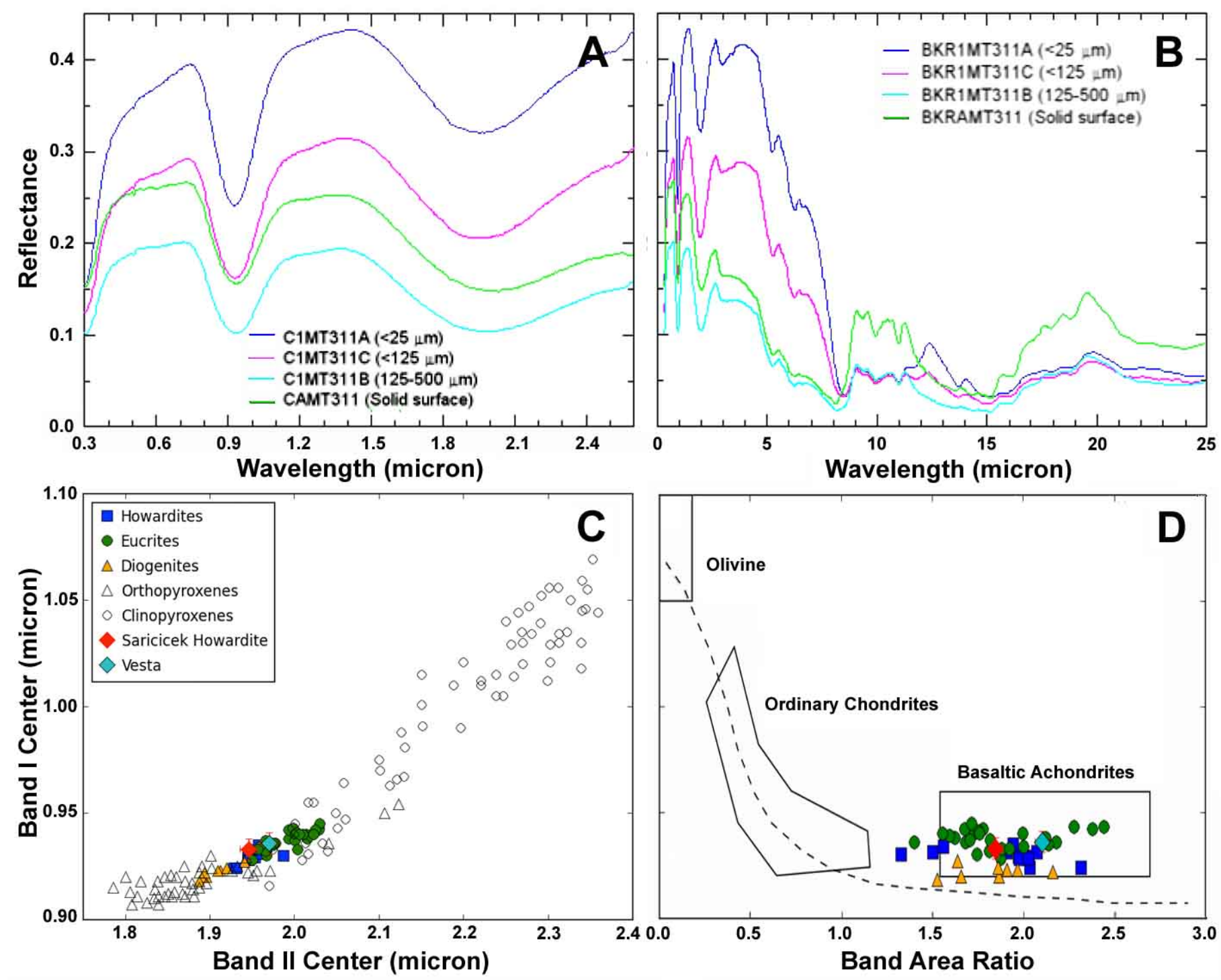

**Fig. 24.** Two sides of meteorite Sariçiçek SC26. (A) Back side with melt rim, regmaglypts, and a surface with a slight yellow hue. (B) Front side with flow lines and a more smooth surface with a slight reddish hue.

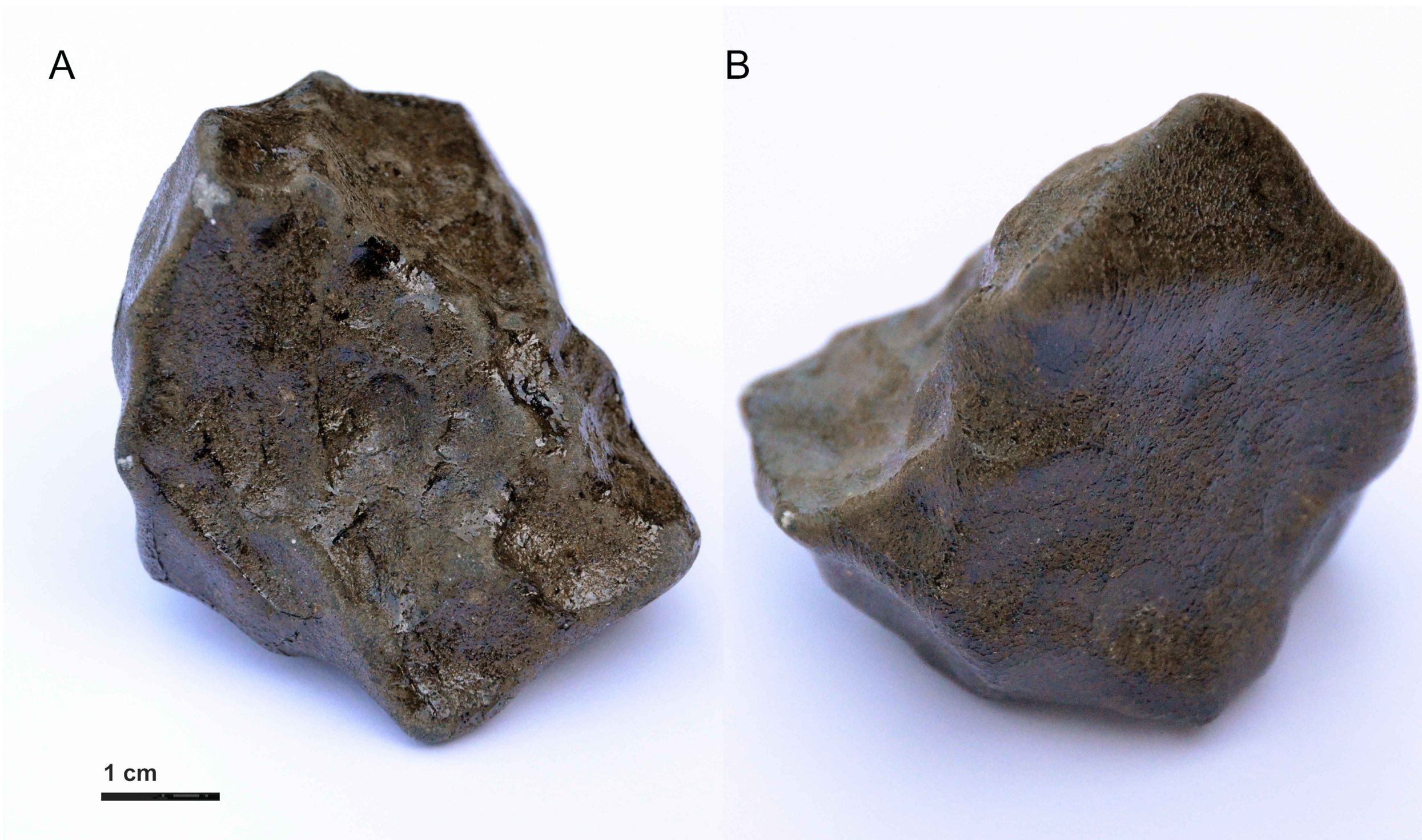

Fig. 25. (A) SEM image of the fusion crust of SC18 studied at Namik Kemal University, showing the trapping of vapor bubbles in the melt. The top 20 μm below the surface has 6 bubbles per 100 $\mu m^2$ with average diameter of 33 μm. The 145 μm layer below that (labeled "Viscous melt") has 80 bubbles per $\mu m^2$ with average diameter 25 μm; (B) Number density distribution of vesicle sizes on a logarithmic scale (right) and in terms of percentage of total void volume on a linear scale (left).

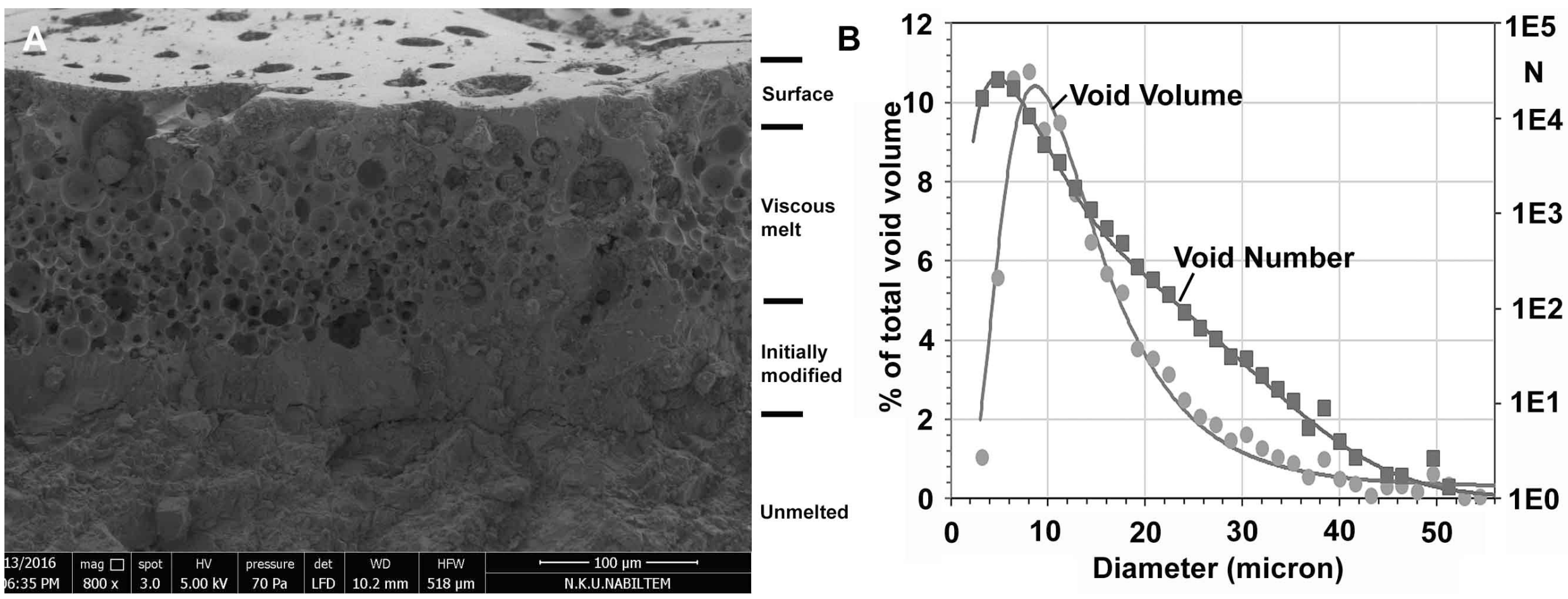

**Fig. 26.** Hydrodynamic model of meteoroid entry, with the meteoroid composed of four monoliths embedded in weaker material.

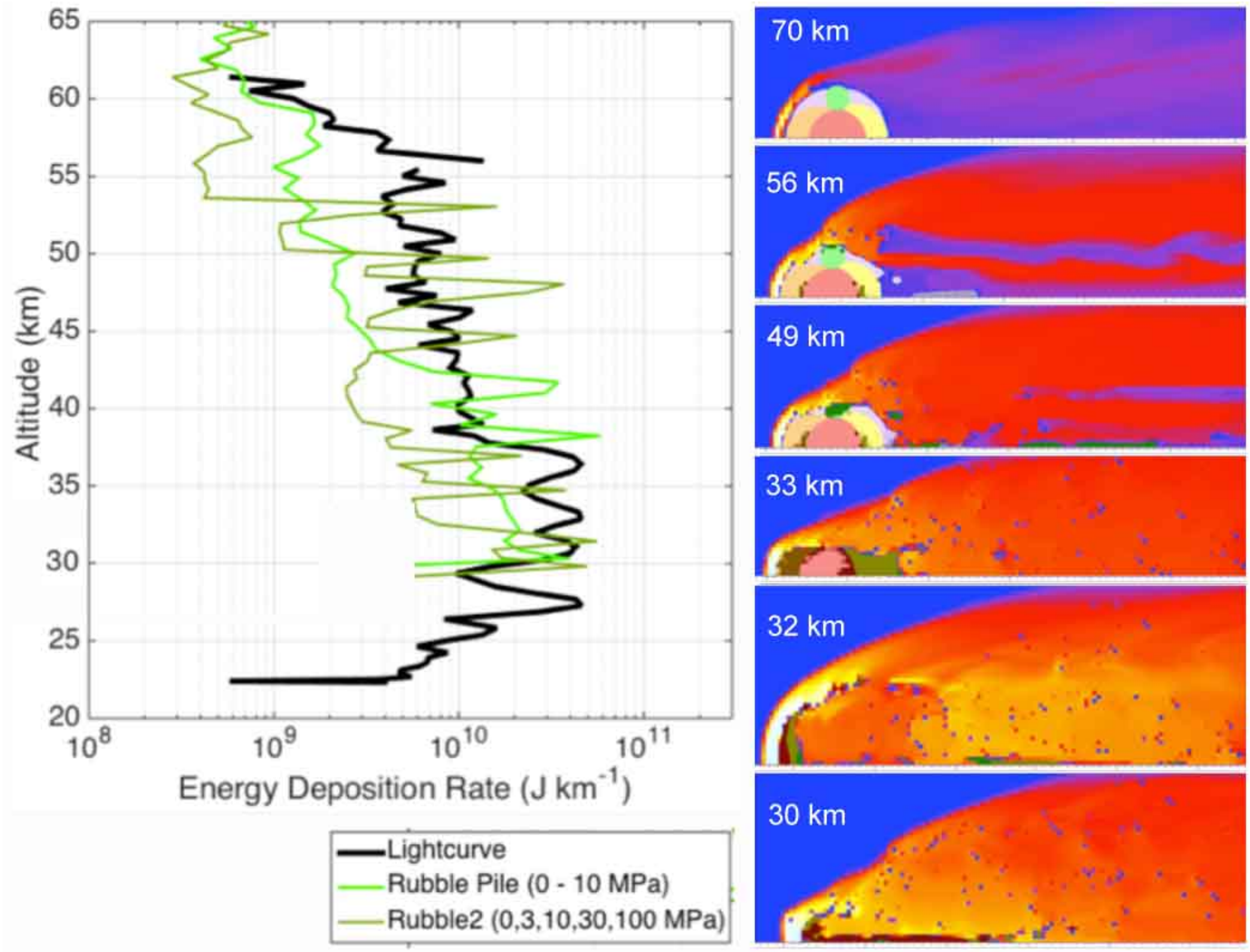

**Fig. 27.** Comparison of Sariçiçek with Chelyabinsk using a rubble pile model containing materials of 3, 10, 30, and 100 MPa strengths.

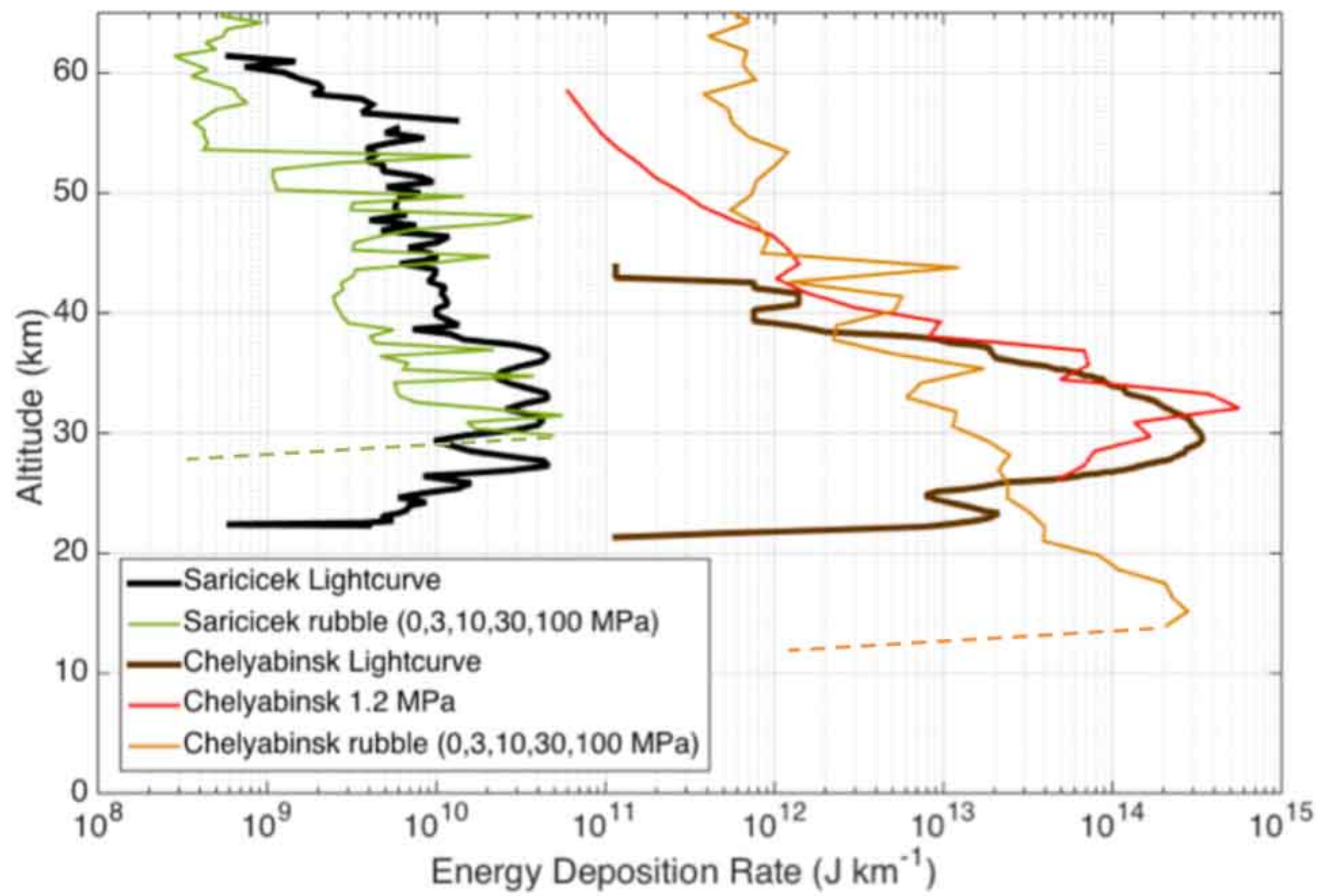

**Fig. 28.** The Sariçiçek bolide and airburst. The fireball trajectory with energy deposition rate on a linear scale (white), with flares at 36.5, 33.0, 31.0 and 27.4 (± 1.2) km altitude, and a model of the resulting airburst 198 s after passing 27.4 km. Top diagram shows the distribution of airburst kinetic energy, while the bottom diagram shows the horizontal wind velocity at that same time (light gray to dark: 0.2 to 2 cm $s^{-1}$).

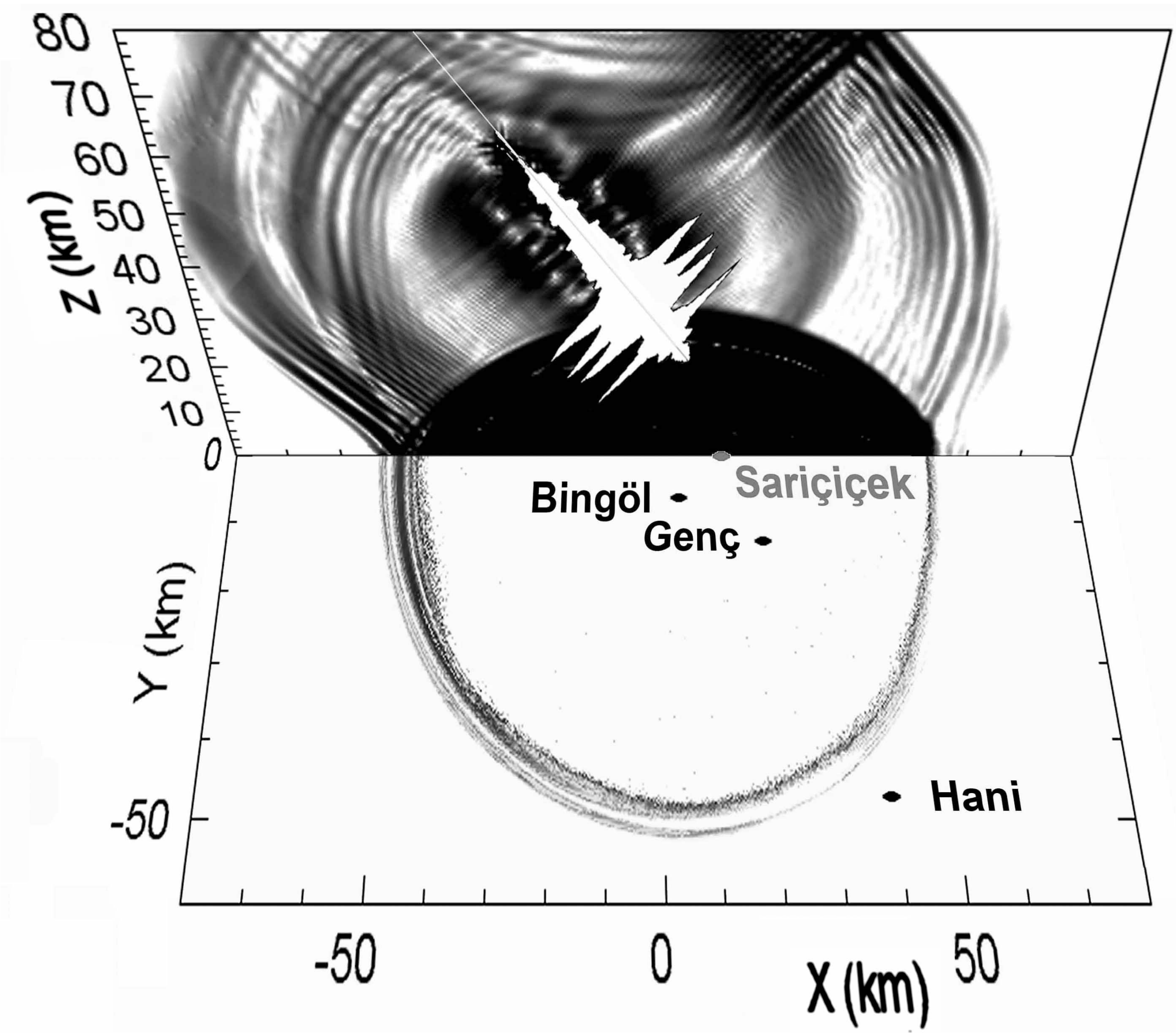

**Fig. 29.** Results from dynamical simulations: fraction of near-Earth Objects of various absolute magnitude (H) originating from given source regions for objects approaching Earth's path on a Sariçiçek-like orbit.

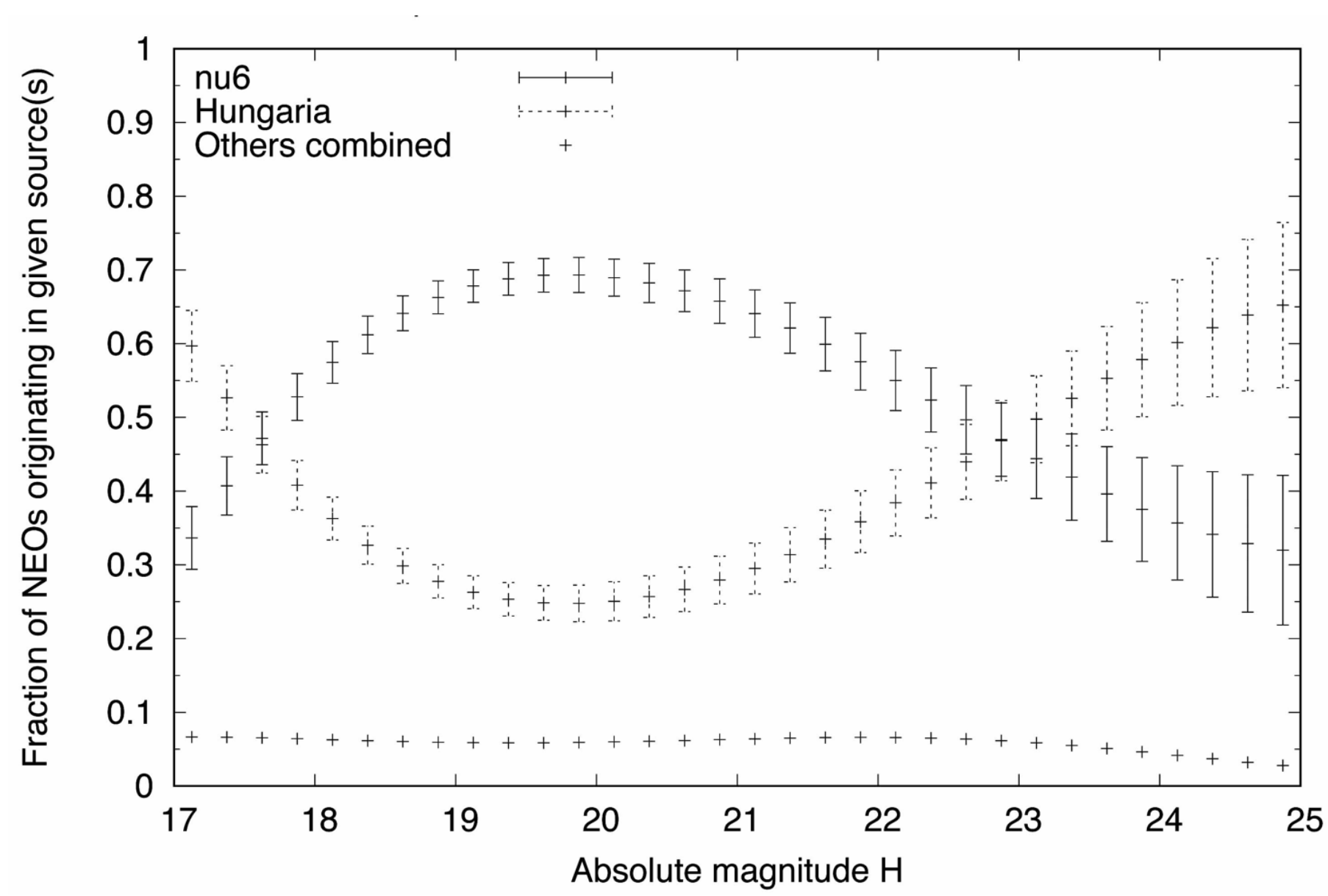

**Fig. 30.** Vesta crater formation ages from data by Kneissl et al. (2014), Krohn et al. (2014), Ruesch et al. (2014), Williams et al. (2014a), and Garry et al. (2014). Results for the lunar-based crater count calibration by Schmedemann et al. (2014) are shown as black dots, while grey dots show those ages in the asteroid-based chronology by Marchi et al. (2012) and O'Brien et al. (2014). Formation ages are "too young" for ejecta to have had time to reach Earth, or "too old" to account for measured HED CRE ages.

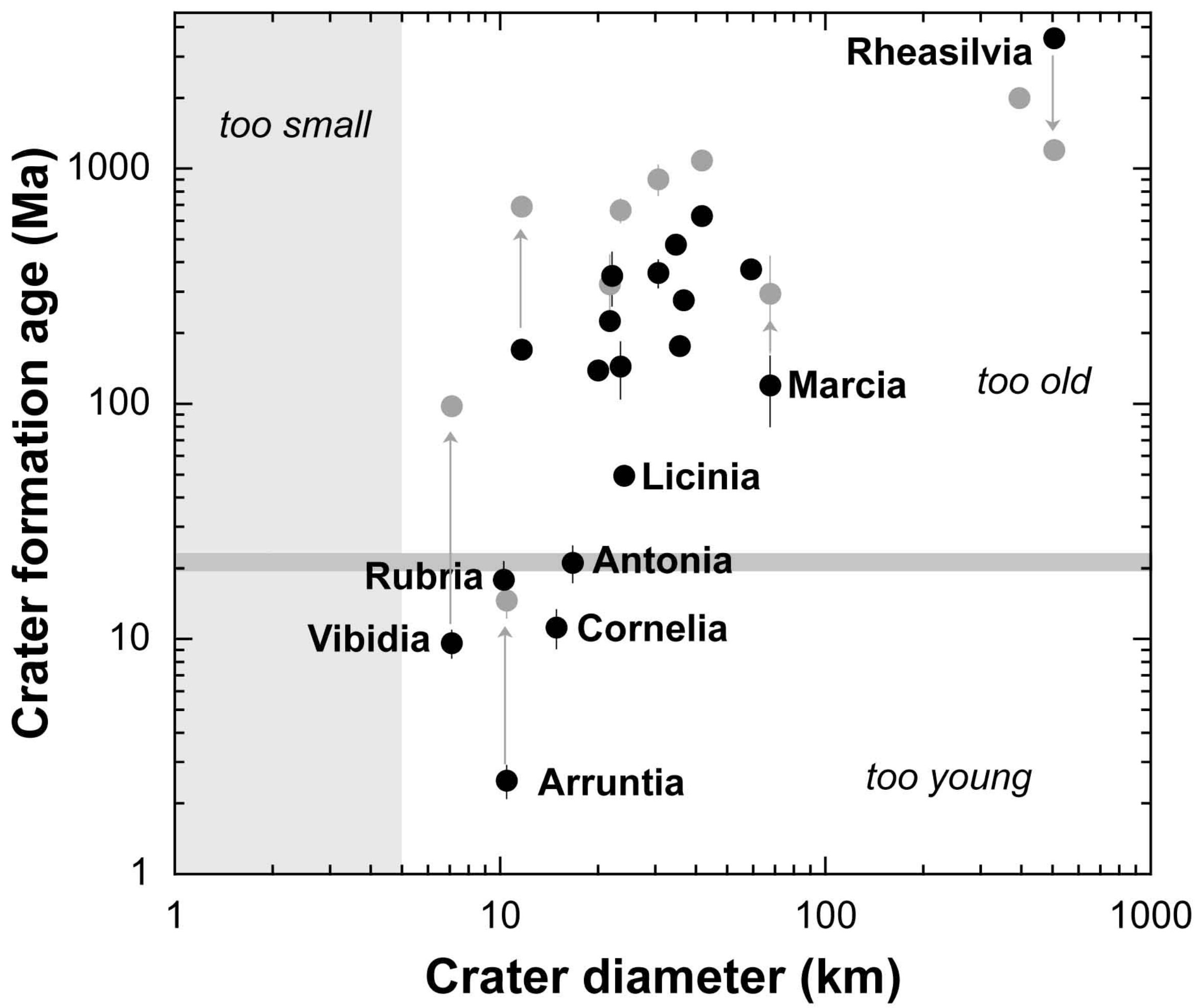

**Fig. 31.** (A) Cosmic-ray exposure age distribution of non-anomalous HED meteorites (Welten et al. 2012; Cartwrigth et al. 2014) and formation age of the Antonia impact crater. (B) The Rheasilvia impact basin (top, colors indicative of topography), with location of Antonia at tip of arrow. Bottom figure colors are indicative of terrain materials: Green = deep 0.90 µm pyroxene absorption band, i.e., rich in diogenites; Blue = blue tilted slope in 0.44–0.75 µm wavelength range, i.e., similar to eucrites and howardites with some carbonaceous chondrite component; Red = red tilted slope, i.e., reddest colors are possibly impact melt (Le Corre et al., 2013).

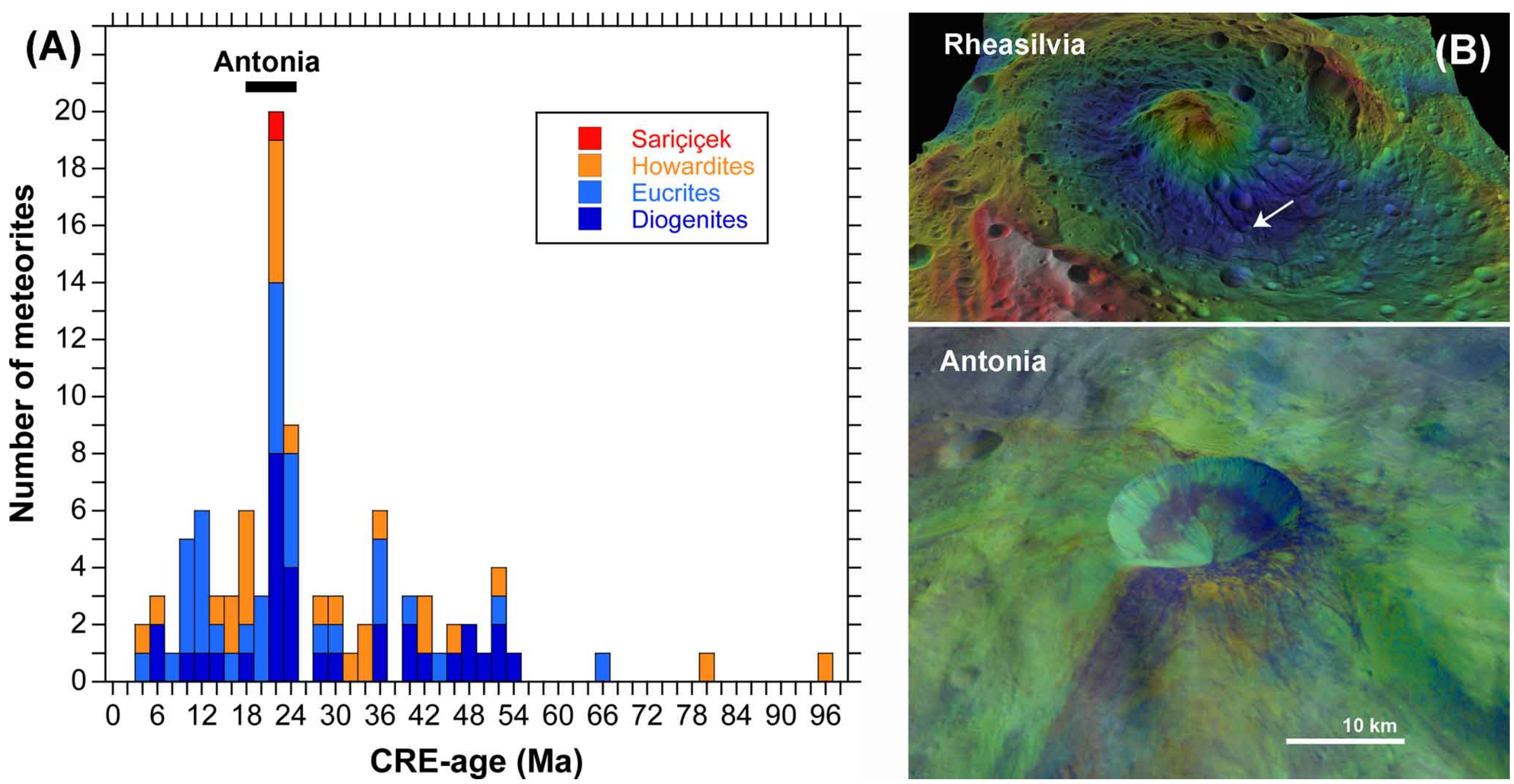

**Fig. 32.** U,Th-He age of SC12 and SC14 as a function of the U concentration in the measured aliquot, compared to the Ar-Ar age distribution of Bogard (2011) and the Antonia terrain age from Kneissl et al. (2014).

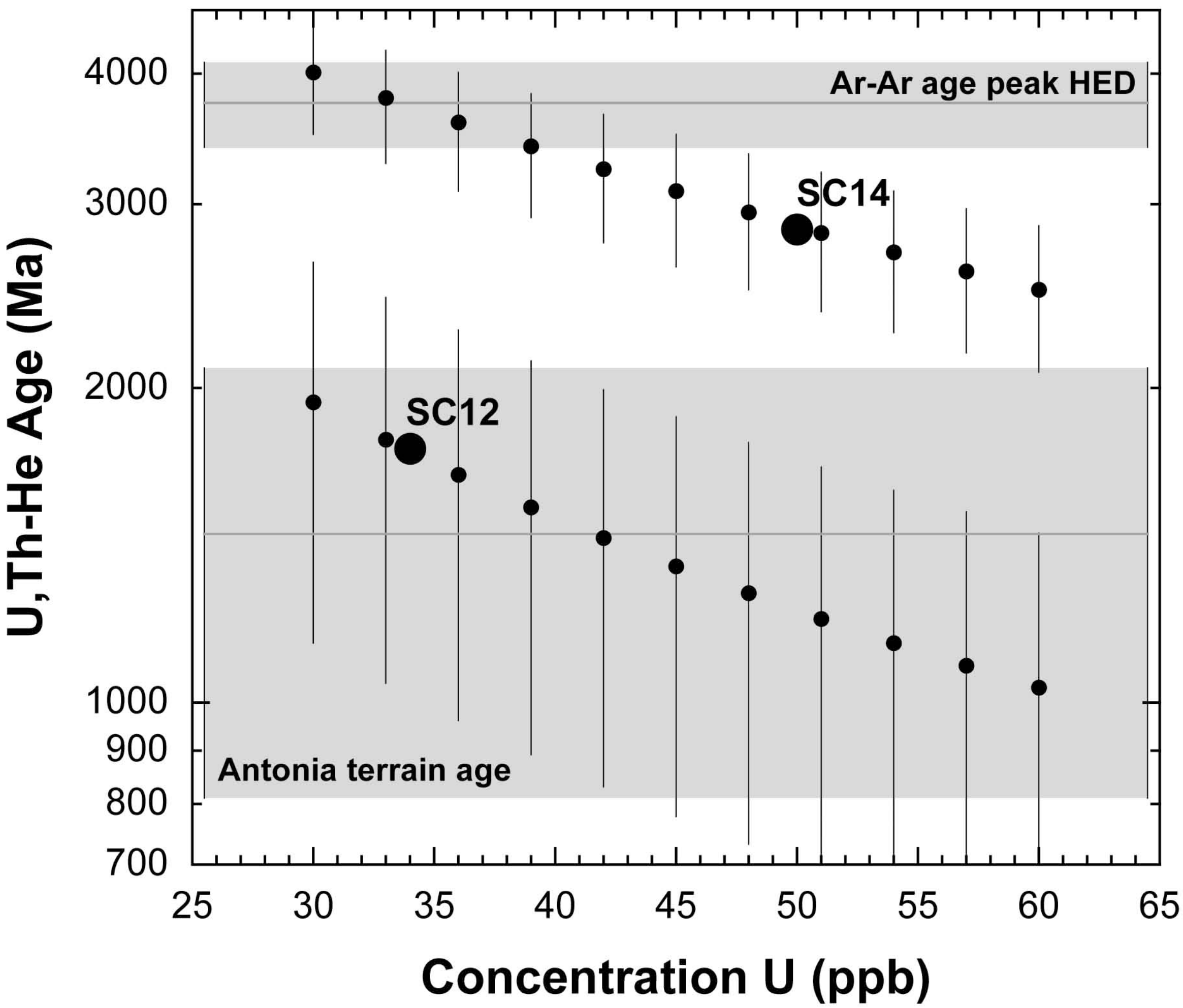